\documentclass[%
 reprint,
 superscriptaddress,
 amsmath, amssymb,
 aps, pra,
]{revtex4-2}

\usepackage{graphicx}% Include figure files
\usepackage{bm}% bold math
\usepackage[
 colorlinks=true,
 allcolors=blue,
 breaklinks=true,
]{hyperref}

\usepackage{dcolumn}
\usepackage[T1]{fontenc}

\begin{document}

\title{Stabilizer Entanglement Distillation and Efficient Fault-Tolerant Encoders}

\author{Yu Shi}
\email{shiyuphys@gmail.com}
\affiliation{Wyant College of Optical Sciences, University of Arizona, Tucson, Arizona 85721, USA}
\affiliation{NSF-ERC Center for Quantum Networks, University of Arizona, Tucson, Arizona 85721, USA}
\affiliation{Department of Electrical and Computer Engineering, University of Maryland, College Park, Maryland 20742, USA}

\author{Ashlesha Patil}
\affiliation{Wyant College of Optical Sciences, University of Arizona, Tucson, Arizona 85721, USA}
\affiliation{NSF-ERC Center for Quantum Networks, University of Arizona, Tucson, Arizona 85721, USA}

\author{Saikat Guha}
\email{saikat@umd.edu}
\affiliation{Wyant College of Optical Sciences, University of Arizona, Tucson, Arizona 85721, USA}
\affiliation{NSF-ERC Center for Quantum Networks, University of Arizona, Tucson, Arizona 85721, USA}
\affiliation{Department of Electrical and Computer Engineering, University of Maryland, College Park, Maryland 20742, USA}

\date{\today}

\begin{abstract}

Entanglement is essential for quantum information processing, but is limited by noise. We address this by developing high-yield entanglement distillation protocols with several advancements. (1) We extend the $2$-to-$1$ recurrence entanglement distillation protocol to higher-rate $n$-to-$\left(n-1\right)$ protocols that can correct any single-qubit errors. These protocols are evaluated through numerical simulations focusing on fidelity and yield. We also outline a method to adapt any classical error-correcting code for entanglement distillation, where the code can correct both bit-flip and phase-flip errors by incorporating Hadamard gates. (2) We propose a constant-depth decoder for stabilizer codes that transforms logical states into physical ones using single-qubit measurements. This decoder is applied to entanglement distillation protocols, reducing circuit depth and enabling protocols derived from high-performance quantum error-correcting codes. We demonstrate this by evaluating the circuit complexity for entanglement distillation protocols based on surface codes and quantum convolutional codes. (3) Our stabilizer entanglement distillation techniques advance quantum computing. We propose a fault-tolerant protocol for constant-depth encoding and decoding of arbitrary states in surface codes, with potential extensions to more general quantum low-density parity-check codes. This protocol is feasible with state-of-the-art reconfigurable atom arrays and surpasses the limits of conventional logarithmic depth encoders. Overall, our study integrates stabilizer formalism, measurement-based quantum computing, and entanglement distillation, advancing both quantum communication and computing.

\end{abstract}

\maketitle

\section{Introduction}

Entanglement is essential for quantum communication~\cite{Wehner2018, Awschalom2021, Azuma2023, Li2023}, quantum metrology~\cite{Giovannetti2011, Gottesman2012, Toth2014, Pezze2018}, and distributed quantum computing~\cite{Gottesman1999, Dur2003, Briegel2009, Li2012, Hu2023}, but it is severely limited by noise~\cite{Briegel1998, Dur1999, Jiang2009, Muralidharan2014, Azuma2015}. Entanglement distillation~\cite{Bennett1996, Bennett1996a, Deutsch1996, Verstraete2001}, or entanglement purification, is a process that transforms previously shared, less entangled Bell states into fewer, but higher-quality states through local operations and classical communication. This technique has been explored across various physical platforms~\cite{Duan2000, Pan2001, Zwerger2012, Zwerger2013, Yan2023}, demonstrated experimentally~\cite{Pan2003, Ourjoumtsev2007, Takahashi2010, Kim2012, Kalb2017, Hu2021, Ecker2021, Huang2022, Liu2022, Yan2022, Cai2024}, and extended to multiqubit systems~\cite{Murao1998, Dur2003a, Aschauer2005a, Hein2006, Kruszynska2006, Sheng2015, deBone2020, Zhou2021, Miguel2023, Vandre2023}. However, the basic $2$-to-$1$ protocols~\cite{Bennett1996, Deutsch1996} are inefficient, invariably sacrificing at least half of the input states, regardless of their initial fidelity. The hashing protocol~\cite{Bennett1996a, Zwerger2014}, while offering adaptive yields based on the input fidelity, requires an infinitely large input, limiting its practical use. This highlights the need for more efficient entanglement distillation protocols capable of generating higher yields from finite input states.

\begin{figure*}[tb]
    \centering
    \includegraphics[width=0.6\paperwidth]{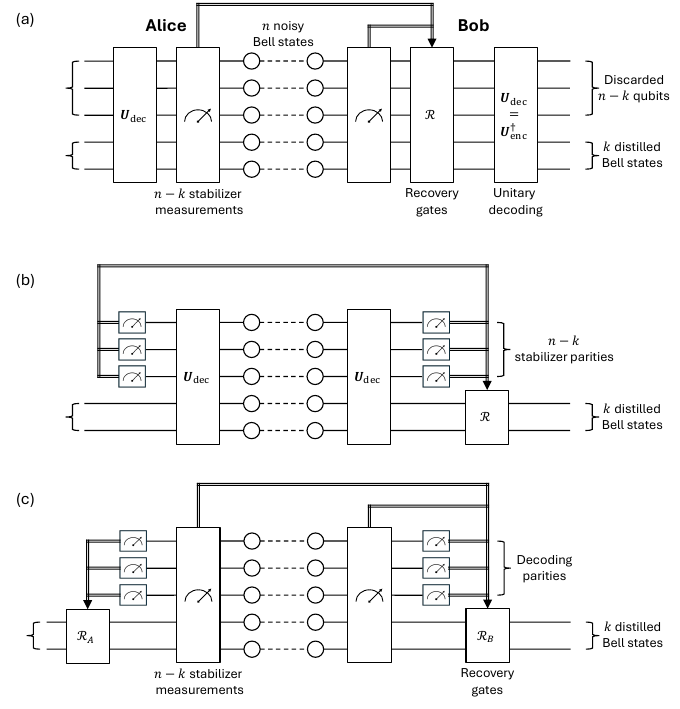}
    \caption{Schematic diagrams of entanglement distillation protocols derived from quantum codes. (a) A complete stage of stabilizer entanglement distillation~\cite{Wilde2007_Arxiv}, featuring stabilizer measurements, recovery by Pauli operators, and unitary decoding. (b) An alternative protocol that eliminates stabilizer measurements~\cite{Aschauer2005, Dur2007}. (c) Our proposed protocol, incorporating stabilizer measurements and single-qubit measurement decoding.}
    \label{fig:protocols}
\end{figure*}

Research on efficient distillation protocols has explored strategies such as high-dimensional systems~\cite{Horodecki1999, Alber2001, Vollbrecht2003, Cheong2007, Sheng2010, Miguel2018}, entanglement-assisted methods~\cite{Vollbrecht2005, Luo2007, Wilde2010, Riera2021, Riera2021a}, symmetry-based searches~\cite{Dehaene2003, Jansen2022}, and heuristic optimizations~\cite{Rozpedek2018, Krastanov2019, Wallnofer2020, Zhao2021, Krastanov2021, Goodenough2024, Addala2025}. A key approach utilizes quantum error correction~\cite{Aschauer2005, Dur2007, Fujii2009, Ruan2018, Jing2020, Rengaswamy2023, Rengaswamy2024}, which provides reliable performance guarantees. Specifically, in the so-called ``stabilizer protocol'' of entanglement distillation~\cite{Wilde2007_Arxiv}, Alice and Bob start with $n$ shared noisy Bell states. As shown in Figure~\ref{fig:protocols}(a), each party independently measures $n-k$ stabilizer generators on their $n$ qubits, projecting them onto logical subspaces. Alice sends her stabilizer parities to Bob, who identifies an error syndrome and corrects the errors by applying appropriate Pauli operators. After error correction, both parties perform unitary decoding, the inverse of the unitary encoding circuit, to transform the $k$ logical Bell states into $k$ physical Bell states. Crucially, it was shown that the same can be accomplished by a direct unitary decoding applied to both Alice's and Bob's qubits of the $n$ noisy Bell states~\cite{Aschauer2005, Dur2007}, as illustrated in Figure~\ref{fig:protocols}(b). This is followed by each party performing $n-k$ single-qubit measurements to reveal stabilizer parities, Alice communicating her parities to Bob, and Bob performing a recovery operation to his $k$ unmeasured qubits, resulting in $k$ distilled physical Bell states. This approach makes initial stabilizer measurements unnecessary. However, stabilizer measurements exhibit a more compact structure than unitary decoding circuits, particularly in quantum low-density parity-check (QLDPC) codes~\cite{Breuckmann2021}, with surface codes being a prominent example ~\cite{Dennis2002, Kitaev2003, Bravyi1998}. These codes feature a constant number of qubits per stabilizer generator and a constant number of generators acting on each qubit, enabling constant-depth stabilizer measurements. The primary challenge lies in developing an efficient entanglement distillation scheme that leverages the compactness of stabilizer measurements.

In this paper, we explore high-yield entanglement distillation protocols and their implementation through stabilizer measurements and efficient decoding via single-qubit measurements. We also extend their application to fault-tolerant quantum computing. The main contributions of this study are summarized as follows:
\begin{enumerate}
    \item We review stabilizer entanglement distillation and apply it to the protocols of Bennett et al.~\cite{Bennett1996} and Deutsch et al.~\cite{Deutsch1996}, mapping these protocols onto a $\left[2,1,2\right]$ classical error detection code. We demonstrate that bilateral Hadamard gates on the distillation output can alternate Pauli-$X$ and Pauli-$Z$ errors while preserving the standard Bell states. Consequently, two sequential applications of these protocols can effectively address any single-qubit errors. Drawing on this insight, we extend the $2$-to-$1$ protocol to higher-rate $n$-to-$\left(n-1\right)$ protocols, corresponding to an error detection process using a $\left[\!\left[n^2,\left(n-1\right)^2,2\right]\!\right]$ quantum code. We evaluate the performance of these protocols through numerical simulations, assessing both input-output fidelity and yield. Additionally, we present a method to adapt any classical error-correcting code to an entanglement distillation protocol using a two-iteration process. The first iteration corrects bit-flip errors, then Hadamard gates convert phase-flip errors to bit-flip errors, which the second iteration corrects. Since the errors in both iterations are classical, the error-correcting capability matches that of the classical code used.

    \begin{table*}
        \caption{\label{tab:general_code_comparison} Comparison of circuit complexity between unitary and measurement-based protocols for stabilizer entanglement distillation. The unitary decoding circuit for general codes is from~\cite{Gottesman1997}; the lower bound for surface codes is discussed in~\cite{Bravyi2006, Aharonov2018}.}
        \begin{ruledtabular}
        \begin{tabular}{lcc}
         & \multicolumn{2}{c}{Circuit Depth\footnote{We measure circuit depth by counting only the two-qubit gates, as these are the key determinants of circuit complexity.}} \\
         & Unitary Protocol & Measurement-Based Protocol \\
        \hline
        General Code & $O\left(n\right)$ & $O\left(n\right)$ \\
        Surface Code & $\Omega\left(\log n\right)$ & $O\left(1\right)$ \\
        \end{tabular}
        \end{ruledtabular}
    \end{table*}

    \begin{table*}
        \caption{\label{tab:convolutional_code_comparison} Comparison of unitary and measurement-based methods for a rate-$\frac{1}{3}$ quantum convolutional code~\cite{Forney2005, Forney2007}. The unitary decoding circuit is sourced from~\cite{Grassl2006}. Numbers represent circuit complexity on a single side, either Alice or Bob, in the entanglement distillation protocol. To our knowledge, there is no established general scaling for circuit complexity in unitary decoding of quantum convolutional codes.}
        \begin{ruledtabular}
        \begin{tabular}{lcc}
         & Unitary Protocol & Measurement-Based Protocol \\
        \colrule
        Circuit Depth & 11 & 4 \\
        Two-Qubit Gates per Frame\footnote{A ``frame'' refers to a set of qubits processed together at one stage of a quantum circuit, akin to windows or blocks of data in classical convolutional coding.} & 14 & 12 \\
        Longest Span of Two-Qubit Gates & 3 frames & 2 frames \\
        \end{tabular}
        \end{ruledtabular}
    \end{table*}

    \item We propose a constant-depth decoder for stabilizer codes using single-qubit measurements, and apply it to stabilizer entanglement distillation. Our protocol, illustrated in Figure~\ref{fig:protocols}(c), involves Alice and Bob performing stabilizer measurements and single-qubit measurement decoding. This compact, parallel approach significantly reduces circuit depth compared to previous protocols. Applying this method to convolutional entanglement distillation~\cite{Wilde2010}, known for its robust error correction in communications, we demonstrate its efficiency with a rate-1/3 quantum convolutional code. Our method outperforms the conventional unitary method~\cite{Grassl2006, Houshmand2013}. For a detailed comparison of circuit complexities, see Tables~\ref{tab:general_code_comparison} and~\ref{tab:convolutional_code_comparison}.

    \item We propose a fault-tolerant protocol for constant-depth encoding and decoding of arbitrary quantum states, applicable to QLDPC codes and surface codes. The protocol incorporates stabilizer measurements and efficient decoding to establish physical-logical Bell states. We enhance fault tolerance through entanglement distillation of these Bell states and employ quantum teleportation for encoding and decoding quantum states. The protocol is feasible with current reconfigurable atom arrays~\cite{Bluvstein2024, Xu2024} and surpasses the conventional depth limits for surface codes, lower bounded by $\Omega\left(\log\left(n\right)\right)$~\cite{Bravyi2006, Aharonov2018}, where $n$ is the number of qubits.
\end{enumerate}
Consequently, this study integrates stabilizer formalism, measurement-based quantum computing, and entanglement distillation, strengthening the connections between quantum communication and computing, advancing these fields, and contributing to practical and efficient quantum information processing.

\section{Review of Stabilizer Entanglement Distillation}

\subsection{Error Model} \label{sec:error_model}

Entanglement distillation involves transforming noisy Bell states shared between two parties, Alice and Bob, into a smaller number of higher-quality Bell states. For clarity, we define the standard Bell state as
\begin{equation} \label{eq:std_Bell_state}
    \left|\Phi^+\right\rangle=\frac{1}{\sqrt2}\left(\left|00\right\rangle+\left|11\right\rangle\right)\;,
\end{equation}
where $\left|xy\right\rangle=\left|x\right\rangle_A\otimes\left|y\right\rangle_B$ represents qubit states on Alice's and Bob's sides, respectively. We model a noisy Bell state as a standard Bell state affected by a noisy quantum channel. This channel applies random errors to the state, denoted by $E_A\otimes E_B\left|\Phi^+\right\rangle$, where $E_{A\left(B\right)}\in\{I, X, Y, Z\}$ are operators acting on Alice’s (Bob’s) qubit. The Bell-state matrix identity (see Appendix~\ref{app:matrix_identity}),
\begin{equation*}
    \left(M\otimes I\right)\left|\Phi^+\right\rangle=\left(I\otimes M^T\right)\left|\Phi^+\right\rangle\;,
\end{equation*}
signifies that an arbitrary operator $M$ applied to Alice’s qubit of the standard Bell state $\left|\Phi^+\right\rangle$ is equivalent to its transpose applied to Bob’s qubit. This allows us to simplify the error model by assuming that only Bob’s qubit is subjected to a noisy channel, while Alice's qubit remains error-free (see Appendix~\ref{app:matrix_identity} for a detailed calculation).
For our model, we consider a depolarizing channel that applies Pauli errors randomly with equal probability. The resulting noisy Bell state is given by
\begin{widetext}
\begin{equation} \label{eq:noisy_Bell_state}
    \rho=\left(1-p\right)\left|\Phi^+\right\rangle\left\langle\Phi^+\right|+\frac{p}{3}\sum_{i=1}^3\left(I\otimes \sigma_i\right)\left|\Phi^+\right\rangle\left\langle\Phi^+\right|\left(I\otimes \sigma_i^\dag\right)\;,
\end{equation}
\end{widetext}
where $p$ is the probability of error, and $\sigma_i\in\{X, Y, Z\}$ is the Pauli operator. This state is equivalent to a mixture of the standard Bell state $\left|\Phi^+\right\rangle$ with the other three orthogonal Bell states.

We study entanglement distillation protocols that distill $n$ independent and identically distributed (i.i.d.) noisy Bell states, represented by $\rho^{\otimes n}$, into $k$ higher-quality states. The initial state can be viewed as a mixture of pure states, represented by
\begin{equation*}
    \rho^{\otimes n} = \sum_{E} \left( I \otimes E \right) \left| \Phi_n^+ \right\rangle \left\langle \Phi_n^+ \right| \left( I \otimes E^\dag \right)\;,
\end{equation*}
where $\left| \Phi_n^+ \right\rangle = \left| \Phi^+ \right\rangle^{\otimes n}$ denotes the tensor product of $n$ standard Bell states. Here, $I$ is the identity operator acting on Alice’s qubits, and $E$ represents an $n$-qubit Pauli operator, expressed as $E=E_1\otimes E_2\otimes\cdots\otimes E_n$, with each $E_i\in\{I, X, Y, Z\}$ acting on Bob’s qubits indexed from $1$ to $n$. The occurrence of $E=I^{\otimes n}$ indicates an error-free event, while $E_i\neq I$ for any $i$ signifies an error on the $i$-th qubit, marking an erroneous event. For brevity, we omit the $\otimes$ symbol for one-partite tensor products in subsequent discussions, for example, writing $E=E_1E_2\cdots E_n$.

\subsection{Entanglement Distillation Protocol} \label{sec:protocol}

To facilitate entanglement distillation, we use a stabilizer code parameterized by $\left[\!\left[n,k,d\right]\!\right]$~\cite{Nielsen2010}. Here, $n$ represents the number of physical qubits, $k$ the number of logical qubits encoded, and $d$ the code distance, defined as the minimum weight of any logical operator. The code features $n-k$ stabilizer generators, which are essentially $n$-qubit Pauli operators, each denoted as $g_i$ for $i=1,2,\ldots,n-k$. These generators are independent and mutually commuting. If a state $\left|\psi\right\rangle$ is an eigenvector of the generator $g_i$, satisfying 
\begin{equation*}
    g_i\left|\psi\right\rangle=\left(-1\right)^{s_i}\left|\psi\right\rangle\;,
\end{equation*}
where $s_i=0$ or $1$, then we say $\left|\psi\right\rangle$ is stabilized by $g_i$ with parity $s_i$. A state stabilized by all generators with parities $\bm{s}=s_1s_2\ldots s_{n-k}$ is referred to as the logical state with parity vector $\bm{s}$. When an error affects the logical state, it flips the parities if it anti-commutes with the corresponding generators. Subsequent stabilizer measurements, characterized by the projector
\begin{equation*}
    P_i^{\left(s_i^\prime\right)}=\frac{I+\left(-1\right)^{s_i^\prime}g_i}{2}\;,
\end{equation*}
can reveal the new parities $\bm{s}^\prime=s_1^\prime s_2^\prime\ldots s_{n-k}^\prime$. By comparing the two sets of parities, $\bm{s}\oplus \bm{s}^\prime$ (bitwise XOR), one can determine the error syndrome and correct it using appropriate Pauli operators. A code with distance $d$ can correct up to $t=\left\lfloor\frac{d-1}{2}\right\rfloor$ physical errors. This capability makes quantum error-correcting codes essential for distilling entanglement, as they can effectively handle up to $t$ errors in the ensemble $\rho^{\otimes n}$.

Considering an instance $I\otimes E\left|\Phi_n^+\right\rangle$ from the ensemble of $n$ noisy Bell states, the entanglement distillation protocol is outlined as follows:
\begin{enumerate}
    \item \textit{Alice performs $n-k$ stabilizer measurements on her qubits.}
    These measurements project the state onto $P_i^{\left(s_i\right)}\otimes E\left|\Phi_n^+\right\rangle$, where $P_i^{\left(s_i\right)}$ is the projector corresponding to the measurement outcome $s_i$. Utilizing operator commutation, projector property $P^2=P$, and Bell-state matrix identity associated with the standard Bell state $\left|\Phi^+\right\rangle$, the postmeasurement state can be represented as
    \begin{equation*}
    P_i^{\left(s_i\right)}\otimes E\left|\Phi_n^+\right\rangle=P_i^{\left(s_i\right)}\otimes E\left(P_i^{\left(s_i\right)}\right)^T\left|\Phi_n^+\right\rangle\;.
    \end{equation*}
    Thus, Alice's measurements project both her qubits and Bob’s qubits onto subspaces stabilized by generators $g_i$ and $g_i^T$, respectively. If $g_i$ contains an odd number of Pauli-$Y$ operators, then $g_i^T=-g_i$; otherwise, $g_i^T=g_i$.

    \item \textit{Alice communicates her measurements, $\bm{s}_A=s_1s_2\ldots s_{n-k}$, to Bob.}
    This allows Bob to know the parities of his qubits before any errors. The parities are given by $\bm{s}_B=s_1^\prime s_2^\prime\ldots s_{n-k}^\prime$. If a generator $g_i$ contains an odd number of Pauli-$Y$ operators, then $s_i^\prime=s_i\oplus1$; otherwise, $s_i^\prime=s_i$.

    \item \textit{Bob performs $n-k$ stabilizer measurements on his qubits, yielding $\bm{t}=t_1t_2\ldots t_{n-k}$. He calculates the error syndrome, $\bm{c}=\bm{s}_B\oplus\bm{t}$, and corrects them with appropriate Pauli operators.}
    Bob’s measurement of $g_i$ with outcome $t_i$ projects the state onto
    \begin{equation*}
    P_i^{\left(s_i\right)}\otimes P_i^{\left(t_i\right)}EP_i^{\left(s_i^\prime\right)}\left|\Phi_n^+\right\rangle\;.
    \end{equation*}
    If $E$ commutes with $g_i$, then $t_i=s_i^\prime$; otherwise, $t_i=s_i^\prime\oplus1$. This step mirrors the process used in quantum error correction.
    
    \item \textit{Alice and Bob decode the logical qubits and extract $k$ higher-quality Bell states.}
\end{enumerate}

This one-way protocol allows Alice and Bob to always distill $k$ Bell states from $n$ inputs. Alternatively, the stabilizer code in error detection mode enables two-way entanglement distillation~\cite{Aschauer2005, Dur2007}. Bob compares his parities with Alice’s. If the parities match, the Bell states are kept; otherwise, they are discarded. The two-way scheme can address up to $d-1$ errors for a code with distance $d$, surpassing the one-way scheme's capability of addressing up to $\left\lfloor\frac{d-1}{2}\right\rfloor$ errors. However, the two-way scheme is probabilistic and requires Bob to communicate his decision back to Alice.

\section{Application I: Extended Recurrence Entanglement Distillation Protocol}

\begin{figure*}[tb]
    \centering
    \includegraphics[width=0.6\paperwidth]{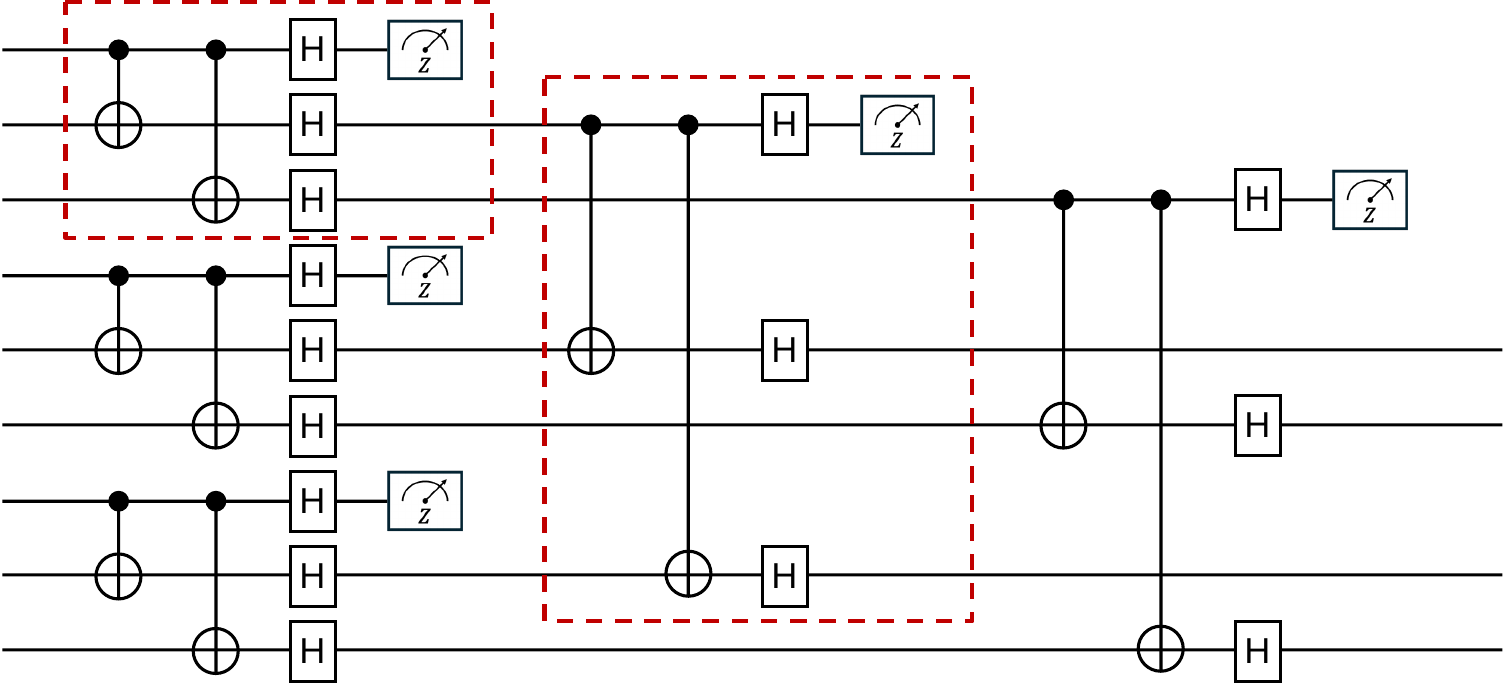}
    \caption{Circuit on Alice's (or Bob's) side for processing Bell states through two iterations in the $3$-to-$2$ recurrence entanglement distillation. Dashed boxes indicate the first and second iterations.}
    \label{fig:circuit_3to2}
\end{figure*}

We apply the stabilizer formalism to the well-known 2-to-1 recurrence entanglement distillation. The protocol by Bennett et al.~\cite{Bennett1996} can be seen as a stabilizer entanglement distillation in error detection mode. 
The code uses a single generator, $X_1X_2$, which detects single-qubit $Y$ and $Z$ errors but not $X$ errors. To address $X$ errors, Bennett et al. used twirling to randomize them, enabling detection in later iterations. Alternatively, Alice and Bob can apply Hadamard gates to their output qubits. The bilateral Hadamard gates preserve the standard Bell state and switch $X$ and $Z$ errors, as
\begin{eqnarray*}
\left(H\otimes H\right)\left|\Phi^+\right\rangle&=&\left|\Phi^+\right\rangle\;,\\
\left(H\otimes H\right)\left(I\otimes X\right)\left|\Phi^+\right\rangle&=&\left(I\otimes Z\right)\left|\Phi^+\right\rangle\;.
\end{eqnarray*}
This allows the remaining $Z$ errors to be addressed in the next iteration, which is more efficient than twirling. The revised protocol is equivalent to the protocol by Deutsch et al.~\cite{Deutsch1996}, where the stabilizer code $Y_1Y_2$ is used to detect $X$ and $Z$ errors, and $R_x\left(\pi/2\right)$ is used to switch $Y$ and $Z$ errors. Since this protocol can detect $X$, $Y$, and $Z$ errors over two sequential iterations, it can detect any single-qubit error according to the discretization of quantum errors~\cite{Nielsen2010}. This explains why the recurrence protocols are effective in various error models, not just the Werner-state model.

Based on the above insight, we can extend the $2$-to-$1$ recurrence protocol to a higher-rate $n$-to-$\left(n-1\right)$ protocol using a stabilizer code $X_1X_2\cdots X_n$. For example, consider the $3$-to-$2$ entanglement distillation circuit illustrated in Figure~~\ref{fig:circuit_3to2}, constructed using the unitary decoding method. The two dashed boxes represent circuits for two sequential iterations. In each box, the projective measurement on the first qubit reveals the parity of $X_1X_2X_3$. By comparing parities, Alice and Bob can detect single-qubit $Y$ or $Z$ errors. Hadamard gates on the remaining qubits convert $X$ errors to $Z$ errors, which are detected in the next iteration. The protocol is similar to the generalized Shor's construction of a quantum code~\cite{Bacon2006}, with one code addressing bit-flip errors and the other addressing phase-flip errors. By permuting the measurements and gates in Figure~\ref{fig:circuit_3to2}, we obtain the stabilizers for two iterations:
\begin{equation*}
\begin{array}{ccccccccc}
    X & X & X & I & I & I & I & I & I \\
    I & I & I & X & X & X & I & I & I \\
    I & I & I & I & I & I & X & X & X \\
    Z & Z & I & Z & Z & I & Z & Z & I \\
    Z & I & Z & Z & I & Z & Z & I & Z
\end{array}\;,
\end{equation*}
which characterize a $\left[\!\left[9,4,2\right]\!\right]$ quantum code capable of detecting a single-qubit error. For the $n$-to-$\left(n-1\right)$ protocol, two iterations correspond to a $\left[\!\left[n^2,\left(n-1\right)^2,2\right]\!\right]$ quantum code. The term ``$n$-to-$(n-1)$ protocol'' reflects that each iteration can be executed independently, requiring only $n$ inputs, corresponding to a classical $\left[n,n-1,2\right]$ error-detecting code. It corrects bit-flip or phase-flip errors, effectively improving output fidelity similarly to the original $2$-to-$1$ protocol.

To prevent correlated multiqubit errors, it is essential to use outputs from different iterations as inputs in subsequent iterations. For example, in Figure~\ref{fig:circuit_3to2}, an $X_1$ error on the control qubit can propagate through the CNOT gates to become $X_1X_2X_3$. Assuming a depolarizing channel with an error probability of $p$ per input qubit, the effective error probability in the first iteration is $\frac{2p}{3}$ due to $Y$ and $Z$ errors. In the second iteration, the effective error probability is also $\frac{2p}{3}$ due to undetected $X$ errors on the output and control qubits. Thus, if the input meets the fidelity threshold initially, it will meet the threshold in subsequent iterations. This result applies to other $n$-to-$\left(n-1\right)$ recurrence protocols as well.

We evaluate the performance of recurrence entanglement distillation protocols based on input-output fidelity and yield. Fidelity is defined as 
\begin{equation*}
    \mathcal{F}=\left\langle\Phi^+\middle|\rho\middle|\Phi^+\right\rangle\;,
\end{equation*}
where $\rho$ is a mixed Bell state. Yield is defined as
\begin{equation*}
    \mathcal{Y}=\frac{k\cdot P_S}{n}\;,
\end{equation*}
where $k$ and $n$ are the logical and physical qubits in an $\left[\!\left[n,k,d\right]\!\right]$ code, and $P_S$ is the success probability. Our simulation uses the QuantumClifford.jl package, which utilizes the stabilizer tableaux formalism~\cite{Aaronson2004}. To evaluate the performance of protocols capable of correcting all types of quantum errors, we simulate two sequential iterations per protocol, corresponding to an $n^2$-to-$\left(n-1\right)^2$ entanglement distillation that corrects both bit-flip and phase flip errors. We use Werner states as inputs, which are mixtures of a standard Bell state and a completely mixed state, as described by Equation~\ref{eq:noisy_Bell_state}. We assume error-free local quantum operations and classical communication, a valid assumption given that channel noise significantly exceeds local noise.

\begin{figure}[tb]
    \centering
    \includegraphics[width=0.4\paperwidth]{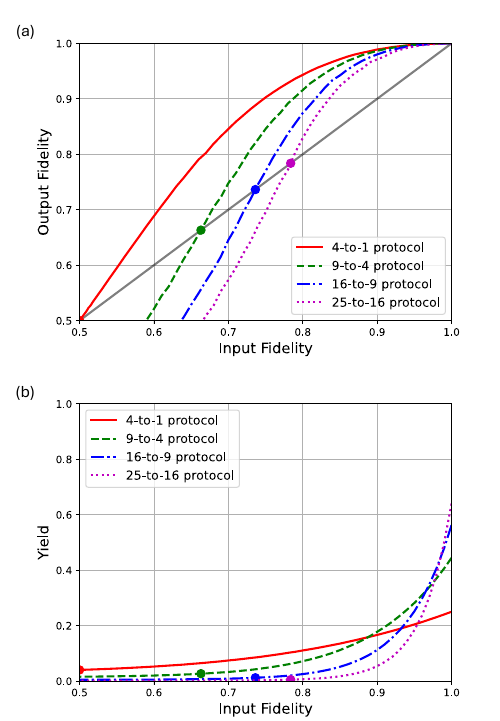}
    \caption{Input-output fidelity and yield of recurrence entanglement distillation protocols. Each undergoes two iterations, forming an $n^2$-to-$\left(n-1\right)^2$ protocol. Circles mark the threshold where output fidelity exceeds input fidelity.}
    \label{fig:performance}
\end{figure}

Figure~\ref{fig:performance}(a) shows the fidelity of distilled Bell states as a function of input fidelity for several distillation protocols. Circles mark the threshold where output fidelity exceeds input fidelity, which increases with the code rate. Figure~\ref{fig:performance}(b) shows yield as a function of input fidelity. The $4$-to-$1$ protocol has the highest yield for input fidelities from $0.5$ to $0.887$. Beyond this range, higher-rate protocols are superior. Near ideal fidelity, yield is determined by code rate. Achieving optimal yield at high fidelity requires large $n$ in $n^2$-to-$\left(n-1\right)^2$ protocols. In Appendix~\ref{app:adaptive_distillation}, we demonstrate that adaptively increasing the block size $n$ based on input fidelity allows for a constant overall yield rate for any desired fidelity.

We can further extend the recurrence protocol by adapting classical codes. In the first iteration, the code corrects bit-flip errors. Hadamard gates then convert the remaining phase-flip errors to bit-flip errors, which the code can correct in the second iteration. Since these are classical codes, the commutation condition for each stabilizer is automatically met. Moreover, each iteration is flexible in its choice of classical code, allowing customization based on the strength of phase-flip and bit-flip errors. For example, if phase-flip errors dominate, we can apply a high-rate classical code in the first (bit-flip) iteration and a lower-rate code with greater error-correcting capability in the second (phase-flip) iteration. This recurrence entanglement distillation supports the direct implementation of well-developed classical error-correcting codes, such as polar codes, LDPC codes, and turbo codes. Although the generalized Shor construction can build quantum codes~\cite{Bacon2006}, this method sacrifices certain desired properties. For example, combining two classical LDPC codes via Shor’s construction does not produce a quantum LDPC code~\cite{Tillich2014}. In contrast, our approach can apply classical LDPC codes in separate iterations, preserving the sparsity of parity-check matrices. While quantum LDPC codes can be constructed using the hypergraph product~\cite{Tillich2014}, the recurrence protocol simplifies the direct use of classical LDPC codes and can readily extend to other types of codes. In practice, noisy Bell states can be non-Werner~\cite{Moehring2007, Dhara2023, Shi2023}. Combining the stabilizer formalism with fidelity and yield aids in designing effective distillation protocols. 

\section{Decoding the Stabilizer Code by Single-Qubit Measurements} \label{sec:decoding}

We propose a decoder that transforms logical Bell states into physical Bell states using single-qubit measurements and conditional Pauli operators. Since stabilizer and single-qubit measurements can be performed in parallel, this significantly reduces the circuit depth for implementing stabilizer entanglement distillation protocols. We first outline how logical Bell states result from stabilizer measurements on physical Bell states, then explain decoding these logical states into physical states via single-qubit measurements. While presented here for entanglement distillation, this measurement-based decoder applies to any scenario involving decoding stabilizer codes, including quantum computing.

To develop the decoder, we represent the stabilizer code using a binary check matrix $\left(\bm{X}\middle|\bm{Z}\right)$~\cite{Nielsen2010}, where $\bm{X}$ and $\bm{Z}$ are submatrices of dimensions $\left(n-k\right)\times n$. The rows represent generators $g_1$ through $g_{n-k}$, and the columns of each submatrix correspond to the $n$ physical qubits. In the $i$th row, represented by vector $\left(\bm{x}_i\middle|\bm{z}_i\right)$, $x_{ij}=1$ indicates a Pauli-$X$ operator on the $j$th qubit, $z_{ij}=1$ indicates a Pauli-$Z$ operator, and $1$ on both indicates a Pauli-$Y$ operator. Two generators $g_i$ and $g_j$ commute if and only if their symplectic inner product is zero, given by
\begin{equation*}
    \bm{x}_i\cdot\bm{z}_j^T+\bm{z}_i\cdot\bm{x}_j^T=0\;.
\end{equation*}
Operations on the check matrix, such as swapping rows (relabeling generators), adding two rows (multiplying generators), and simultaneously swapping columns in both submatrices (relabeling qubits), preserve the properties of the stabilizer code.

There is a standard form for the stabilizer code~\cite{Gottesman1997, Nielsen2010}, derived by applying Gaussian elimination to the binary check matrix and represented by\vspace{5mm}
\begin{equation} \label{eq:standard_form}
\begin{array}{r} r\left\{\right. \\ n-k-r\left\{\right. \end{array}
\left[ \begin{array}{ccc|ccc}
\raisebox{0ex}[1.5ex]{$\overbrace{\bm{I}}^{r}$} & 
\raisebox{0ex}[1.5ex]{$\overbrace{\bm{A}_1}^{n-k-r}$} & 
\raisebox{0ex}[1.5ex]{$\overbrace{\bm{A}_2}^{k}$} & 
\raisebox{0ex}[1.5ex]{$\overbrace{\bm{B}}^{r}$} & 
\raisebox{0ex}[1.5ex]{$\overbrace{\bm{0}}^{n-k-r}$} & 
\raisebox{0ex}[1.5ex]{$\overbrace{\bm{C}}^{k}$} \\
\bm{0} & \bm{0} & \bm{0} & \bm{D} & \bm{I} & \bm{E}
\end{array} \right]\;.
\end{equation}
This matrix has a block structure with dimensions indicated. $\bm{I}$ is the identity matrix, and other submatrices are determined by Gaussian elimination. Using the rows of this check matrix as new generators, we define $k$ logical operators. The logical $X$ operators are given by
\begin{equation} \label{eq:logical_x}
    L_X = \left[ \begin{array}{ccc|ccc}
    \bm{0} & \bm{E}^T & \bm{I} & \bm{C}^T & \bm{0} & \bm{0}
    \end{array} \right]\;,
\end{equation}
where each submatrix has $k$ rows, with column sizes $\left(r,n-k-r,k\right)$ for both left and right blocks. The $i$th row, $L_X^i$, represents the $X$ operator for the $i$th logical qubit. The logical $Z$ operators are given by
\begin{equation} \label{eq:logical_z}
    L_Z = \left[ \begin{array}{ccc|ccc}
    \bm{0} & \bm{0} & \bm{0} & \bm{A}_2^T & \bm{0} & \bm{I}
    \end{array} \right]\;,
\end{equation}
which follows the same dimensions. The $i$th row, $L_Z^{(i)}$, represents the $Z$ operator for the $i$th logical qubit. These logical operators commute with the stabilizer generators. $X_L^{(i)}$ and $Z_L^{(i)}$ anti-commute, while $X_L^{(i)}$ and $Z_L^{(j)}$ commute for $i\neq j$.

The $n$ Bell states naturally form logical Bell states in the standard form. The standard Bell state, as given in Equation~\ref{eq:std_Bell_state}, is stabilized by operators $X\otimes X$ and $Z\otimes Z$, which act in the Hilbert space $H_A\otimes H_B$. The stabilizers for $n$ Bell states can be enumerated as
\begin{equation} \label{eq:Bell_states}
    \begin{matrix}
    X_1\otimes X_1\;, & Z_1\otimes Z_1 \\
    X_2\otimes X_2\;, & Z_2\otimes Z_2 \\
    \vdots & \vdots \\
    X_n\otimes X_n\;, & Z_n\otimes Z_n \\
    \end{matrix}\;,
\end{equation}
where subscripts denote qubit indices. Aligning these Bell stabilizers with the generators of a quantum code, we have the following observations based on Equations~\ref{eq:standard_form},~\ref{eq:logical_x}, and~\ref{eq:logical_z}:
\begin{enumerate}
    \item \label{ob:1} The tensor product of code generators $g_i\otimes g_i$ commutes with all Bell stabilizers $X_j\otimes X_j$ and $Z_j\otimes Z_j$.
    
    \item \label{ob:2} Among the first $r$ generators, $g_i$ is independent of $X_j$ for $j<i$. For the remaining $n-k-r$ generators, $g_i$ is independent of all $X$s and independent of $Z_j$ for $r<j<i$.
    
    \item \label{ob:3} The logical operator $X_L^{(i)}$ is independent of $X_j$ for $j\le r$ and $n-k<j<n-k+i$, and independent of $Z_j$ for $r<j\le n$. The logical operator $Z_L^{(i)}$ is independent of all $X$s and independent of $Z_j$ for $r<j<n-k+i$.

    \item \label{ob:4} Among the first $r$ generators, $g_i$ anti-commutes with $Z_i$ but commutes with all other $Z_j$ for $j\le r$ and $j\neq i$. For the remaining $n-k-r$ generators, $g_i$ anti-commutes with $X_i$ but commutes with all other $X_j$ for $r<j\le n-k$ and $j\neq i$.
\end{enumerate}
Observation~\ref{ob:1} allows for an equivalent representation of the $n$ Bell states using the code generators. Multiplying the Bell stabilizers in Equation~\ref{eq:Bell_states} according to the standard form yields
\begin{equation} \label{eq:pre-meas_Bell_states}
    \begin{matrix}
    g_1 \otimes g_1\;, & Z_1 \otimes Z_1 \\
    \vdots & \vdots \\
    g_r \otimes g_r\;, & Z_r \otimes Z_r \\
    X_{r+1} \otimes X_{r+1}\;, & g_{r+1} \otimes g_{r+1} \\
    \vdots & \vdots \\
    X_{n-k} \otimes X_{n-k}\;, & g_{n-k} \otimes g_{n-k} \\
    X_L^{(1)} \otimes X_L^{(1)}\;, & Z_L^{(1)} \otimes Z_L^{(1)} \\
    \vdots & \vdots \\
    X_L^{(k)} \otimes X_L^{(k)}\;, & Z_L^{(k)} \otimes Z_L^{(k)} \\
    \end{matrix}\;.
\end{equation}
Observation~\ref{ob:2} enables the transformation of $X_i\otimes X_i$ into $g_i\otimes g_i$ for $i=1$ to $r$, and $Z_i\otimes Z_i$ into $g_i\otimes g_i$ for $i=r+1$ to $n-k$. Observation~\ref{ob:3} enables the transformation of $X_i\otimes X_i$ and $Z_i\otimes Z_i$ into logical operators for $i=n-k+1$ to $n$.
During the derivation, we assume each $g_i$ contains an even number of $Y$s, resulting in a bilateral stabilizer of $g_i\otimes g_i$. For cases with an odd number of $Y$s, the bilateral stabilizer is given by $-g_i\otimes g_i$.

Alice’s stabilizer measurements update the Bell stabilizers in Equation~\ref{eq:pre-meas_Bell_states}. For example, $g_i\otimes I$ anti-commutes with $Z_i\otimes Z_i$ for $i=1$ to $r$ and with $X_i\otimes X_i$ for $i=r+1$ to $n-k$ according to Observation~\ref{ob:4}. Measuring $g_i\otimes I$ replaces the anti-commutative operator with $\left(-1\right)^{s_i}g_i\otimes I$, where $s_i$ is the measured parity. The $n-k$ stabilizer measurements result in new stabilizers, symmetrically represented by
\begin{equation} \label{eq:logical_Bell_states}
    \begin{matrix}
    \left(-1\right)^{s_1}g_i\otimes I\;, & \left(-1\right)^{s_1}I\otimes g_1\\
    \vdots & \vdots\\
    \left(-1\right)^{s_{n-k}}g_{n-k}\otimes I\;, & \left(-1\right)^{s_{n-k}}I\otimes g_{n-k}\\
    X_L^1\otimes X_L^1\;, & Z_L^1\otimes Z_L^1\\
    \vdots & \vdots\\
    X_L^k\otimes X_L^k\;, & Z_L^k\otimes Z_L^k\\
    \end{matrix}\;.
\end{equation}
Here, $\left(-1\right)^{s_i}I\otimes g_i$ results from multiplying $g_i\otimes g_i$ with $\left(-1\right)^{s_i}g_i\otimes I$, representing the generator on Bob’s side.

The standard form in Equation~\ref{eq:logical_Bell_states} facilitates efficient decoding of logical Bell states into physical Bell states via single-qubit measurements. According to Equations~~\ref{eq:logical_x} and~\ref{eq:logical_z}, the logical operators commute with $Z_j$ for $j=1$ to $r$ and with $X_j$ for $j=r+1$ to $n-k$. To decode the logical states, Alice and Bob separately measure the first $r$ qubits in the $Z$ basis and the next $n-k-r$ qubits in the $X$ basis.
For example, as per Equation~\ref{eq:logical_x}, the $i$th logical operator $X_L^i$ is given by the vector
\begin{equation*}
    \left[ \begin{array}{ccc|ccc}\bm{0} & \bm{e}_i^\prime & \bm{\delta}_i & \bm{c}_i^\prime & \bm{0} & \bm{0}\end{array} \right]\;,
\end{equation*}
where $\bm{c}_i^\prime$, $\bm{e}_i^\prime$, and $\bm{\delta}_i$ are the $i$th rows of submatrices $\bm{C}^T$, $\bm{E}^T$ and $\bm{I}$, respectively. Thus, $X_L^i$ can be expressed as
\begin{equation} \label{eq:logx_Pauli}
    X_L^i=\bigotimes_{j:C_{ij}^T=1}{Z_j} \bigotimes_{j:E_{ij}^T=1}{X_{r+j}} \bigotimes X_{n-k+i}\;,
\end{equation}
which is the tensor product of $Z_j$ for $j$ with $C_{ij}^T=1$ among the first $r$ qubits, $X_j$ with $E_{ij}^T=1$ among the next $n-k-r$ qubits, and the $X$ operator on the $\left(n-k+i\right)$th qubit. We denote Alice’s single-qubit measurement outcomes as decoding parities, represented by the binary vector $\bm{a}=a_1a_2\ldots a_{n-k}$. Replacing the Pauli operators in Equation~\ref{eq:logx_Pauli} with these measurements gives
\begin{equation} \label{eq:physical_x}
    X_L^{(i)}=\left(-1\right)^{\alpha_x^{(i)}}X_{n-k+i}\;,
\end{equation}
where $\alpha_x^{(i)}=\sum_{j:C_{ij}^T=1} a_j+\sum_{j:E_{ij}^T=1} a_{r+j}$. Similarly, the logical operator $Z_L^{(i)}$ is given by
\begin{equation} \label{eq:physical_z}
    Z_L^{(i)}=\left(-1\right)^{\alpha_z^{(i)}}Z_{n-k+i}\;,
\end{equation}
where $\alpha_z^{(i)}=\sum_{j:\left(A_2^T\right)_{ij}=1} a_j$ and $\left(A_2^T\right)_{ij}$ is the element of submatrix $\bm{A}_2^T$ from Equation~\ref{eq:logical_z}. The single-qubit measurements anti-commute with the code generators, updating the stabilizers in Equation~\ref{eq:logical_Bell_states} to
\begin{widetext}
\begin{equation*}
    \begin{matrix}  
    \left(-1\right)^{a_1} Z_1 \otimes I\;, & \left(-1\right)^{b_1} I \otimes Z_r \\
    \vdots & \vdots \\
    \left(-1\right)^{a_r} Z_r \otimes I\;, & \left(-1\right)^{b_r} I \otimes Z_r \\
    \left(-1\right)^{a_{r+1}} X_{r+1} \otimes I\;, & \left(-1\right)^{b_{r+1}} I \otimes X_{r+1} \\
    \vdots & \vdots \\
    \left(-1\right)^{a_{n-k}} X_{n-k} \otimes I\;, & \left(-1\right)^{b_{n-k}} I \otimes X_{n-k} \\ 
    \left(-1\right)^{\alpha_x^{(1)}+\beta_x^{(1)}} X_{n-k+1} \otimes X_{n-k+1}\;, & \left(-1\right)^{\alpha_z^{(1)}+\beta_z^{(1)}} Z_{n-k+1} \otimes Z_{n-k+1} \\
    \vdots & \vdots \\
    \left(-1\right)^{\alpha_x^{(k)}+\beta_x^{(k)}} X_n \otimes X_n\;, & \left(-1\right)^{\alpha_z^{(k)}+\beta_z^{(k)}} Z_n \otimes Z_n \\
    \end{matrix}\;.
\end{equation*}
\end{widetext}
Here, $\bm{b}=b_1b_2\ldots b_{n-k}$ denotes Bob’s decoding parities from single-qubit measurements, which are independent of Alice’s. $\beta_x^{(i)}$ and $\beta_z^{(i)}$ are computed in the same way as $\alpha_x^{(i)}$ and $\alpha_z^{(i)}$, respectively. After the measurements, Alice and Bob discard their first $n-k$ qubits and retain the last $k$ qubits, forming the decoded physical Bell states. They adjust the phases $\alpha_x^{(i)}$, $\alpha_z^{(i)}$, $\beta_x^{(i)}$, and $\beta_z^{(i)}$ to $0$ by applying Pauli operators to the corresponding qubits, yielding $k$ standard Bell states.

\begin{figure}[tb]
    \centering
    \includegraphics[width=0.4\paperwidth]{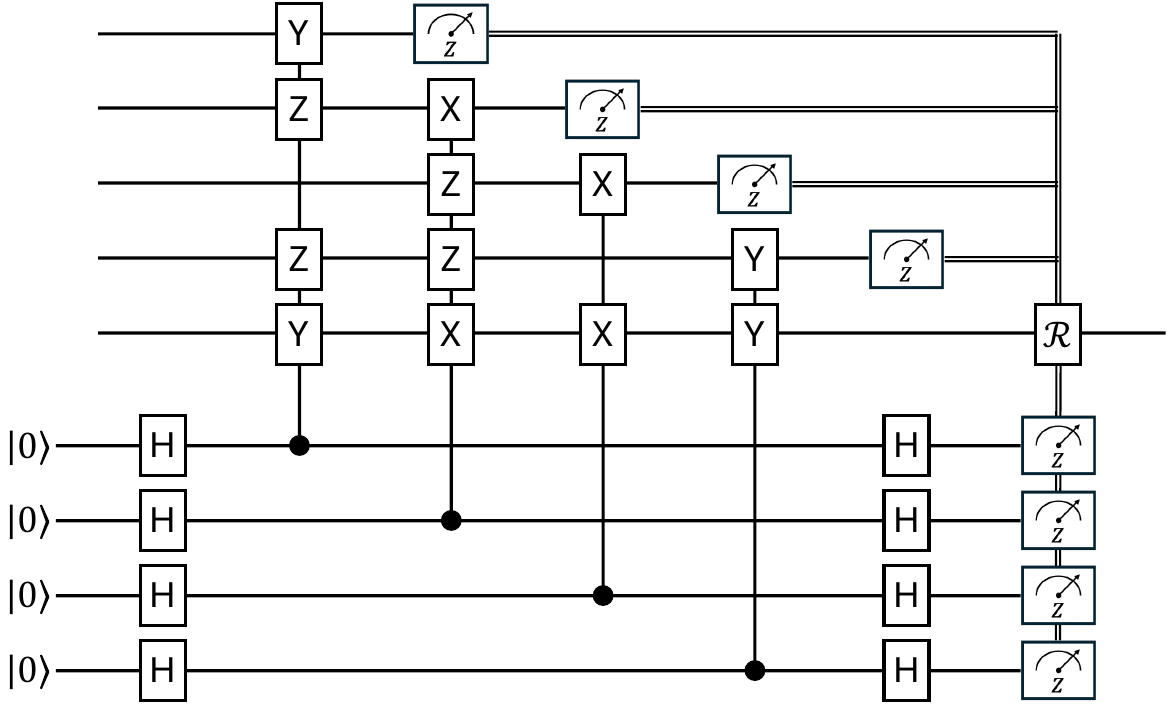}
    \caption{Circuit implementing stabilizer measurements, decoding, and error correction using the $\left[\!\left[5,1,3\right]\!\right]$ code. $\mathcal{R}$ represents a recovery Pauli operator conditioned on the measurement outcomes.}
    \label{fig:circuit_integrated}
\end{figure}

By integrating stabilizer measurements, error correction, and our decoder (see Appendix~\ref{app:integration} for more details), we outline the measurement-based stabilizer entanglement distillation protocol as follows:
\begin{enumerate}
    \item Alice measures stabilizer parities $\bm{s}_A=s_1s_2\ldots s_{n-k}$ and decoding parties $\bm{a}=a_1a_2\ldots a_{n-k}$. She communicates $\bm{s}_A$ to Bob, discards her first $n-k$ qubits, and corrects the phase of her last $k$ qubits based on $\bm{a}$.

    \item Bob measures stabilizer parities $\bm{s}_B=s_1^\prime s_2^\prime\ldots s_{n-k}^\prime$ and decoding parities $\bm{b}=b_1b_2\ldots b_{n-k}$. After receiving $\bm{s}_A$, he calculates the error syndrome $\bm{c}=\bm{s}_A\oplus \bm{s}_B$, discards his first $n-k$ qubits, and corrects the phase of his last $k$ qubits based on $\bm{b}$ and $\bm{c}$.
\end{enumerate}
This results in $k$ distilled Bell states shared between Alice and Bob. Note that two-way entanglement distillation, where the stabilizer code operates in error detection mode, can be implemented similarly using this protocol. For example, consider the $\left[\!\left[5,1,3\right]\!\right]$ code. Its stabilizer generators are
\begin{equation*}
\begin{array}{ccccccccc}
    Y & Z & I & Z & Y \\
    I & X & Z & Z & X \\
    Z & Z & X & I & X \\
    Z & I & Z & Y & Y
\end{array}\;.
\end{equation*}
We can integrate stabilizer measurements and single-qubit measurement decoding into a circuit as shown in Figure~\ref{fig:circuit_integrated}. This circuit yields both stabilizer parities and decoding parities needed for error correction.

We compare the circuit complexity of entanglement distillation between a unitary protocol and our measurement-based protocol. For general stabilizer codes, both circuits scale similarly in the number of two-qubit gates, with an upper bound of $O\left(n\left(n-k\right)\right)$. This is expected, as the unitary decoding circuit derives from the standard form of stabilizer generators~\cite{Gottesman1997}. However, stabilizer measurements offer more flexibility. Well-designed codes, such as quantum convolutional codes (Section~\ref{sec:convolutional}) and quantum low-density parity-check (QLDPC) codes (Section~\ref{sec:encoding}), support parallel stabilizer measurements, reducing circuit depth. The measurement-based decoder applies broadly, including in quantum computing, as it relies only on the standard form of logical operators. The logical $X$ and $Z$ operators share commutative Pauli operators on the first $n-k$ qubits, allowing single-qubit measurements on these qubits to preserve logical information. Replacing these Pauli operators with their measurement outcomes decodes the logical operators into single-qubit operators. However, the phase to be corrected is calculated by the sum of single-qubit measurement outcomes, where errors can accumulate. In Section~\ref{sec:encoding}, we propose a fault-tolerant protocol for encoding and decoding arbitrary quantum states by combining entanglement distillation, constant-depth decoders, and quantum teleportation.

\section{Application II: Entanglement Distillation using Quantum Convolutional Codes} \label{sec:convolutional}

Convolutional entanglement distillation implements the stabilizer entanglement distillation protocol using convolutional codes~\cite{Ollivier2003, Ollivier2004, Forney2005, Forney2007, Grassl2007, Wilde2010}. These codes are widely used in modern communication systems due to their robust error correction capabilities~\cite{Johannesson2015} and are key components of Turbo codes~\cite{Poulin2009, Wilde2014, Bolt2016}. They allow flexible adjustments in code rates to balance error correction strength and efficiency, making them ideal for addressing fluctuating noise. Quantum convolutional codes inherit these benefits and show potential for entanglement distillation. However, they are prone to catastrophic error propagation~\cite{{Poulin2009}, Grassl2006, Houshmand2013}, where a single error can spread indefinitely. To avoid this, encoding circuits are selectively constructed using a restricted set of quantum gates, often requiring many gates and deeper circuits. This increased demand for resources can challenge current quantum technologies. Our measurement-based approach provides an efficient non-catastrophic scheme, making convolutional entanglement distillation viable.

Quantum convolutional codes use a stream of stabilizers, represented by the extended check matrix $\bm{G}\left(D\right)=\left(\bm{X}\left(D\right)\middle| \bm{Z}\left(D\right)\right)$~\cite{Grassl2006}, where $\bm{X}\left(D\right)$ and $\bm{Z}\left(D\right)$ are $\left(n-k\right)\times n$ submatrices. The entries of the submatrices are polynomials in the delay operator $D$, given by
\begin{eqnarray*}
    X_{ij}\left(D\right) &=& \sum_{k=0}^{m}{x_{ij}\left(k\right)D^k}\;,\nonumber\\
    Z_{ij}\left(D\right) &=& \sum_{k=0}^{m}{z_{ij}\left(k\right)D^k}\;,
\end{eqnarray*}
where $x_{ij}\left(k\right)$ and $z_{ij}\left(k\right)$ are binary numbers, and $m$ is the constraint length. The rows of $\bm{G}\left(D\right)$ represent $n-k$ independent, shift-invariant stabilizer generators, and the columns of each submatrix correspond to the $n$ physical qubits in a frame. The tuple $\left(x_{ij}\left(k\right)\middle| z_{ij}\left(k\right)\right)$ denotes the Pauli operator on the $j$th qubit of the $i$th generator in the $k$th frame. The operator $D^k$ delays the Pauli operators by $k$ frames or $kn$ qubits. For example, the row $\left[\begin{array}{cc|cc} 1 & 1+D & D & 1+D^2\end{array}\right]$ represents the generator $X_1 Y_2 Z_3 X_4 Z_6$ and its shifts.

Like block codes, operations on the check matrix correspond to generator operations, such as swapping rows (relabeling generators), adding two rows (multiplying generators), and swapping columns in both submatrices (relabeling qubits). Additionally, multiplying row $g_i\left(D\right)$ by $D^k$ delays the generator by $k$ frames. Both $g_i\left(D\right)$ and $D^kg_i\left(D\right)$ independently act as generators of the convolutional code. The inverse operator $D^{-k}$ moves the generator forward by $k$ frames. We limit the forward movement to be finite but allow unrestricted delays. Combining shift and addition operations produces convoluted generators, represented by $f\left(D\right)g_i\left(D\right)$, where $f\left(D\right)=\sum_{i=d}^{\infty}{f_iD^i}$ is a Laurent series, $f_i$ is a binary number, and $d$ is a lower bounded integer. Note that $f\left(D\right)$ can be expressed as $\frac{p\left(D\right)}{q\left(D\right)}$, where $p\left(D\right)$ and $q\left(D\right)$ are polynomials. For example, $\frac{1}{1+D}=1+D+D^2+\cdots$ is a Laurent series. Similarly, $f\left(D^{-1}\right)$ also forms a Laurent series. Two shift-invariant generators $g_i(D)=\left(\bm{x}_i\left(D\right)\middle| \bm{z}_i\left(D\right)\right)$ and $g_j(D)=\left(\bm{x}_j\left(D\right)\middle| \bm{z}_j\left(D\right)\right)$ commute if $g_i(D)$ also commutes with the generator $D^l g_j(D)$. Setting their symplectic inner product to zero for all $l$, the condition for two shift-invariant generators to commute is given by~\cite{Grassl2006}
\begin{equation} \label{eq:commute}
    \bm{x}_i\left(D\right)\bm{z}_j^T\left(D^{-1}\right)+\bm{z}_i\left(D\right)\bm{x}_j^T\left(D^{-1}\right)=0\;.
\end{equation}

Deriving the standard form of the check matrix $\bm{G}\left(D\right)$ for convolutional codes involves Gaussian elimination and column permutations, similar to block codes. The standard form is given by
\begin{equation*}
    \left[ \begin{array}{ccc|ccc}
    \bm{I} & \bm{A}_1\left(D\right) & \bm{A}_2\left(D\right) & \bm{B}\left(D\right) & \bm{0} & \bm{C}\left(D\right) \\
    \bm{0} & \bm{0} & \bm{0} & \bm{D}\left(D\right) & \bm{I} & \bm{E}\left(D\right)\\
    \end{array} \right]\;,
\end{equation*}
where the submatrices conform to the dimensions specified in Equation~\ref{eq:standard_form}. Here, $\bm{I}$ is the identity matrix, and other submatrices contain Laurent series in $D$. Logical operators are defined as
\begin{eqnarray} \label{eq:convolitional_logical}
    L_X\left(D\right) &=& \left[ \begin{array}{ccc|ccc}
    \bm{0} & \bm{E}^T\left(D^{-1}\right) & \bm{I} & \bm{C}^T\left(D^{-1}\right) & \bm{0} & \bm{0}
    \end{array} \right]\;,\nonumber\\
    L_Z\left(D\right) &=& \left[ \begin{array}{ccc|ccc}
    \bm{0} & \bm{0} & \bm{0} & \bm{A}_2^T\left(D^{-1}\right) & \bm{0} & \bm{I}
    \end{array}\right]\;.
\end{eqnarray}
We can verify the commutativity of these operators with the stabilizers using Equation~\ref{eq:commute}. For convenience, we denote $\bm{A}^\prime\left(D\right)=\bm{A}_2^T\left(D^{-1}\right)$, $\bm{C}^\prime\left(D\right)=\bm{C}^T\left(D^{-1}\right)$, and $\bm{E}^\prime\left(D\right)=\bm{E}^T\left(D^{-1}\right)$. Generally, the entries of these submatrices are Laurent series, but we assume them to be finite Laurent polynomials. This assumption holds when Gaussian elimination is performed with row operations using finite Laurent polynomial multipliers (see the example in Appendix~\ref{app:conv_standard_form}). For cases involving entries as Laurent series, pre-shared Bell states can facilitate decoding with asymptotically zero overhead. These Bell states can be obtained from previous entanglement distillation. For more details, see Appendix~\ref{app:Laurent_series}.

Given that the submatrices $\bm{A}^\prime\left(D\right)$, $\bm{C}^\prime\left(D\right)$, and $\bm{E}^\prime\left(D\right)$ consist of polynomial entries, the measurement-based decoding mirrors the block code method in Section~\ref{sec:decoding}. For each frame of $n$ qubits, we measure the first $r$ qubits in the $Z$ basis and the next $n-k-r$ qubits in the $X$ basis. The remaining $k$ qubits form the decoded qubits, with their phases corrected based on measurement outcomes, as given by Equations~\ref{eq:physical_x} and~\ref{eq:physical_z}. Note that since the entries are polynomials, the phases are also influenced by measurements in nearby frames.

\begin{figure*}[tb]
    \centering
    \includegraphics[width=0.66\paperwidth]{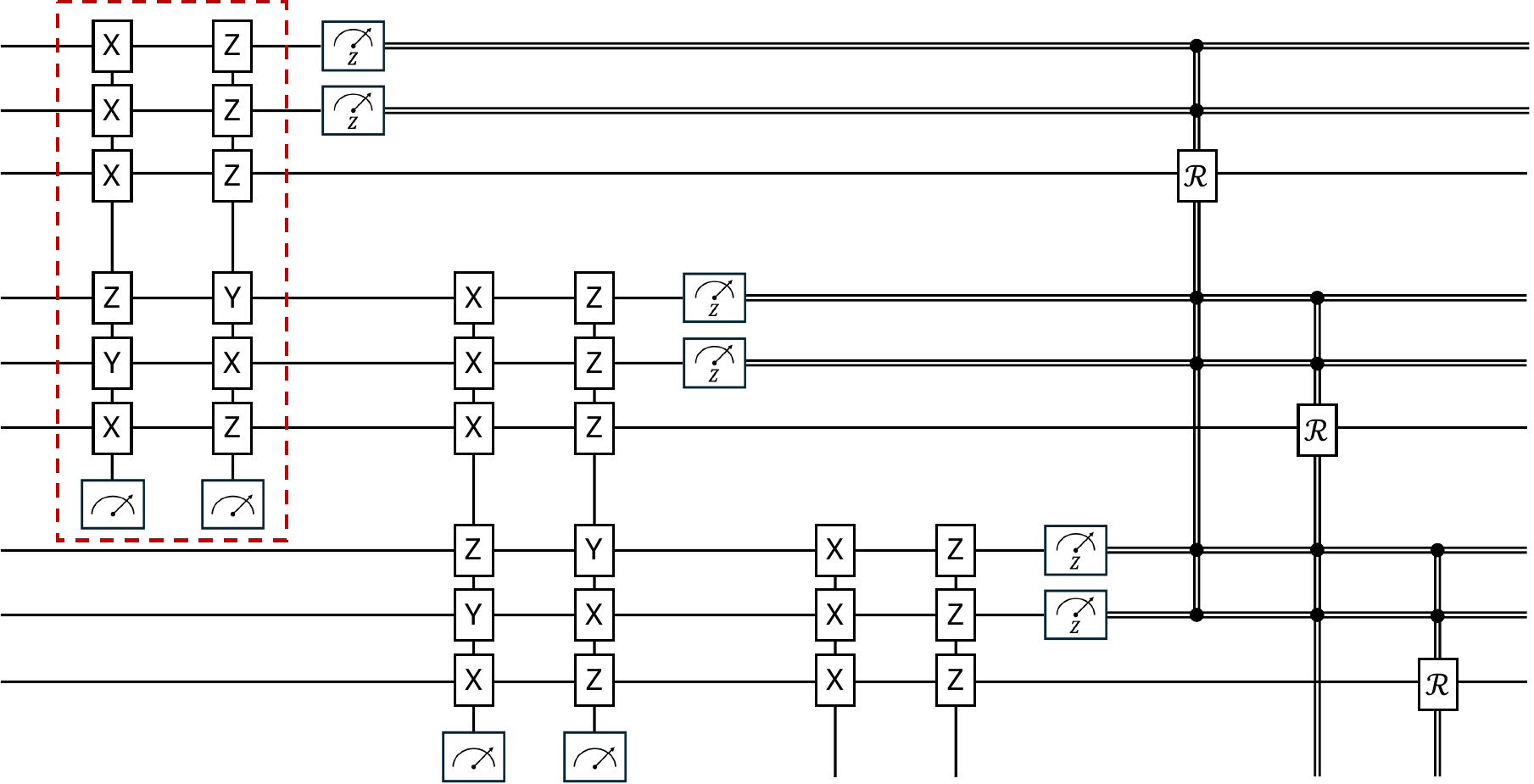}
    \caption{Alice’s circuit for the entanglement distillation protocol using the rate-$\frac{1}{3}$ quantum convolutional code. Each input qubit forms a Bell state with a corresponding qubit on Bob’s side. Measured qubits are discarded, and unmeasured output qubits form the distilled Bell states at one-third the input rate. The dashed box indicates two stabilizer measurements per frame. $\mathcal{R}$ represents a recovery Pauli operator conditioned on single-qubit measurement outcomes. Bob’s circuit mirrors Alice’s, with recovery gates adjusted based on his measurement outcomes and stabilizer parities from both sides.}
    \label{fig:conv_circuit}
\end{figure*}

We use a rate-$\frac{1}{3}$ quantum convolutional code from~\cite{Forney2007} to demonstrate the measurement-based protocol. This code encodes one logical qubit into three physical qubits per frame. The stabilizer generators are
\begin{equation*}
\begin{array}{ccc}
    X_1X_2X_3 & Z_4Y_5X_6 & \\
    Z_1Z_2Z_3 & Y_4X_5Z_6 & \\
    & X_4X_5X_6 & Z_7Y_8X_9 \\
    & Z_4Z_5Z_6 & Y_7X_8Z_9 \\
    & & \ddots
\end{array}\;,
\end{equation*}
where the qubits are rearranged from the original configuration by permuting $\left(1,2,3\right)\rightarrow\left(2,3,1\right)$. The check matrix is given by
\begin{equation*}
    \bm{G}\left(D\right)=\left[ \begin{array}{ccc|ccc}
    1 & 1+D & 1+D & D   & D & 0 \\
    D & D   & 0   & 1+D & 1 & 1+D
    \end{array} \right]\;.
\end{equation*}
Gaussian elimination (see Appendix~\ref{app:conv_standard_form} for detailed steps) yields the standard form
\begin{widetext}
\begin{equation} \label{eq:conv_standard_form}
    \left[ \begin{array}{ccc|ccc}
    1 & 0 & D^{-1}+1 & D^{-2} & D^{-2}+D^{-1}+1 & D^{-2}+1 \\
    0 & 1 & D^{-1}+1 & D^{-2}+D^{-1}+1 & D^{-2}+1 & D^{-2}+D^{-1}
    \end{array} \right]\;.
\end{equation}
\end{widetext}
According to Equation~\ref{eq:convolitional_logical}, the logical operators are defined as
\begin{eqnarray}
    X_L &=& \left[ \begin{array}{ccc|ccc}
    0 & 0 & 1 & 1+D^2 & D+D^2 & 0
    \end{array} \right]\;,\nonumber \\
    Z_L &=& \left[ \begin{array}{ccc|ccc}
    0 & 0 & 0 & 1+D & 1+D & 1
    \end{array} \right]\;.
\end{eqnarray}
The logical qubit depends on qubits in the current and next two frames. Decoding involves measuring first two qubits in the $Z$ basis and identifying the third qubit as the decoded physical qubit in each frame. In the $k$th frame, the post-measurement logical operators are given by
\begin{eqnarray*}
    X_L\left(k\right) &=& \left(-1\right)^{x\left(k\right)}X_{3k}\;,\\
    Z_L\left(k\right) &=& \left(-1\right)^{z\left(k\right)}Z_{3k}\;,
\end{eqnarray*}
with phases determined by
\begin{eqnarray*}
    x\left(k\right) &=& z_{3k-2}+z_{3k+2}+z_{3k+4}+z_{3k+5}\;, \\
    z\left(k\right) &=& z_{3k-2}+z_{3k-1}+z_{3k+1}+z_{3k+2}\;,
\end{eqnarray*}
where $z_i$ is the measurement outcome of the $i$th qubit. These phases are then corrected using Pauli operators.

Figure~\ref{fig:conv_circuit} shows Alice's circuit for the entanglement distillation protocol using the rate-$\frac{1}{3}$ quantum convolutional code. Each input qubit forms a Bell state with a corresponding qubit on Bob's side. Unmeasured output qubits result in distilled Bell states at one-third the input rate. The setup includes stabilizer measurements, measurement-based decoding, and conditional phase recovery gates. Alice sends her stabilizer measurement outcomes to Bob. Bob's circuit mirrors Alice's but includes additional error correction. His phase recovery gates are adjusted based on his single-qubit measurement outcomes and the error syndrome~\cite{Schalkwijk1975, Johannesson2015}, which is derived by comparing stabilizer parities from both Alice and Bob.

We compare the circuit complexity of our method with the unitary method described from ~\cite{Grassl2006} for the rate-$\frac{1}{3}$ convolutional code. Our circuit on one side has a depth of $4$, using $12$ two-qubit gates per frame, with the longest two-qubit gate spanning $2$ frames. The unitary decoding circuit, constrained by the non-catastrophic requirement, has a depth of $11$ and $14$ two-qubit gates per frame, with the longest gate spanning $3$ frames. Thus, our method is simpler. While this code is a relatively simple convolutional code, more complex convolutional codes with higher encoding rates and larger code distances may be preferable for improved performance, where our method could offer even greater advantages. However, comparing general convolutional codes remains an open question, requiring theoretical complexity analysis of their unitary encoding circuits.

\section{Application III: Efficient Fault-Tolerant Encoding and Decoding for QLDPC Codes} \label{sec:encoding}

We propose an efficient fault-tolerant protocol for encoding and decoding quantum states in quantum computing, using stabilizer entanglement distillation and quantum teleportation. This approach is particularly effective for quantum low-density parity-check (QLDPC) codes~\cite{Breuckmann2021}, which feature a constant number of qubits per stabilizer generator and a constant number of generators acting on each qubit. This enables parallel stabilizer measurements, significantly reducing the circuit depth compared to unitary encoding methods. In this section, we first present a constant-depth encoder and decoder for surface codes~\cite{Dennis2002, Kitaev2003, Bravyi1998}, a well-studied class of QLDPC codes, and then demonstrate its fault tolerance. Finally, we discuss the potential and challenges of extending these encoder and decoder protocols to broader QLDPC codes with higher encoding rates.

\begin{figure}[tb]
    \centering
    \includegraphics[width=0.4\paperwidth]{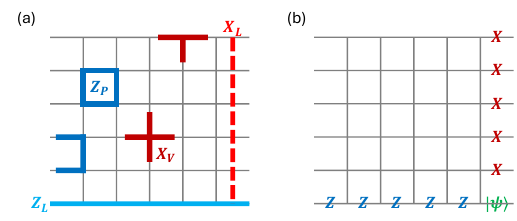}
    \caption{Planar code on a $6\times 6$ lattice where each edge represents a physical qubit. (a) Vertex and boundary vertex stabilizers consist of $X$ operators. Plaquette and boundary plaquette stabilizers consist of $Z$ operators. Dashed and solid lines indicate logical operators $X_L$ and $Z_L$, respectively. (b) Decoding involves single-qubit measurements targeting logical $X_L$ and $Z_L$ operators, excluding the intersecting qubit, which is identified as the decoded qubit.}
    \label{fig:planar_code}
\end{figure}

The efficient encoder and decoder protocol for surface codes involves generating a resource physical-logical Bell state, represented by
\begin{equation*}
    \left|\Phi_R\right\rangle=\frac{1}{\sqrt2}\left(\left|0\right\rangle_A\left|0_L\right\rangle_B+\left|1\right\rangle_A\left|1_L\right\rangle_B\right)\;,
\end{equation*}
where $0$ and $1$ represent physical states, and $0_L$ and $1_L$ represent logical states. Even though the physical qubits may reside in the same quantum computer, they are divided into two groups labeled Alice and Bob for clarity. The system uses a surface code on an $L \times L$ square lattice, with Alice's and Bob's qubits placed on separate but interconnected layers. The protocol is outlined as follows:
\begin{enumerate}
    \item All qubits start in the state $\left|0\right\rangle^{\otimes n}$. Alice and Bob apply transversal gates to generate $n$ Bell states between their qubits.

    \item Alice and Bob measure the code stabilizers on their qubits, projecting the physical Bell states into a logical Bell state.

    \item Alice communicates her stabilizer parities to Bob, who performs error correction based on the error syndrome.

    \item Alice uses single-qubit measurement decoding on her logical qubit, yielding the resource physical-logical Bell state $\left|\Phi_R\right\rangle$.

    \item To encode a state, Alice teleports the physical state to Bob by Bell measurement and classical communication. To decode a state, Bob teleports the logical state back to Alice through logical Bell measurement and classical communication.
\end{enumerate}
Logical operators, such as Pauli, Hadamard, phase, and CNOT gates, can be implemented transversally~\cite{Bravyi2020, Fowler2009, Moussa2016}. Therefore, the entire protocol can operate within constant depth, improving on the well-known lower bound of $\Omega\left(\log\left(n\right)\right)$ for surface code unitary encoding circuits without classical communication~\cite{Bravyi2006, Aharonov2018}, where $n$ is the number of physical qubits. For example, Figure~\ref{fig:planar_code} illustrates stabilizers, logical operators, and decoding measurements for a planar code with parameters $\left[\!\left[2L^2-2L+1, 1, L\right]\!\right]$. Note that decoding measurements are not unique. Single-qubit measurements can target any logical $X_L$ and $Z_L$ operators, except for the intersecting qubit, which is identified as the decoded qubit~\cite{Mazurek2014}.

\begin{figure}[tb]
    \centering
    \includegraphics[width=0.4\paperwidth]{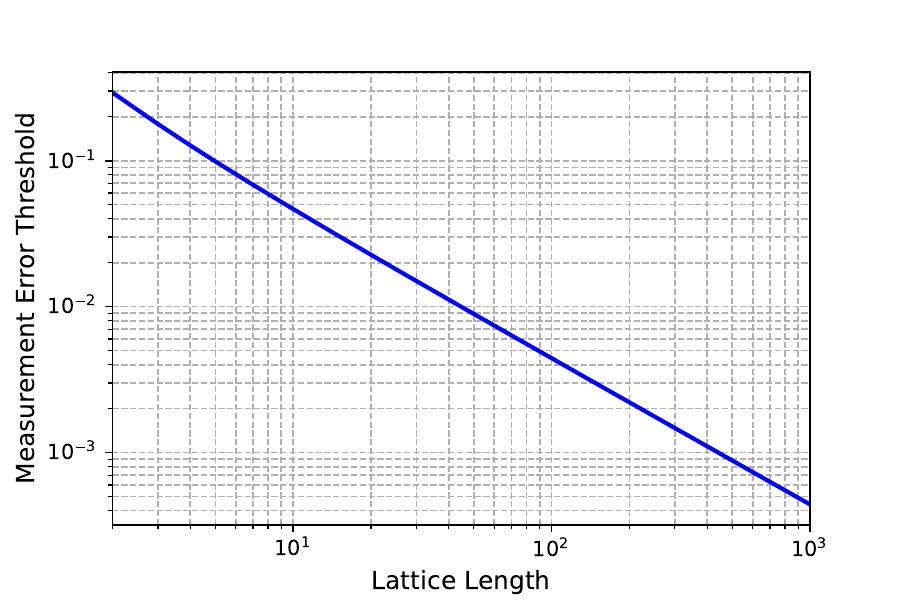}
    \caption{Measurement error threshold as a function of lattice length $L$ for the efficient fault-tolerant encoding and decoding of a $L \times L$ planar code.}
    \label{fig:measurement_error}
\end{figure}

In the presence of noise, all operations except measurement-based decoding can be applied fault-tolerantly, as in conventional quantum computing. The decoding challenge arises from accumulated errors in the phase corrections, which are based on the sum of single-qubit measurement outcomes. We model these errors using a noisy channel on the resource states, with an error probability of
\begin{equation*}
    P_L=\frac{3}{4}-\frac{1}{2}\left(1-2p\right)^{L-1}-\frac{1}{4}\left(1-2p\right)^{2\left(L-1\right)}\;,
\end{equation*}
where $p$ is the error probability of a single-qubit measurement, and $L$ is the lattice length (see Appendix~\ref{app:measurement_error} for a detailed derivation). We can suppress the error by distilling the resource state using a constant-depth circuit. To achieve fault tolerance, the resource state needs only a fidelity of $1-p_{th}$, where $p_{th}$ is the error threshold for fault-tolerant quantum computing with the surface code. This fidelity is attainable with a constant number of distillations, maintaining constant circuit depth. However, successful distillation necessitates $P_L<0.5$, defining a threshold for measurement error. Figure~\ref{fig:measurement_error} shows this threshold as a function of lattice size, following the power law $p\propto L^{-1}$ at large $L$. For an $L=23$ lattice, encoding a logical qubit with $1013$ physical qubits, the measurement error threshold is $0.0196$, which is feasible with current technology.

Noisy stabilizer measurements present another challenge, potentially introducing errors during correction and leading to logical errors. This issue can be addressed by multiple measurement rounds~\cite{Bombin2015, Shor1996} or by preparing high-quality entangled states~\cite{Bravyi2020, Steane1997, Knill2005}. Our protocol uses a constant number of stabilizer measurement rounds, with remaining errors corrected through entanglement distillation of the resource state $\left|\Phi_R\right\rangle$. Typically, $\Theta(d)$ rounds of stabilizer measurements are required to suppress logical errors to $p^{\Omega{(d)}}$~\cite{Bravyi2020, Moussa2016}, where $d$ is the surface code distance and $p$ is the measurement error probability. However, for the encoder, this requirement can be relaxed, as logical errors on Bob's part of $\left|\Phi_R\right\rangle$ can be transferred to Alice's qubit via the Bell-state identity (see Appendix~\ref{app:matrix_identity}). These errors can then be interpreted as errors on the input states, primarily determined by very few physical gates. Consequently, the fidelity of the resource state only needs to meet the fault-tolerance threshold, supporting a constant number of rounds for stabilizer measurements and distillation.

Quantum LDPC codes with constant encoding rates offer advantages over 2D surface codes~\cite{Gottesman2014, Breuckmann2021}. Our protocol is compatible with these codes at each step, though some challenges remain. One issue is handling non-local stabilizer measurements, as qubits within the same stabilizer may be spatially separated. According to the BPT bound~\cite{Bravyi2010}, the best QLDPC codes with local stabilizers in $D$ dimensions are constrained by parameters $\left[\!\left[n, k = n^{1-\frac{2}{D}}, d = n^{1-\frac{1}{D}}\right]\!\right]$~\cite{Bravyi2009, Haah2021}. Surface codes are optimal for $D=2$ but achieve only a rate of $1/n$, while asymptotically constant-rate QLDPC codes become feasible in $D\geq 3$~\cite{Portnoy2023, Lin2024, Freedman2021}. Multi-layer architectures enable higher-dimensional connections~\cite{Bravyi2024, Tremblay2022}, and reconfigurable neutral atoms support long-range connectivity~\cite{Xu2024}. However, implementing non-local stabilizer measurements with these atoms may require atom rearrangement, potentially increasing gate times. The single-shot property, which enables quantum error correction with a single round of syndrome measurements, is often needed for the efficient implementation of QLDPC codes. Certain QLDPC codes, including specific topological codes~\cite{Bombin2015, Quintavalle2021}, quantum expander codes~\cite{Fawzi2018, Xu2024}, and quantum Tanner codes~\cite{Gu2024}, are currently known to exhibit this property. Another challenge is the limitation of transversal gates, as each block encodes multiple qubits, complicating selective logical operations. Research on logical gates for QLDPC codes is ongoing. For example, code deformation~\cite{Krishna2021} provides a universal gate set through state injection within a single block. Another approach uses state teleportation into ancilla codes, enabling logical operations by separating computation and memory~\cite{Xu2024, Cohen2022, Nautrup2017, Horsman2012, Breuckmann2017}. However, these logical gates may increase circuit depth. The challenges discussed above are general to quantum computing with QLDPC codes rather than specific to our protocol~\cite{Gottesman2022}. Our protocol remains more efficient than unitary encoding, benefiting from QLDPC code sparsity and classical communication. For general stabilizer codes, both unitary encoding and our measurement-based encoding have a circuit depth of $O(n)$ and a gate count of $O(n^2)$~\cite{Gottesman1997}.

We discuss the advancements of our results with previous literature. Our encoding and decoding scheme operates on quantum platforms that support multi-dimensional qubit interactions, recently enabled by reconfigurable atom arrays~\cite{Bluvstein2024, Xu2024}. {\L}odyga et al.~\cite{Lodyga2015} proposed an encoding and decoding scheme for surface codes using stabilizer and single-qubit measurements, asserting that logical operators must intersect at a single physical qubit or at adjacent qubits. However, our findings indicate this condition is not necessary, as our scheme naturally arises from the standard form of stabilizer codes. Unlike their approach, which relies on a specific 3D lattice, our scheme uses entanglement distillation for fault tolerance and is adaptable to various geometries, enhancing its applicability. Bravyi et al.~\cite{Bravyi2020} proposed a scheme for preparing logical state $0_L$ and logical Bell state of surface codes using measurements within a constant circuit depth. Their scheme requires a prepared $3D$ Raussendorf cluster state~\cite{Raussendorf2005} as well. Notably, when preparing the logical $0_L$ state via stabilizer measurements, starting with all physical qubits in the state $\left|0\right\rangle^{\otimes n}$ results in a maximally mixed logical state. However, by initializing qubits according to the standard form, with some in $\left|0\right\rangle$ and others in $\left|+\right\rangle$, the logical $0_L$ state can be prepared with a single round of stabilizer measurements.

\section{Conclusion}

In this paper, we first review stabilizer entanglement distillation and apply it to the well-known $2$-to-$1$ distillation protocols of Bennett et al. and Deutsch et al., mapping these protocols onto a $\left[2,1,2\right]$ classical error detection code. We show that bilateral Hadamard gates on the distillation output can alternate Pauli-$X$ and Pauli-$Z$ errors while preserving the standard Bell states. Therefore, two sequential applications of these protocols can effectively address any single-qubit errors. Building on this insight, we extend the $2$-to-$1$ distillation protocol to higher-rate $n$-to-$\left(n-1\right)$ protocols, which achieve higher yield for high-fidelity input Bell states. We also present a method to adapt any classical error-correcting code to an entanglement distillation protocol using the two-iteration process. The first iteration corrects bit-flip errors, then Hadamard gates convert phase-flip errors to bit-flip errors, which the second iteration corrects. The errors in both iterations are classical, so the error-correcting capability matches that of the classical code used.

Furthermore, we propose a constant-depth decoder for stabilizer codes and apply it to a stabilizer entanglement distillation protocol based on quantum convolutional codes, achieving reduced circuit complexity compared to unitary decoding. We also develop an efficient fault-tolerant encoding and decoding protocol for surface codes, discussing its potential and challenges in extending to broader QLDPC codes with higher encoding rates. This protocol combines techniques from stabilizer entanglement distillation, constant-depth decoders, and quantum teleportation. These operations are feasible with state-of-the-art reconfigurable atom arrays, and the encoder surpasses the conventional depth limits for surface codes, which are lower bounded by $\Omega\left(\log\left(n\right)\right)$.

Our work directly applies to entanglement routing protocols in quantum networks~\cite{Patil2024, Milligen2024}. The $n$-to-$\left(n-1\right)$ protocol with small values of $n$ could potentially be demonstrated experimentally in the near future. Measurement-based decoding reduces circuit depth and minimizes latency, thereby enhancing the performance of quantum repeaters. Overall, our work integrates stabilizer formalism, measurement-based quantum computing, and entanglement distillation, advancing both quantum communication and computing, and contributing to the development of practical and efficient quantum information processing techniques.

\begin{acknowledgments}
This research was supported by the National Science Foundation (NSF), under the following awards: ``Quantum Networks to Connect Quantum Technology (QuanNeCQT)'' awarded under Grant 2134891, ``CIF-QODED: Quantum codes Optimized for the Dynamics between Encoded Computation and Decoding using Classical Coding Techniques'' awarded under Grant 2106189, and the ``NSF Engineering Research Center for Quantum Networks (CQN)'' awarded under Grant 1941583.
\end{acknowledgments}

\appendix

\section{Bell-State Matrix Identity} \label{app:matrix_identity}

We show the Bell-state matrix identity for $n$ standard Bell states. The standard Bell state is defined by
\begin{equation*}
    \left|\Phi^+\right\rangle = \frac{1}{\sqrt{2}} \left(\left|00\right\rangle + \left|11\right\rangle\right)\;.
\end{equation*}
Thus, we express the tensor product of $n$ Bell states as
\begin{equation*}
    \left|\Phi_n^+\right\rangle = \frac{1}{\sqrt{2^n}} \sum_{x=0}^{2^n-1} \left|x\right\rangle_A \left|x\right\rangle_B\;.
\end{equation*}
Applying an operator $M = \sum_{pq} M_{pq} \left|p\right\rangle \left\langle q\right|$ to Alice’s qubits yields
\begin{align*}
\left(M \otimes I\right) \left|\Phi_n^+\right\rangle &= \frac{1}{\sqrt{2^n}} \sum_x \sum_p M_{xp} \left|p\right\rangle_A \left|x\right\rangle_B \\
&= \frac{1}{\sqrt{2^n}} \sum_p \left|p\right\rangle_A \left(\sum_x \left(M^T\right)_{px} \left|x\right\rangle_B\right) \\
&= \frac{1}{\sqrt{2^n}} \sum_p \left|p\right\rangle_A M^T \left|p\right\rangle_B\;,
\end{align*}
where we use $M_{xp} = \left(M^T\right)_{px}$. Therefore, we have the Bell-state matrix identity
\begin{equation*}
    \left(M \otimes I\right) \left|\Phi_n^+\right\rangle = \left(I \otimes M^T\right) \left|\Phi_n^+\right\rangle\;.
\end{equation*}
For Pauli operators, $X^T=X$, $Z^T=Z$, and $Y^T=-Y$.

We demonstrate the equivalence of the bilateral and single-lateral depolarizing channels on the standard Bell state using the Bell-state identity. Starting with the bilateral depolarizing channel, we assume a channel on both side that applies a Pauli operator from $\{X,Y,Z\}$, each with probability $p_i=p/3$, or an identity operator with probability $p_0=1-p$. The Kraus operators for this channel are given by $\{\sqrt{p_ip_j}E_j\otimes E_i\}$, where $E_0=I$ and $E_i\in\{X,Y,Z\}$ for $i=1,2,3$. Applying these operators to the Bell state $\left|\Psi^+\right\rangle$, we use the Bell matrix identity to express them as $\{\sqrt{p_ip_j}I\otimes E_iE_j^T\}$. For simplicity, we rewrite the equivalent Kraus operators as $\{\sqrt{p_ip_j}E_iE_j\}$, omitting the operator on Alice’s side and applying only to Bob’s qubit. Any minus sign from Pauli-$Y$ operator is absorbed when applied as a superoperator. The Kraus operators fall into three cases:
\begin{enumerate}
    \item $i=j=0$: The Kraus operator is $\{\left(1-p\right)I\}$.
    \item Either $i$ or $j=0$: The Kraus operators are $\{\sqrt{\frac{2}{3}p\left(1-p\right)}E_k\}$ for $k=1,2,3$.
    \item Both $i$ and $j\neq0$: The Kraus operators are $\{\frac{\sqrt3}{3}pI,\frac{\sqrt2}{3}pE_k\}$ for $k=1,2,3$.
\end{enumerate}
Summing these terms yields a single-lateral depolarizing channel with Kraus operators $\{\sqrt{1-p^\prime}I,\sqrt{\frac{p^\prime}{3}}E_i\}$, where $p^\prime=2p-\frac{4}{3}p^2$. Thus, the bilateral depolarizing channel is equivalent to a single-lateral depolarizing channel with an adjusted error probability. The single-lateral model, simplifying simulations, was used. Figure~\ref{fig:performance}, as a function of input and output fidelity, automatically captures the adjusted error probability, remaining unaffected by the choice of channel model.

\section{Adaptive Distillation with Constant Yield Rate} \label{app:adaptive_distillation}

In this appendix, we demonstrate that iterative applications of the $\left[\!\left[n^2,\left(n-1\right)^2,2\right]\!\right]$ distillation protocol, with adaptive block size $n$ based on input fidelity, achieve a constant overall yield rate. We compute the yield rate $R$ for bit-flip error correction. For both bit-flip and phase-flip errors, the yield rate is given by $R^2$. The Bell states are characterized by their infidelity (error rate) $p=1-\mathcal{F}$, where $\mathcal{F}$ denotes the fidelity. In the following calculations, we consider the regime where the infidelity is asymptotically small ($p\to0$), allowing us to retain only the first-order term in $p$ while neglecting higher-order contributions. Since the protocol improves fidelity at each round, this assumption holds throughout. Achieving any finite infidelity requires only a finite number of distillation rounds, ensuring constant overhead for the yield rate.

Applying the $\left[n, n-1, 2\right]$ protocol to $n$ Bell states with initial infidelity $p_{\rm in}$, the output infidelity is given by $p_{\rm{out}}=\frac{n(n-1)}{2} p_{\rm{in}}^2$ and the distillation success probability is $P_S=1 - n p_{\rm{in}}$, both calculated using a first-order approximation. The protocol requires $p_{\rm{out}} < p_{\rm{in}}$, which imposes the condition $n \leq \sqrt{2} p_{\rm{in}}^{-1/2}$ on the block size. To satisfy this condition, we set $n = \sqrt{2} p_{\rm{in}}^{-\alpha}$, where $\alpha$ is a constant parameter chosen within the range $0 \leq \alpha \leq \frac{1}{2}$.

In a sequence of distillations, each iteration $k$ is characterized by input infidelity $p_{k-1}$ and output infidelity $p_k$, with the output of iteration $k$ serving as the input for iteration $k+1$. The initial infidelity is $p_0$, used as input to the first iteration. The recursive parameters for iteration $k$ are as follows:
\begin{eqnarray*}
    n_k &=& \sqrt{2} p_{k-1}^{-\alpha}\;, \\
    r_k &=& \frac{n_k - 1}{n_k}\;, \\
    p_k &=& \frac{1}{2} n_k^2 p_{k-1}^2\;, \\
    P_k &=& 1 - n_k p_{k-1}\;,
\end{eqnarray*}
where $n_k$, $r_k$, $p_k$, and $P_k$ represent the block size, encoding rate, output infidelity, and success probability, respectively. Solving these recursive equations yields:
\begin{eqnarray*}
    n_k &=& \sqrt{2} p_0^{-\alpha (2 - 2\alpha)^{k-1}}\;, \\
    r_k &=& 1 - \frac{\sqrt{2}}{2} p_0^{\alpha (2 - 2\alpha)^{k-1}}\;, \\
    p_k &=& p_0^{(2 - 2\alpha)^k}\;, \\
    P_k &=& 1 - \sqrt{2} p_0^{\frac{1}{2} (2 - 2\alpha)^k}\;.
\end{eqnarray*}
The overall yield rate is given by $R = \prod_{k} r_k P_k$. Taking its logarithm and applying the approximation $\ln(1 + x) \approx x$ for small $x$, we derive the first-order result
\begin{equation*}
    -\ln{R} = \sum_{k} \left( \sqrt{2} p_0^{\frac{1}{2} (2 - 2\alpha)^k} + \frac{\sqrt{2}}{2} p_0^{\alpha (2 - 2\alpha)^{k-1}} \right)\;.
\end{equation*}
For $0 < \alpha < \frac{1}{2}$ at large $k$, $(2 - 2\alpha)^k > k$. We can bound the summation as
\begin{eqnarray*}
c &=& \sum_{k} \left( \sqrt{2} p_0^{\frac{1}{2} (2 - 2\alpha)^k} + \frac{\sqrt{2}}{2} p_0^{\alpha (2 - 2\alpha)^{k-1}} \right) \\
    &<& \sum_{k} \left( \sqrt{2} p_0^{\frac{1}{2}k} + \frac{\sqrt{2}}{2} p_0^{\alpha (k-1)} \right) \\
    &=& \frac{\sqrt{2p_0}}{1 - \sqrt{p_0}} + \frac{\sqrt{2}}{2 (1 - p_0^\alpha)}\;.
\end{eqnarray*}
Thus, the overall yield rate of the adaptive entanglement distillation is lower bounded by $R > e^{-c}$, ensuring a constant yield rate. Further optimization of the protocol is possible; for example, selecting block sizes as $n_k\sim k^2$ also satisfies the threshold requirements and maintains a constant lower bound for the overall yield rate.

\section{Integrating Stabilizer Measurements, Error Correction, and Decoding} \label{app:integration}

The entanglement distillation protocol can be enhanced by integrating stabilizer measurements, error correction, and decoding. We consider Bob's side, with Alice following a similar strategy.

First, we explain the integration of error correction and decoding. Bob computes the error syndrome $\bm{c}=\bm{s}_A\oplus \bm{s}_B$, where $\bm{s}_A$ are the stabilizer parities received from Alice, and $\bm{s}_B$ are from Bob’s stabilizer measurements. He determines the recovery operation as
\begin{equation*}
    \mathcal{F}^{\left(\bm{c}\right)}=\bigotimes_{j=1}^{n}F_j^{\left(\bm{c}\right)}\;,
\end{equation*}
with $F_j^{\left(\bm{c}\right)}$ as the Pauli operator on the $j$th qubit. For measurement-based decoding, Bob measures his first $r$ qubits in the $Z$ basis and the next $n-k-r$ qubits in the $X$ basis, resulting in decoding parities $\bm{b}=b_1b_2\ldots b_{n-k}$. He then applies Pauli operators to the last $k$ qubits to adjust their phases, given by
\begin{equation*}
    \mathcal{D}^{\left(\bm{b}\right)}=\bigotimes_{j=1}^{r}{P_Z^{\left(b_j\right)}\bigotimes_{j=r+1}^{n-k}{P_X^{\left(b_j\right)}\bigotimes_{j=n-k+1}^{n}D_j^{\left(\bm{b}\right)}}}\;,
\end{equation*}
where $P_X^{\left(b_j\right)}$ and $P_Z^{\left(b_j\right)}$ are projectors with measurement outcome $b_j$, and $D_j^{\left(\bm{b}\right)}$ is a Pauli operator on the $j$th qubit to adjust its phase. Conventionally, Bob performs error correction and decoding sequentially as  $\mathcal{D}^{\left(\bm{b}\right)}\circ\mathcal{F}^{\left(\bm{c}\right)}$. However, he can reorder these operations as
\begin{equation*}
    \bigotimes_{j=1}^{r}{F_j^{\left(\bm{c}\right)}P_Z^{\left(b_j^\prime\right)}\bigotimes_{j=r+1}^{n-k}{F_j^{\left(\bm{c}\right)}P_X^{\left(b_j^\prime\right)}\bigotimes_{j=n-k+1}^{n}{D_j^{\left(\bm{b}\right)}F_j^{\left(\bm{c}\right)}}}}\;.
\end{equation*}
If the projector commutes with $F_j^{\left(\bm{c}\right)}$, then $b_j^\prime=b_j$; otherwise, $b_j^\prime=b_j\oplus1$. Once $\bm{b}^\prime$ is computed, Bob discards the first $n-k$ qubits. Thus, error correction and decoding can be integrated into a single operation
\begin{equation*}
    \mathcal{R}^{\left(\bm{b},\bm{c}\right)}=\bigotimes_{j=1}^{r}{P_Z^{\left(b_j\right)}\bigotimes_{j=r+1}^{n-k}{P_X^{\left(b_j\right)}\bigotimes_{j=n-k+1}^{n}R_j^{\left(\bm{b}^\prime,\bm{c}\right)}}}\;,
\end{equation*}
where $R_j^{\left(\bm{b}^\prime,\bm{c}\right)}=D_j^{\left(\bm{b}^\prime\right)}F_j^{\left(\bm{c}\right)}$ represents a Pauli operator on the $j$th qubit.

We can integrate stabilizer measurements with single-qubit measurement decoding. Consider a stabilizer code in its standard form. The $i$th generator, for $i<r$, consists of a single bit flip on the $i$th qubit. After measuring $g_i$, the following stabilizer and single-qubit measurements use only Pauli-$Z$ operators on the $i$th qubit. This allows us to replace $Z_i$ in these stabilizers with its measurement outcome, reducing the number of two-qubit gates. Refer to Figure~\ref{fig:circuit_integrated} for an example of entanglement distillation using the $\left[\!\left[5,1,3\right]\!\right]$ code.

\section{Decoding Quantum Convolutional Codes with Laurent Series} \label{app:Laurent_series}

Given the standard form of logical operators for quantum convolutional codes in Equation~\ref{eq:convolitional_logical}, the matrix entries are generally Laurent series. A Laurent series $f\left(D^{-1}\right)$ can be expressed as $\frac{p\left(D\right)}{q\left(D\right)}$, where $p\left(D\right)$ and $q\left(D\right)$ are polynomials. Thus, the logical operators can be represented as
\begin{eqnarray*}
    L_X\left(D\right) &=& \frac{1}{Q\left(D\right)}\left[ \begin{array}{ccc|ccc}
    \bm{0} & \bm{E}^\prime\left(D\right) & \bm{Q}\left(D\right) & \bm{C}^\prime\left(D\right) & \bm{0} & \bm{0}
    \end{array} \right]\;,\nonumber\\
    L_Z\left(D\right) &=& \frac{1}{Q\left(D\right)}\left[ \begin{array}{ccc|ccc}
    \bm{0} & \bm{0} & \bm{0} & \bm{A}^\prime\left(D\right) & \bm{0} & \bm{Q}\left(D\right)
    \end{array}\right]\;,
\end{eqnarray*}
where $Q\left(D\right)$ is the least common multiple of the denominators of all entries, also a polynomial. For convenience, we define
\begin{eqnarray*}
    \bm{Q}\left(D\right) &=& Q\left(D\right)\bm{I}\;, \\
    \bm{A}^\prime\left(D\right) &=& Q\left(D\right)\bm{A}_2^T\left(D^{-1}\right)\;, \\
    \bm{C}^\prime\left(D\right) &=& Q\left(D\right)\bm{C}^T\left(D^{-1}\right)\;, \\
    \bm{E}^\prime\left(D\right) &=& Q\left(D\right)\bm{E}^T\left(D^{-1}\right)\;.
\end{eqnarray*}
Multiplying the matrix of the logical operators by $Q\left(D\right)$, we obtain a new set of logical operators
\begin{eqnarray*}
    L_X^\prime\left(D\right) &=& \left[ \begin{array}{ccc|ccc}
    \bm{0} & \bm{E}^\prime\left(D\right) & \bm{Q}\left(D\right) & \bm{C}^\prime\left(D\right) & \bm{0} & \bm{0}
    \end{array} \right]\;,\nonumber\\
    L_Z^\prime\left(D\right) &=& \left[ \begin{array}{ccc|ccc}
    \bm{0} & \bm{0} & \bm{0} & \bm{A}^\prime\left(D\right) & \bm{0} & \bm{Q}\left(D\right)
    \end{array}\right]\;,
\end{eqnarray*}
where all entries are finite Laurent polynomials.

Single-qubit measurement decoding applies to these logical operators, decoupling different logical qubits while qubits within the same logical qubit remain coupled. Their relationship is determined by the polynomial $Q\left(D\right)$. Given $Q\left(D\right)=\sum_{l=0}^{m}q_lD^l$, the $i$th logical operators can be represented by
\begin{eqnarray*}
    X_L^{(i)} &=& \bigotimes_{l:q_l=1}^{m}X_{\left(l+1\right)n-k+i}\;,\\
    Z_L^{(i)} &=& \bigotimes_{l:q_l=1}^{m}Z_{\left(l+1\right)n-k+i}\;.
\end{eqnarray*}
These operators, associated with a catastrophic encoding circuit~\cite{Ollivier2004}, require further decoding. For example, with $Q\left(D\right)=1+D+D^2$, the logical qubits are encoded by the following circuit
\begin{center}
    \includegraphics[width=0.25\paperwidth]{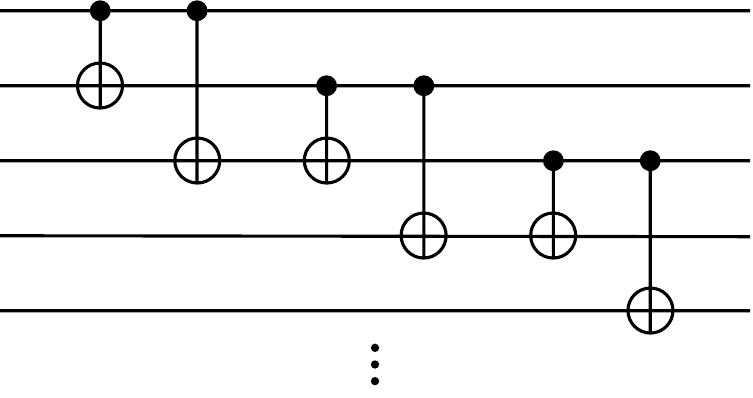}
\end{center}
Here, a single-qubit error operator $X$ can propagate indefinitely through the CNOT gates.

To avoid catastrophic error propagation in decoding logical qubits, we use an entanglement-assisted method. For simplicity, we illustrate this with $Q\left(D\right)=1+D+D^2$, which can be extended to any polynomials. We relabel the relevant qubits and define the logical operators as
\begin{eqnarray*}
    X_L\left(k\right) &=& X_kX_{k+1}X_{k+2}\;, \\
    Z_L\left(k\right) &=& Z_kZ_{k+1}Z_{k+2}\;, \\
\end{eqnarray*}
where $k$ indexes the frames.

Considering logical Bell states shared between Alice and Bob, their stabilizers are
\begin{equation*}
    \begin{matrix}
    X_1X_2X_3\otimes X_1X_2X_3\;, & Z_1Z_2Z_3\otimes Z_1Z_2Z_3 \\
    X_2X_3X_4\otimes X_2X_3X_4\;, & Z_2Z_3Z_4\otimes Z_2Z_3Z_4 \\
    X_3X_4X_5\otimes X_3X_4X_5\;, & Z_3Z_4Z_5\otimes Z_3Z_4Z_5 \\
    \vdots & \vdots
    \end{matrix}\;.
\end{equation*}
The parities of the operators $X_1\otimes X_1$, $Z_1\otimes Z_1$, $X_2\otimes X_2$, and $Z_2\otimes Z_2$ are unknown. Assuming Alice and Bob have two pre-shared standard Bell states, indexed as $-1$ and $0$, the total stabilizers are
\begin{widetext}
\begin{equation*}
    \begin{matrix}
    X_{-1}\otimes X_{-1}\;, & Z_{-1}\otimes Z_{-1} \\
    X_0\otimes X_0\;, & Z_0\otimes Z_0 \\
    X_{-1}X_1X_0X_2X_3\otimes X_{-1}X_1X_0X_2X_3\;, & Z_{-1}Z_1Z_0Z_2Z_3\otimes Z_{-1}Z_1Z_0Z_2Z_3 \\
    X_0X_2X_3X_4\otimes X_0X_2X_3X_4\;, & Z_0Z_2Z_3Z_4\otimes Z_0Z_2Z_3Z_4 \\
    X_3X_4X_5\otimes X_3X_4X_5\;, & Z_3Z_4Z_5\otimes Z_3Z_4Z_5 \\
    \vdots & \vdots
    \end{matrix}\;.
\end{equation*}
\end{widetext}
Alice and Bob perform local parity measurements of $X_{-1}X_1$, $Z_{-1}Z_1$, $X_0X_2$, and $Z_0Z_2$, resulting in the post-measurement stabilizers
\begin{widetext}
\begin{equation*}
    \begin{matrix}
    \left(-1\right)^{a_x^{(1)}+a_x^{(2)}+b_x^{(1)}+b_x^{(2)}}X_3\otimes X_3\;, & \left(-1\right)^{a_z^{(1)}+a_z^{(2)}+b_z^{(1)}+b_z^{(2)}}Z_3\otimes Z_3 \\
    \left(-1\right)^{a_x^{(2)}+b_x^{(2)}}X_3X_4\otimes X_3X_4\;, & \left(-1\right)^{a_z^{(2)}+b_z^{(2)}}Z_3Z_4\otimes Z_3Z_4 \\
    X_3X_4X_5\otimes X_3X_4X_5\;, & Z_3Z_4Z_5\otimes Z_3Z_4Z_5\\
    \vdots & \vdots
    \end{matrix}\;.
\end{equation*}
\end{widetext}
Here, $a_x^{(1)}$ and $b_x^{(1)}$ are the measurement outcomes of $X_{-1}X_1$ by Alice and Bob, respectively. $a_x^{(2)}$ and $b_x^{(2)}$ are the outcomes of $X_0X_2$, $a_z^{(1)}$ and $b_z^{(1)}$ are the outcomes of $Z_{-1}Z_1$, and $a_z^{(2)}$ and $b_z^{(2)}$ are the outcomes of $Z_0Z_2$. By recursively calculating the parities for subsequent physical Bell states, these logical Bell states are decoded.

Errors in parity measurements can propagate indefinitely. For example, if an error occurs on $a_x^1$, it introduces an incorrect recovery phase for $X_3\otimes X_3$, which then propagates through subsequent operators such as $X_4\otimes X_4$ and $X_5\otimes X_5$, leading to catastrophic errors. To address this, we segment the sequence into blocks of $N$ frames. Measuring parities independently for each block confines error propagation to a length of $N$. For a general polynomial $Q\left(D\right)=\sum_{l=0}^{m}q_lD^l$, the entanglement-assisted approach consumes $m$ pre-shared Bell states and the first $m$ Bell states per block, yielding $N-m$ decoded physical Bell states. The overhead, defined as the ratio of consumed to output Bell states, is $\frac{2m}{N-m}$, approaching zero as $N$ increases.

\section{Deriving the Standard Form for the Rate-1/3 Convolutional Code} \label{app:conv_standard_form}

The check matrix of the rate-$1/3$ quantum convolutional code from~\cite{Forney2007} is given by
\begin{equation*}
    \left[ \begin{array}{ccc|ccc}
    1+D & 1 & 1+D & 0 & D & D \\
    0 & D & D & 1+D & 1+D & 1
    \end{array} \right]\;.
\end{equation*}
We derive its standard form through the following steps:
\begin{widetext}
\begin{enumerate}
    \item $\rm Permute\ Columns\ \left(1,2,3\right)\ \longrightarrow\ \left(2,1,3\right):$
        \begin{equation*}
            \left[ \begin{array}{ccc|ccc}
            1 & 1+D & 1+D & D & 0 & D \\
            D & 0 & D & 1+D & 1+D & 1
            \end{array} \right]\;.
        \end{equation*}
        
    \item $\rm Row\ 2\ \longrightarrow\ D^{-1}\times Row\ 2 + Row\ 1:$
	\begin{equation*}
    	\left[ \begin{array}{ccc|ccc}
    	1 & 1+D & 1+D & D & 0 & D \\
    	0 & 1+D & D & D^{-1}+1+D & D^{-1}+1 & D^{-1}+D \\
    	\end{array} \right]\;.
	\end{equation*}

    \item $\rm Permute\ Columns\ \left(1,2,3\right)\ \longrightarrow\ \left(1,3,2\right):$
	\begin{equation*}
	\left[\begin{array}{ccc|ccc}
	1 & 1+D & 1+D & D & D & 0 \\
	0 & D & 1+D & D^{-1}+1+D & D^{-1}+D & D^{-1}+1 \\
	\end{array}\right]\;.
	\end{equation*}

    \item $\rm Row\ 2\ \longrightarrow\ D^{-1}\times Row\ 2:$
	\begin{equation*}
	\left[\begin{array}{ccc|ccc}
	1 & 1+D & 1+D & D & D & 0 \\
	0 & 1 & D^{-1}+1 & D^{-2}+D^{-1}+1 & D^{-2}+1 & D^{-2}+D^{-1} \\
	\end{array}\right]\;.
	\end{equation*}

    \item $\rm Row\ 1\ \longrightarrow\ Row\ 1 + \left(1+D\right) \times Row\ 2:$
	\begin{equation*}
	\left[\begin{array}{ccc|ccc}
	1 & 0 & D^{-1}+1 & D^{-2} & D^{-2}+D^{-1}+1 & D^{-2}+1 \\
	0 & 1 & D^{-1}+1 & D^{-2}+D^{-1}+1 & D^{-2}+1 & D^{-2}+D^{-1} \\
	\end{array}\right]\;.
	\end{equation*}
\end{enumerate}
\end{widetext}

This process results in the standard form as specified in Equation~\ref{eq:conv_standard_form}. Note that during Gaussian elimination, multiplication factors representing finite shifts are preferred. Using a Laurent series as a multiplication factor, such as $\frac{1}{1+D}=1+D+D^2+\cdots$, can result in a standard form consisting of Laurent series entries, increasing decoding complexity. Obtaining a standard form solely through row Gaussian elimination is challenging, as it requires column entries to be coprime; otherwise, only the greatest common divisor of the polynomials is obtained. Further elimination would require column operations involving two-qubit gates, which we aim to avoid in measurement-based decoding. We leave it as an open problem to develop a general algorithm for achieving a standard form with finite entries.

\section{Modeling Errors in Measurement-Based Decoding} \label{app:measurement_error}

In measurement-based decoding, Pauli operators in logical operators are replaced by their measurement outcomes
\begin{eqnarray*}
    X_L^{(i)} &=& \left(-1\right)^{\alpha_x^{(i)}}X_{n-k+i}\;,\\
    Z_L^{(i)} &=& \left(-1\right)^{\alpha_z^{(i)}}Z_{n-k+i}\;,
\end{eqnarray*}
where $\alpha_x^{(i)}=\sum_{j:C_{ij}^T=1} a_j+\sum_{j:E_{ij}^T=1} a_{r+j}$ and $\alpha_z^{(i)}=\sum_{j:\left(A_2^T\right)_{ij}=1} a_j$, as defined in Equations~\ref{eq:physical_x} and~\ref{eq:physical_z}. Noise from single-qubit measurements leads to phase errors. $\alpha_x^{(i)}$ errors yield a $Z$ error, $\alpha_z^{(i)}$ errors yield an $X$ error, and errors in both $\alpha_x^{(i)}$ and $\alpha_z^{(i)}$ yield a $Y$ error on the decoded qubit.

We can derive the error probability for a qubit decoded from a planar code on an $L \times L$ lattice using induction. To decode the logical $X_L$ operator, $L-1$ qubits are measured in the $X$ basis along a horizontal line. The error probability of the parity sum for $k$ qubits is given by
\begin{equation*}
    q_k = \left(1-p\right)\cdot q_{k-1} + p\cdot\left(1-q_{k-1}\right)\;,
\end{equation*}
where $p$ is the error probability of a single-qubit measurement, and $q_{k-1}$ is the error probability of the parity sum for $k-1$ qubits. We obtain
\begin{equation*}
    q_k = \frac{1}{2} - \frac{1}{2}(1-2p)^k\;.
\end{equation*}
The effective error arises when parities on logical $X_L$ or $Z_L$ operators are incorrect, with a probability of $2q_k - q_k^2$. Let $k=L-1$ for decoding a planar code on an $L \times L$ lattice, the error probability is given by 
\begin{equation*}
    P_L = \frac{3}{4} - \frac{1}{2}(1-2p)^{L-1} - \frac{1}{4}(1-2p)^{2(L-1)}\;.
\end{equation*}

\bibliography{main.bib}

\begin{thebibliography}{135}%
\makeatletter
\providecommand \@ifxundefined [1]{%
 \@ifx{#1\undefined}
}%
\providecommand \@ifnum [1]{%
 \ifnum #1\expandafter \@firstoftwo
 \else \expandafter \@secondoftwo
 \fi
}%
\providecommand \@ifx [1]{%
 \ifx #1\expandafter \@firstoftwo
 \else \expandafter \@secondoftwo
 \fi
}%
\providecommand \natexlab [1]{#1}%
\providecommand \enquote  [1]{``#1''}%
\providecommand \bibnamefont  [1]{#1}%
\providecommand \bibfnamefont [1]{#1}%
\providecommand \citenamefont [1]{#1}%
\providecommand \href@noop [0]{\@secondoftwo}%
\providecommand \href [0]{\begingroup \@sanitize@url \@href}%
\providecommand \@href[1]{\@@startlink{#1}\@@href}%
\providecommand \@@href[1]{\endgroup#1\@@endlink}%
\providecommand \@sanitize@url [0]{\catcode `\\12\catcode `\$12\catcode `\&12\catcode `\#12\catcode `\^12\catcode `\_12\catcode `\%12\relax}%
\providecommand \@@startlink[1]{}%
\providecommand \@@endlink[0]{}%
\providecommand \url  [0]{\begingroup\@sanitize@url \@url }%
\providecommand \@url [1]{\endgroup\@href {#1}{\urlprefix }}%
\providecommand \urlprefix  [0]{URL }%
\providecommand \Eprint [0]{\href }%
\providecommand \doibase [0]{https://doi.org/}%
\providecommand \selectlanguage [0]{\@gobble}%
\providecommand \bibinfo  [0]{\@secondoftwo}%
\providecommand \bibfield  [0]{\@secondoftwo}%
\providecommand \translation [1]{[#1]}%
\providecommand \BibitemOpen [0]{}%
\providecommand \bibitemStop [0]{}%
\providecommand \bibitemNoStop [0]{.\EOS\space}%
\providecommand \EOS [0]{\spacefactor3000\relax}%
\providecommand \BibitemShut  [1]{\csname bibitem#1\endcsname}%
\let\auto@bib@innerbib\@empty
%</preamble>
\bibitem [{\citenamefont {Wehner}\ \emph {et~al.}(2018)\citenamefont {Wehner}, \citenamefont {Elkouss},\ and\ \citenamefont {Hanson}}]{Wehner2018}%
  \BibitemOpen
  \bibfield  {author} {\bibinfo {author} {\bibfnamefont {S.}~\bibnamefont {Wehner}}, \bibinfo {author} {\bibfnamefont {D.}~\bibnamefont {Elkouss}},\ and\ \bibinfo {author} {\bibfnamefont {R.}~\bibnamefont {Hanson}},\ }\bibfield  {title} {\bibinfo {title} {Quantum internet: A vision for the road ahead},\ }\href {https://doi.org/10.1126/science.aam9288} {\bibfield  {journal} {\bibinfo  {journal} {Science}\ }\textbf {\bibinfo {volume} {362}},\ \bibinfo {pages} {eaam9288} (\bibinfo {year} {2018})}\BibitemShut {NoStop}%
\bibitem [{\citenamefont {Awschalom}\ \emph {et~al.}(2021)\citenamefont {Awschalom}, \citenamefont {Berggren}, \citenamefont {Bernien}, \citenamefont {Bhave}, \citenamefont {Carr}, \citenamefont {Davids}, \citenamefont {Economou}, \citenamefont {Englund}, \citenamefont {Faraon}, \citenamefont {Fejer}, \citenamefont {Guha}, \citenamefont {Gustafsson}, \citenamefont {Hu}, \citenamefont {Jiang}, \citenamefont {Kim}, \citenamefont {Korzh}, \citenamefont {Kumar}, \citenamefont {Kwiat}, \citenamefont {Lon\ifmmode~\check{c}\else \v{c}\fi{}ar}, \citenamefont {Lukin}, \citenamefont {Miller}, \citenamefont {Monroe}, \citenamefont {Nam}, \citenamefont {Narang}, \citenamefont {Orcutt}, \citenamefont {Raymer}, \citenamefont {Safavi-Naeini}, \citenamefont {Spiropulu}, \citenamefont {Srinivasan}, \citenamefont {Sun}, \citenamefont {Vu\ifmmode \check{c}\else \v{c}\fi{}kovi\ifmmode~\acute{c}\else \'{c}\fi{}}, \citenamefont {Waks}, \citenamefont {Walsworth}, \citenamefont {Weiner},\ and\ \citenamefont {Zhang}}]{Awschalom2021}%
  \BibitemOpen
  \bibfield  {author} {\bibinfo {author} {\bibfnamefont {D.}~\bibnamefont {Awschalom}}, \bibinfo {author} {\bibfnamefont {K.~K.}\ \bibnamefont {Berggren}}, \bibinfo {author} {\bibfnamefont {H.}~\bibnamefont {Bernien}}, \bibinfo {author} {\bibfnamefont {S.}~\bibnamefont {Bhave}}, \bibinfo {author} {\bibfnamefont {L.~D.}\ \bibnamefont {Carr}}, \bibinfo {author} {\bibfnamefont {P.}~\bibnamefont {Davids}}, \bibinfo {author} {\bibfnamefont {S.~E.}\ \bibnamefont {Economou}}, \bibinfo {author} {\bibfnamefont {D.}~\bibnamefont {Englund}}, \bibinfo {author} {\bibfnamefont {A.}~\bibnamefont {Faraon}}, \bibinfo {author} {\bibfnamefont {M.}~\bibnamefont {Fejer}}, \bibinfo {author} {\bibfnamefont {S.}~\bibnamefont {Guha}}, \bibinfo {author} {\bibfnamefont {M.~V.}\ \bibnamefont {Gustafsson}}, \bibinfo {author} {\bibfnamefont {E.}~\bibnamefont {Hu}}, \bibinfo {author} {\bibfnamefont {L.}~\bibnamefont {Jiang}}, \bibinfo {author} {\bibfnamefont {J.}~\bibnamefont {Kim}}, \bibinfo {author} {\bibfnamefont {B.}~\bibnamefont
  {Korzh}}, \bibinfo {author} {\bibfnamefont {P.}~\bibnamefont {Kumar}}, \bibinfo {author} {\bibfnamefont {P.~G.}\ \bibnamefont {Kwiat}}, \bibinfo {author} {\bibfnamefont {M.}~\bibnamefont {Lon\ifmmode~\check{c}\else \v{c}\fi{}ar}}, \bibinfo {author} {\bibfnamefont {M.~D.}\ \bibnamefont {Lukin}}, \bibinfo {author} {\bibfnamefont {D.~A.}\ \bibnamefont {Miller}}, \bibinfo {author} {\bibfnamefont {C.}~\bibnamefont {Monroe}}, \bibinfo {author} {\bibfnamefont {S.~W.}\ \bibnamefont {Nam}}, \bibinfo {author} {\bibfnamefont {P.}~\bibnamefont {Narang}}, \bibinfo {author} {\bibfnamefont {J.~S.}\ \bibnamefont {Orcutt}}, \bibinfo {author} {\bibfnamefont {M.~G.}\ \bibnamefont {Raymer}}, \bibinfo {author} {\bibfnamefont {A.~H.}\ \bibnamefont {Safavi-Naeini}}, \bibinfo {author} {\bibfnamefont {M.}~\bibnamefont {Spiropulu}}, \bibinfo {author} {\bibfnamefont {K.}~\bibnamefont {Srinivasan}}, \bibinfo {author} {\bibfnamefont {S.}~\bibnamefont {Sun}}, \bibinfo {author} {\bibfnamefont {J.}~\bibnamefont {Vu\ifmmode \check{c}\else
  \v{c}\fi{}kovi\ifmmode~\acute{c}\else \'{c}\fi{}}}, \bibinfo {author} {\bibfnamefont {E.}~\bibnamefont {Waks}}, \bibinfo {author} {\bibfnamefont {R.}~\bibnamefont {Walsworth}}, \bibinfo {author} {\bibfnamefont {A.~M.}\ \bibnamefont {Weiner}},\ and\ \bibinfo {author} {\bibfnamefont {Z.}~\bibnamefont {Zhang}},\ }\bibfield  {title} {\bibinfo {title} {Development of quantum interconnects (quics) for next-generation information technologies},\ }\href {https://doi.org/10.1103/PRXQuantum.2.017002} {\bibfield  {journal} {\bibinfo  {journal} {PRX Quantum}\ }\textbf {\bibinfo {volume} {2}},\ \bibinfo {pages} {017002} (\bibinfo {year} {2021})}\BibitemShut {NoStop}%
\bibitem [{\citenamefont {Azuma}\ \emph {et~al.}(2023)\citenamefont {Azuma}, \citenamefont {Economou}, \citenamefont {Elkouss}, \citenamefont {Hilaire}, \citenamefont {Jiang}, \citenamefont {Lo},\ and\ \citenamefont {Tzitrin}}]{Azuma2023}%
  \BibitemOpen
  \bibfield  {author} {\bibinfo {author} {\bibfnamefont {K.}~\bibnamefont {Azuma}}, \bibinfo {author} {\bibfnamefont {S.~E.}\ \bibnamefont {Economou}}, \bibinfo {author} {\bibfnamefont {D.}~\bibnamefont {Elkouss}}, \bibinfo {author} {\bibfnamefont {P.}~\bibnamefont {Hilaire}}, \bibinfo {author} {\bibfnamefont {L.}~\bibnamefont {Jiang}}, \bibinfo {author} {\bibfnamefont {H.-K.}\ \bibnamefont {Lo}},\ and\ \bibinfo {author} {\bibfnamefont {I.}~\bibnamefont {Tzitrin}},\ }\bibfield  {title} {\bibinfo {title} {Quantum repeaters: From quantum networks to the quantum internet},\ }\href {https://doi.org/10.1103/RevModPhys.95.045006} {\bibfield  {journal} {\bibinfo  {journal} {Rev. Mod. Phys.}\ }\textbf {\bibinfo {volume} {95}},\ \bibinfo {pages} {045006} (\bibinfo {year} {2023})}\BibitemShut {NoStop}%
\bibitem [{\citenamefont {Li}\ \emph {et~al.}(2023)\citenamefont {Li}, \citenamefont {Xue}, \citenamefont {Li}, \citenamefont {Chen}, \citenamefont {Li}, \citenamefont {Wang}, \citenamefont {Yu}, \citenamefont {Wei}, \citenamefont {Sun},\ and\ \citenamefont {Lu}}]{Li2023}%
  \BibitemOpen
  \bibfield  {author} {\bibinfo {author} {\bibfnamefont {Z.}~\bibnamefont {Li}}, \bibinfo {author} {\bibfnamefont {K.}~\bibnamefont {Xue}}, \bibinfo {author} {\bibfnamefont {J.}~\bibnamefont {Li}}, \bibinfo {author} {\bibfnamefont {L.}~\bibnamefont {Chen}}, \bibinfo {author} {\bibfnamefont {R.}~\bibnamefont {Li}}, \bibinfo {author} {\bibfnamefont {Z.}~\bibnamefont {Wang}}, \bibinfo {author} {\bibfnamefont {N.}~\bibnamefont {Yu}}, \bibinfo {author} {\bibfnamefont {D.~S.~L.}\ \bibnamefont {Wei}}, \bibinfo {author} {\bibfnamefont {Q.}~\bibnamefont {Sun}},\ and\ \bibinfo {author} {\bibfnamefont {J.}~\bibnamefont {Lu}},\ }\bibfield  {title} {\bibinfo {title} {Entanglement-assisted quantum networks: Mechanics, enabling technologies, challenges, and research directions},\ }\href {https://doi.org/10.1109/COMST.2023.3294240} {\bibfield  {journal} {\bibinfo  {journal} {IEEE Communications Surveys \& Tutorials}\ }\textbf {\bibinfo {volume} {25}},\ \bibinfo {pages} {2133} (\bibinfo {year} {2023})}\BibitemShut {NoStop}%
\bibitem [{\citenamefont {Giovannetti}\ \emph {et~al.}(2011)\citenamefont {Giovannetti}, \citenamefont {Lloyd},\ and\ \citenamefont {Maccone}}]{Giovannetti2011}%
  \BibitemOpen
  \bibfield  {author} {\bibinfo {author} {\bibfnamefont {V.}~\bibnamefont {Giovannetti}}, \bibinfo {author} {\bibfnamefont {S.}~\bibnamefont {Lloyd}},\ and\ \bibinfo {author} {\bibfnamefont {L.}~\bibnamefont {Maccone}},\ }\bibfield  {title} {\bibinfo {title} {Advances in quantum metrology},\ }\href {https://doi.org/10.1038/nphoton.2011.35} {\bibfield  {journal} {\bibinfo  {journal} {Nature Photonics}\ }\textbf {\bibinfo {volume} {5}},\ \bibinfo {pages} {222} (\bibinfo {year} {2011})}\BibitemShut {NoStop}%
\bibitem [{\citenamefont {Gottesman}\ \emph {et~al.}(2012)\citenamefont {Gottesman}, \citenamefont {Jennewein},\ and\ \citenamefont {Croke}}]{Gottesman2012}%
  \BibitemOpen
  \bibfield  {author} {\bibinfo {author} {\bibfnamefont {D.}~\bibnamefont {Gottesman}}, \bibinfo {author} {\bibfnamefont {T.}~\bibnamefont {Jennewein}},\ and\ \bibinfo {author} {\bibfnamefont {S.}~\bibnamefont {Croke}},\ }\bibfield  {title} {\bibinfo {title} {Longer-baseline telescopes using quantum repeaters},\ }\href {https://doi.org/10.1103/PhysRevLett.109.070503} {\bibfield  {journal} {\bibinfo  {journal} {Phys. Rev. Lett.}\ }\textbf {\bibinfo {volume} {109}},\ \bibinfo {pages} {070503} (\bibinfo {year} {2012})}\BibitemShut {NoStop}%
\bibitem [{\citenamefont {Tóth}\ and\ \citenamefont {Apellaniz}(2014)}]{Toth2014}%
  \BibitemOpen
  \bibfield  {author} {\bibinfo {author} {\bibfnamefont {G.}~\bibnamefont {Tóth}}\ and\ \bibinfo {author} {\bibfnamefont {I.}~\bibnamefont {Apellaniz}},\ }\bibfield  {title} {\bibinfo {title} {Quantum metrology from a quantum information science perspective},\ }\href {https://doi.org/10.1088/1751-8113/47/42/424006} {\bibfield  {journal} {\bibinfo  {journal} {Journal of Physics A: Mathematical and Theoretical}\ }\textbf {\bibinfo {volume} {47}},\ \bibinfo {pages} {424006} (\bibinfo {year} {2014})}\BibitemShut {NoStop}%
\bibitem [{\citenamefont {Pezz\`e}\ \emph {et~al.}(2018)\citenamefont {Pezz\`e}, \citenamefont {Smerzi}, \citenamefont {Oberthaler}, \citenamefont {Schmied},\ and\ \citenamefont {Treutlein}}]{Pezze2018}%
  \BibitemOpen
  \bibfield  {author} {\bibinfo {author} {\bibfnamefont {L.}~\bibnamefont {Pezz\`e}}, \bibinfo {author} {\bibfnamefont {A.}~\bibnamefont {Smerzi}}, \bibinfo {author} {\bibfnamefont {M.~K.}\ \bibnamefont {Oberthaler}}, \bibinfo {author} {\bibfnamefont {R.}~\bibnamefont {Schmied}},\ and\ \bibinfo {author} {\bibfnamefont {P.}~\bibnamefont {Treutlein}},\ }\bibfield  {title} {\bibinfo {title} {Quantum metrology with nonclassical states of atomic ensembles},\ }\href {https://doi.org/10.1103/RevModPhys.90.035005} {\bibfield  {journal} {\bibinfo  {journal} {Rev. Mod. Phys.}\ }\textbf {\bibinfo {volume} {90}},\ \bibinfo {pages} {035005} (\bibinfo {year} {2018})}\BibitemShut {NoStop}%
\bibitem [{\citenamefont {Gottesman}\ and\ \citenamefont {Chuang}(1999)}]{Gottesman1999}%
  \BibitemOpen
  \bibfield  {author} {\bibinfo {author} {\bibfnamefont {D.}~\bibnamefont {Gottesman}}\ and\ \bibinfo {author} {\bibfnamefont {I.~L.}\ \bibnamefont {Chuang}},\ }\bibfield  {title} {\bibinfo {title} {Demonstrating the viability of universal quantum computation using teleportation and single-qubit operations},\ }\href {https://doi.org/10.1038/46503} {\bibfield  {journal} {\bibinfo  {journal} {Nature}\ }\textbf {\bibinfo {volume} {402}},\ \bibinfo {pages} {390} (\bibinfo {year} {1999})}\BibitemShut {NoStop}%
\bibitem [{\citenamefont {D\"ur}\ and\ \citenamefont {Briegel}(2003)}]{Dur2003}%
  \BibitemOpen
  \bibfield  {author} {\bibinfo {author} {\bibfnamefont {W.}~\bibnamefont {D\"ur}}\ and\ \bibinfo {author} {\bibfnamefont {H.-J.}\ \bibnamefont {Briegel}},\ }\bibfield  {title} {\bibinfo {title} {Entanglement purification for quantum computation},\ }\href {https://doi.org/10.1103/PhysRevLett.90.067901} {\bibfield  {journal} {\bibinfo  {journal} {Phys. Rev. Lett.}\ }\textbf {\bibinfo {volume} {90}},\ \bibinfo {pages} {067901} (\bibinfo {year} {2003})}\BibitemShut {NoStop}%
\bibitem [{\citenamefont {Briegel}\ \emph {et~al.}(2009)\citenamefont {Briegel}, \citenamefont {Browne}, \citenamefont {D{\"u}r}, \citenamefont {Raussendorf},\ and\ \citenamefont {Van~den Nest}}]{Briegel2009}%
  \BibitemOpen
  \bibfield  {author} {\bibinfo {author} {\bibfnamefont {H.~J.}\ \bibnamefont {Briegel}}, \bibinfo {author} {\bibfnamefont {D.~E.}\ \bibnamefont {Browne}}, \bibinfo {author} {\bibfnamefont {W.}~\bibnamefont {D{\"u}r}}, \bibinfo {author} {\bibfnamefont {R.}~\bibnamefont {Raussendorf}},\ and\ \bibinfo {author} {\bibfnamefont {M.}~\bibnamefont {Van~den Nest}},\ }\bibfield  {title} {\bibinfo {title} {Measurement-based quantum computation},\ }\href {https://doi.org/10.1038/nphys1157} {\bibfield  {journal} {\bibinfo  {journal} {Nature Physics}\ }\textbf {\bibinfo {volume} {5}},\ \bibinfo {pages} {19} (\bibinfo {year} {2009})}\BibitemShut {NoStop}%
\bibitem [{\citenamefont {Li}\ and\ \citenamefont {Benjamin}(2012)}]{Li2012}%
  \BibitemOpen
  \bibfield  {author} {\bibinfo {author} {\bibfnamefont {Y.}~\bibnamefont {Li}}\ and\ \bibinfo {author} {\bibfnamefont {S.~C.}\ \bibnamefont {Benjamin}},\ }\bibfield  {title} {\bibinfo {title} {High threshold distributed quantum computing with three-qubit nodes},\ }\href {https://doi.org/10.1088/1367-2630/14/9/093008} {\bibfield  {journal} {\bibinfo  {journal} {New Journal of Physics}\ }\textbf {\bibinfo {volume} {14}},\ \bibinfo {pages} {093008} (\bibinfo {year} {2012})}\BibitemShut {NoStop}%
\bibitem [{\citenamefont {Hu}\ \emph {et~al.}(2023)\citenamefont {Hu}, \citenamefont {Guo}, \citenamefont {Liu}, \citenamefont {Li},\ and\ \citenamefont {Guo}}]{Hu2023}%
  \BibitemOpen
  \bibfield  {author} {\bibinfo {author} {\bibfnamefont {X.-M.}\ \bibnamefont {Hu}}, \bibinfo {author} {\bibfnamefont {Y.}~\bibnamefont {Guo}}, \bibinfo {author} {\bibfnamefont {B.-H.}\ \bibnamefont {Liu}}, \bibinfo {author} {\bibfnamefont {C.-F.}\ \bibnamefont {Li}},\ and\ \bibinfo {author} {\bibfnamefont {G.-C.}\ \bibnamefont {Guo}},\ }\bibfield  {title} {\bibinfo {title} {Progress in quantum teleportation},\ }\href {https://doi.org/10.1038/s42254-023-00588-x} {\bibfield  {journal} {\bibinfo  {journal} {Nature Reviews Physics}\ }\textbf {\bibinfo {volume} {5}},\ \bibinfo {pages} {339} (\bibinfo {year} {2023})}\BibitemShut {NoStop}%
\bibitem [{\citenamefont {Briegel}\ \emph {et~al.}(1998)\citenamefont {Briegel}, \citenamefont {D\"ur}, \citenamefont {Cirac},\ and\ \citenamefont {Zoller}}]{Briegel1998}%
  \BibitemOpen
  \bibfield  {author} {\bibinfo {author} {\bibfnamefont {H.-J.}\ \bibnamefont {Briegel}}, \bibinfo {author} {\bibfnamefont {W.}~\bibnamefont {D\"ur}}, \bibinfo {author} {\bibfnamefont {J.~I.}\ \bibnamefont {Cirac}},\ and\ \bibinfo {author} {\bibfnamefont {P.}~\bibnamefont {Zoller}},\ }\bibfield  {title} {\bibinfo {title} {Quantum repeaters: The role of imperfect local operations in quantum communication},\ }\href {https://doi.org/10.1103/PhysRevLett.81.5932} {\bibfield  {journal} {\bibinfo  {journal} {Phys. Rev. Lett.}\ }\textbf {\bibinfo {volume} {81}},\ \bibinfo {pages} {5932} (\bibinfo {year} {1998})}\BibitemShut {NoStop}%
\bibitem [{\citenamefont {D\"ur}\ \emph {et~al.}(1999)\citenamefont {D\"ur}, \citenamefont {Briegel}, \citenamefont {Cirac},\ and\ \citenamefont {Zoller}}]{Dur1999}%
  \BibitemOpen
  \bibfield  {author} {\bibinfo {author} {\bibfnamefont {W.}~\bibnamefont {D\"ur}}, \bibinfo {author} {\bibfnamefont {H.-J.}\ \bibnamefont {Briegel}}, \bibinfo {author} {\bibfnamefont {J.~I.}\ \bibnamefont {Cirac}},\ and\ \bibinfo {author} {\bibfnamefont {P.}~\bibnamefont {Zoller}},\ }\bibfield  {title} {\bibinfo {title} {Quantum repeaters based on entanglement purification},\ }\href {https://doi.org/10.1103/PhysRevA.59.169} {\bibfield  {journal} {\bibinfo  {journal} {Phys. Rev. A}\ }\textbf {\bibinfo {volume} {59}},\ \bibinfo {pages} {169} (\bibinfo {year} {1999})}\BibitemShut {NoStop}%
\bibitem [{\citenamefont {Jiang}\ \emph {et~al.}(2009)\citenamefont {Jiang}, \citenamefont {Taylor}, \citenamefont {Nemoto}, \citenamefont {Munro}, \citenamefont {Van~Meter},\ and\ \citenamefont {Lukin}}]{Jiang2009}%
  \BibitemOpen
  \bibfield  {author} {\bibinfo {author} {\bibfnamefont {L.}~\bibnamefont {Jiang}}, \bibinfo {author} {\bibfnamefont {J.~M.}\ \bibnamefont {Taylor}}, \bibinfo {author} {\bibfnamefont {K.}~\bibnamefont {Nemoto}}, \bibinfo {author} {\bibfnamefont {W.~J.}\ \bibnamefont {Munro}}, \bibinfo {author} {\bibfnamefont {R.}~\bibnamefont {Van~Meter}},\ and\ \bibinfo {author} {\bibfnamefont {M.~D.}\ \bibnamefont {Lukin}},\ }\bibfield  {title} {\bibinfo {title} {Quantum repeater with encoding},\ }\href {https://doi.org/10.1103/PhysRevA.79.032325} {\bibfield  {journal} {\bibinfo  {journal} {Phys. Rev. A}\ }\textbf {\bibinfo {volume} {79}},\ \bibinfo {pages} {032325} (\bibinfo {year} {2009})}\BibitemShut {NoStop}%
\bibitem [{\citenamefont {Muralidharan}\ \emph {et~al.}(2014)\citenamefont {Muralidharan}, \citenamefont {Kim}, \citenamefont {L\"utkenhaus}, \citenamefont {Lukin},\ and\ \citenamefont {Jiang}}]{Muralidharan2014}%
  \BibitemOpen
  \bibfield  {author} {\bibinfo {author} {\bibfnamefont {S.}~\bibnamefont {Muralidharan}}, \bibinfo {author} {\bibfnamefont {J.}~\bibnamefont {Kim}}, \bibinfo {author} {\bibfnamefont {N.}~\bibnamefont {L\"utkenhaus}}, \bibinfo {author} {\bibfnamefont {M.~D.}\ \bibnamefont {Lukin}},\ and\ \bibinfo {author} {\bibfnamefont {L.}~\bibnamefont {Jiang}},\ }\bibfield  {title} {\bibinfo {title} {Ultrafast and fault-tolerant quantum communication across long distances},\ }\href {https://doi.org/10.1103/PhysRevLett.112.250501} {\bibfield  {journal} {\bibinfo  {journal} {Phys. Rev. Lett.}\ }\textbf {\bibinfo {volume} {112}},\ \bibinfo {pages} {250501} (\bibinfo {year} {2014})}\BibitemShut {NoStop}%
\bibitem [{\citenamefont {Azuma}\ \emph {et~al.}(2015)\citenamefont {Azuma}, \citenamefont {Tamaki},\ and\ \citenamefont {Lo}}]{Azuma2015}%
  \BibitemOpen
  \bibfield  {author} {\bibinfo {author} {\bibfnamefont {K.}~\bibnamefont {Azuma}}, \bibinfo {author} {\bibfnamefont {K.}~\bibnamefont {Tamaki}},\ and\ \bibinfo {author} {\bibfnamefont {H.-K.}\ \bibnamefont {Lo}},\ }\bibfield  {title} {\bibinfo {title} {All-photonic quantum repeaters},\ }\href {https://doi.org/10.1038/ncomms7787} {\bibfield  {journal} {\bibinfo  {journal} {Nature Communications}\ }\textbf {\bibinfo {volume} {6}},\ \bibinfo {pages} {6787} (\bibinfo {year} {2015})}\BibitemShut {NoStop}%
\bibitem [{\citenamefont {Bennett}\ \emph {et~al.}(1996{\natexlab{a}})\citenamefont {Bennett}, \citenamefont {Brassard}, \citenamefont {Popescu}, \citenamefont {Schumacher}, \citenamefont {Smolin},\ and\ \citenamefont {Wootters}}]{Bennett1996}%
  \BibitemOpen
  \bibfield  {author} {\bibinfo {author} {\bibfnamefont {C.~H.}\ \bibnamefont {Bennett}}, \bibinfo {author} {\bibfnamefont {G.}~\bibnamefont {Brassard}}, \bibinfo {author} {\bibfnamefont {S.}~\bibnamefont {Popescu}}, \bibinfo {author} {\bibfnamefont {B.}~\bibnamefont {Schumacher}}, \bibinfo {author} {\bibfnamefont {J.~A.}\ \bibnamefont {Smolin}},\ and\ \bibinfo {author} {\bibfnamefont {W.~K.}\ \bibnamefont {Wootters}},\ }\bibfield  {title} {\bibinfo {title} {Purification of noisy entanglement and faithful teleportation via noisy channels},\ }\href {https://doi.org/10.1103/PhysRevLett.76.722} {\bibfield  {journal} {\bibinfo  {journal} {Phys. Rev. Lett.}\ }\textbf {\bibinfo {volume} {76}},\ \bibinfo {pages} {722} (\bibinfo {year} {1996}{\natexlab{a}})}\BibitemShut {NoStop}%
\bibitem [{\citenamefont {Bennett}\ \emph {et~al.}(1996{\natexlab{b}})\citenamefont {Bennett}, \citenamefont {DiVincenzo}, \citenamefont {Smolin},\ and\ \citenamefont {Wootters}}]{Bennett1996a}%
  \BibitemOpen
  \bibfield  {author} {\bibinfo {author} {\bibfnamefont {C.~H.}\ \bibnamefont {Bennett}}, \bibinfo {author} {\bibfnamefont {D.~P.}\ \bibnamefont {DiVincenzo}}, \bibinfo {author} {\bibfnamefont {J.~A.}\ \bibnamefont {Smolin}},\ and\ \bibinfo {author} {\bibfnamefont {W.~K.}\ \bibnamefont {Wootters}},\ }\bibfield  {title} {\bibinfo {title} {Mixed-state entanglement and quantum error correction},\ }\href {https://doi.org/10.1103/PhysRevA.54.3824} {\bibfield  {journal} {\bibinfo  {journal} {Phys. Rev. A}\ }\textbf {\bibinfo {volume} {54}},\ \bibinfo {pages} {3824} (\bibinfo {year} {1996}{\natexlab{b}})}\BibitemShut {NoStop}%
\bibitem [{\citenamefont {Deutsch}\ \emph {et~al.}(1996)\citenamefont {Deutsch}, \citenamefont {Ekert}, \citenamefont {Jozsa}, \citenamefont {Macchiavello}, \citenamefont {Popescu},\ and\ \citenamefont {Sanpera}}]{Deutsch1996}%
  \BibitemOpen
  \bibfield  {author} {\bibinfo {author} {\bibfnamefont {D.}~\bibnamefont {Deutsch}}, \bibinfo {author} {\bibfnamefont {A.}~\bibnamefont {Ekert}}, \bibinfo {author} {\bibfnamefont {R.}~\bibnamefont {Jozsa}}, \bibinfo {author} {\bibfnamefont {C.}~\bibnamefont {Macchiavello}}, \bibinfo {author} {\bibfnamefont {S.}~\bibnamefont {Popescu}},\ and\ \bibinfo {author} {\bibfnamefont {A.}~\bibnamefont {Sanpera}},\ }\bibfield  {title} {\bibinfo {title} {Quantum privacy amplification and the security of quantum cryptography over noisy channels},\ }\href {https://doi.org/10.1103/PhysRevLett.77.2818} {\bibfield  {journal} {\bibinfo  {journal} {Phys. Rev. Lett.}\ }\textbf {\bibinfo {volume} {77}},\ \bibinfo {pages} {2818} (\bibinfo {year} {1996})}\BibitemShut {NoStop}%
\bibitem [{\citenamefont {Verstraete}\ \emph {et~al.}(2001)\citenamefont {Verstraete}, \citenamefont {Dehaene},\ and\ \citenamefont {DeMoor}}]{Verstraete2001}%
  \BibitemOpen
  \bibfield  {author} {\bibinfo {author} {\bibfnamefont {F.}~\bibnamefont {Verstraete}}, \bibinfo {author} {\bibfnamefont {J.}~\bibnamefont {Dehaene}},\ and\ \bibinfo {author} {\bibfnamefont {B.}~\bibnamefont {DeMoor}},\ }\bibfield  {title} {\bibinfo {title} {Local filtering operations on two qubits},\ }\href {https://doi.org/10.1103/PhysRevA.64.010101} {\bibfield  {journal} {\bibinfo  {journal} {Phys. Rev. A}\ }\textbf {\bibinfo {volume} {64}},\ \bibinfo {pages} {010101} (\bibinfo {year} {2001})}\BibitemShut {NoStop}%
\bibitem [{\citenamefont {Duan}\ \emph {et~al.}(2000)\citenamefont {Duan}, \citenamefont {Giedke}, \citenamefont {Cirac},\ and\ \citenamefont {Zoller}}]{Duan2000}%
  \BibitemOpen
  \bibfield  {author} {\bibinfo {author} {\bibfnamefont {L.-M.}\ \bibnamefont {Duan}}, \bibinfo {author} {\bibfnamefont {G.}~\bibnamefont {Giedke}}, \bibinfo {author} {\bibfnamefont {J.~I.}\ \bibnamefont {Cirac}},\ and\ \bibinfo {author} {\bibfnamefont {P.}~\bibnamefont {Zoller}},\ }\bibfield  {title} {\bibinfo {title} {Entanglement purification of gaussian continuous variable quantum states},\ }\href {https://doi.org/10.1103/PhysRevLett.84.4002} {\bibfield  {journal} {\bibinfo  {journal} {Phys. Rev. Lett.}\ }\textbf {\bibinfo {volume} {84}},\ \bibinfo {pages} {4002} (\bibinfo {year} {2000})}\BibitemShut {NoStop}%
\bibitem [{\citenamefont {Pan}\ \emph {et~al.}(2001)\citenamefont {Pan}, \citenamefont {Simon}, \citenamefont {Brukner},\ and\ \citenamefont {Zeilinger}}]{Pan2001}%
  \BibitemOpen
  \bibfield  {author} {\bibinfo {author} {\bibfnamefont {J.-W.}\ \bibnamefont {Pan}}, \bibinfo {author} {\bibfnamefont {C.}~\bibnamefont {Simon}}, \bibinfo {author} {\bibfnamefont {{\v{C}}.}~\bibnamefont {Brukner}},\ and\ \bibinfo {author} {\bibfnamefont {A.}~\bibnamefont {Zeilinger}},\ }\bibfield  {title} {\bibinfo {title} {Entanglement purification for quantum communication},\ }\href {https://doi.org/10.1038/35074041} {\bibfield  {journal} {\bibinfo  {journal} {Nature}\ }\textbf {\bibinfo {volume} {410}},\ \bibinfo {pages} {1067} (\bibinfo {year} {2001})}\BibitemShut {NoStop}%
\bibitem [{\citenamefont {Zwerger}\ \emph {et~al.}(2012)\citenamefont {Zwerger}, \citenamefont {D\"ur},\ and\ \citenamefont {Briegel}}]{Zwerger2012}%
  \BibitemOpen
  \bibfield  {author} {\bibinfo {author} {\bibfnamefont {M.}~\bibnamefont {Zwerger}}, \bibinfo {author} {\bibfnamefont {W.}~\bibnamefont {D\"ur}},\ and\ \bibinfo {author} {\bibfnamefont {H.~J.}\ \bibnamefont {Briegel}},\ }\bibfield  {title} {\bibinfo {title} {Measurement-based quantum repeaters},\ }\href {https://doi.org/10.1103/PhysRevA.85.062326} {\bibfield  {journal} {\bibinfo  {journal} {Phys. Rev. A}\ }\textbf {\bibinfo {volume} {85}},\ \bibinfo {pages} {062326} (\bibinfo {year} {2012})}\BibitemShut {NoStop}%
\bibitem [{\citenamefont {Zwerger}\ \emph {et~al.}(2013)\citenamefont {Zwerger}, \citenamefont {Briegel},\ and\ \citenamefont {D\"ur}}]{Zwerger2013}%
  \BibitemOpen
  \bibfield  {author} {\bibinfo {author} {\bibfnamefont {M.}~\bibnamefont {Zwerger}}, \bibinfo {author} {\bibfnamefont {H.~J.}\ \bibnamefont {Briegel}},\ and\ \bibinfo {author} {\bibfnamefont {W.}~\bibnamefont {D\"ur}},\ }\bibfield  {title} {\bibinfo {title} {Universal and optimal error thresholds for measurement-based entanglement purification},\ }\href {https://doi.org/10.1103/PhysRevLett.110.260503} {\bibfield  {journal} {\bibinfo  {journal} {Phys. Rev. Lett.}\ }\textbf {\bibinfo {volume} {110}},\ \bibinfo {pages} {260503} (\bibinfo {year} {2013})}\BibitemShut {NoStop}%
\bibitem [{\citenamefont {Yan}\ \emph {et~al.}(2023)\citenamefont {Yan}, \citenamefont {Zhou}, \citenamefont {Zhong},\ and\ \citenamefont {Sheng}}]{Yan2023}%
  \BibitemOpen
  \bibfield  {author} {\bibinfo {author} {\bibfnamefont {P.-S.}\ \bibnamefont {Yan}}, \bibinfo {author} {\bibfnamefont {L.}~\bibnamefont {Zhou}}, \bibinfo {author} {\bibfnamefont {W.}~\bibnamefont {Zhong}},\ and\ \bibinfo {author} {\bibfnamefont {Y.-B.}\ \bibnamefont {Sheng}},\ }\bibfield  {title} {\bibinfo {title} {Advances in quantum entanglement purification},\ }\href {https://doi.org/10.1007/s11433-022-2065-x} {\bibfield  {journal} {\bibinfo  {journal} {Science China Physics, Mechanics {\&} Astronomy}\ }\textbf {\bibinfo {volume} {66}},\ \bibinfo {pages} {250301} (\bibinfo {year} {2023})}\BibitemShut {NoStop}%
\bibitem [{\citenamefont {Pan}\ \emph {et~al.}(2003)\citenamefont {Pan}, \citenamefont {Gasparoni}, \citenamefont {Ursin}, \citenamefont {Weihs},\ and\ \citenamefont {Zeilinger}}]{Pan2003}%
  \BibitemOpen
  \bibfield  {author} {\bibinfo {author} {\bibfnamefont {J.-W.}\ \bibnamefont {Pan}}, \bibinfo {author} {\bibfnamefont {S.}~\bibnamefont {Gasparoni}}, \bibinfo {author} {\bibfnamefont {R.}~\bibnamefont {Ursin}}, \bibinfo {author} {\bibfnamefont {G.}~\bibnamefont {Weihs}},\ and\ \bibinfo {author} {\bibfnamefont {A.}~\bibnamefont {Zeilinger}},\ }\bibfield  {title} {\bibinfo {title} {Experimental entanglement purification of arbitrary unknown states},\ }\href {https://doi.org/10.1038/nature01623} {\bibfield  {journal} {\bibinfo  {journal} {Nature}\ }\textbf {\bibinfo {volume} {423}},\ \bibinfo {pages} {417} (\bibinfo {year} {2003})}\BibitemShut {NoStop}%
\bibitem [{\citenamefont {Ourjoumtsev}\ \emph {et~al.}(2007)\citenamefont {Ourjoumtsev}, \citenamefont {Dantan}, \citenamefont {Tualle-Brouri},\ and\ \citenamefont {Grangier}}]{Ourjoumtsev2007}%
  \BibitemOpen
  \bibfield  {author} {\bibinfo {author} {\bibfnamefont {A.}~\bibnamefont {Ourjoumtsev}}, \bibinfo {author} {\bibfnamefont {A.}~\bibnamefont {Dantan}}, \bibinfo {author} {\bibfnamefont {R.}~\bibnamefont {Tualle-Brouri}},\ and\ \bibinfo {author} {\bibfnamefont {P.}~\bibnamefont {Grangier}},\ }\bibfield  {title} {\bibinfo {title} {Increasing entanglement between gaussian states by coherent photon subtraction},\ }\href {https://doi.org/10.1103/PhysRevLett.98.030502} {\bibfield  {journal} {\bibinfo  {journal} {Phys. Rev. Lett.}\ }\textbf {\bibinfo {volume} {98}},\ \bibinfo {pages} {030502} (\bibinfo {year} {2007})}\BibitemShut {NoStop}%
\bibitem [{\citenamefont {Takahashi}\ \emph {et~al.}(2010)\citenamefont {Takahashi}, \citenamefont {Neergaard-Nielsen}, \citenamefont {Takeuchi}, \citenamefont {Takeoka}, \citenamefont {Hayasaka}, \citenamefont {Furusawa},\ and\ \citenamefont {Sasaki}}]{Takahashi2010}%
  \BibitemOpen
  \bibfield  {author} {\bibinfo {author} {\bibfnamefont {H.}~\bibnamefont {Takahashi}}, \bibinfo {author} {\bibfnamefont {J.~S.}\ \bibnamefont {Neergaard-Nielsen}}, \bibinfo {author} {\bibfnamefont {M.}~\bibnamefont {Takeuchi}}, \bibinfo {author} {\bibfnamefont {M.}~\bibnamefont {Takeoka}}, \bibinfo {author} {\bibfnamefont {K.}~\bibnamefont {Hayasaka}}, \bibinfo {author} {\bibfnamefont {A.}~\bibnamefont {Furusawa}},\ and\ \bibinfo {author} {\bibfnamefont {M.}~\bibnamefont {Sasaki}},\ }\bibfield  {title} {\bibinfo {title} {Entanglement distillation from gaussian input states},\ }\href {https://doi.org/10.1038/nphoton.2010.1} {\bibfield  {journal} {\bibinfo  {journal} {Nature Photonics}\ }\textbf {\bibinfo {volume} {4}},\ \bibinfo {pages} {178} (\bibinfo {year} {2010})}\BibitemShut {NoStop}%
\bibitem [{\citenamefont {Kim}\ \emph {et~al.}(2012)\citenamefont {Kim}, \citenamefont {Lee}, \citenamefont {Kwon},\ and\ \citenamefont {Kim}}]{Kim2012}%
  \BibitemOpen
  \bibfield  {author} {\bibinfo {author} {\bibfnamefont {Y.-S.}\ \bibnamefont {Kim}}, \bibinfo {author} {\bibfnamefont {J.-C.}\ \bibnamefont {Lee}}, \bibinfo {author} {\bibfnamefont {O.}~\bibnamefont {Kwon}},\ and\ \bibinfo {author} {\bibfnamefont {Y.-H.}\ \bibnamefont {Kim}},\ }\bibfield  {title} {\bibinfo {title} {Protecting entanglement from decoherence using weak measurement and quantum measurement reversal},\ }\href {https://doi.org/10.1038/nphys2178} {\bibfield  {journal} {\bibinfo  {journal} {Nature Physics}\ }\textbf {\bibinfo {volume} {8}},\ \bibinfo {pages} {117} (\bibinfo {year} {2012})}\BibitemShut {NoStop}%
\bibitem [{\citenamefont {Kalb}\ \emph {et~al.}(2017)\citenamefont {Kalb}, \citenamefont {Reiserer}, \citenamefont {Humphreys}, \citenamefont {Bakermans}, \citenamefont {Kamerling}, \citenamefont {Nickerson}, \citenamefont {Benjamin}, \citenamefont {Twitchen}, \citenamefont {Markham},\ and\ \citenamefont {Hanson}}]{Kalb2017}%
  \BibitemOpen
  \bibfield  {author} {\bibinfo {author} {\bibfnamefont {N.}~\bibnamefont {Kalb}}, \bibinfo {author} {\bibfnamefont {A.~A.}\ \bibnamefont {Reiserer}}, \bibinfo {author} {\bibfnamefont {P.~C.}\ \bibnamefont {Humphreys}}, \bibinfo {author} {\bibfnamefont {J.~J.~W.}\ \bibnamefont {Bakermans}}, \bibinfo {author} {\bibfnamefont {S.~J.}\ \bibnamefont {Kamerling}}, \bibinfo {author} {\bibfnamefont {N.~H.}\ \bibnamefont {Nickerson}}, \bibinfo {author} {\bibfnamefont {S.~C.}\ \bibnamefont {Benjamin}}, \bibinfo {author} {\bibfnamefont {D.~J.}\ \bibnamefont {Twitchen}}, \bibinfo {author} {\bibfnamefont {M.}~\bibnamefont {Markham}},\ and\ \bibinfo {author} {\bibfnamefont {R.}~\bibnamefont {Hanson}},\ }\bibfield  {title} {\bibinfo {title} {Entanglement distillation between solid-state quantum network nodes},\ }\href {https://doi.org/10.1126/science.aan0070} {\bibfield  {journal} {\bibinfo  {journal} {Science}\ }\textbf {\bibinfo {volume} {356}},\ \bibinfo {pages} {928} (\bibinfo {year} {2017})}\BibitemShut {NoStop}%
\bibitem [{\citenamefont {Hu}\ \emph {et~al.}(2021)\citenamefont {Hu}, \citenamefont {Huang}, \citenamefont {Sheng}, \citenamefont {Zhou}, \citenamefont {Liu}, \citenamefont {Guo}, \citenamefont {Zhang}, \citenamefont {Xing}, \citenamefont {Huang}, \citenamefont {Li},\ and\ \citenamefont {Guo}}]{Hu2021}%
  \BibitemOpen
  \bibfield  {author} {\bibinfo {author} {\bibfnamefont {X.-M.}\ \bibnamefont {Hu}}, \bibinfo {author} {\bibfnamefont {C.-X.}\ \bibnamefont {Huang}}, \bibinfo {author} {\bibfnamefont {Y.-B.}\ \bibnamefont {Sheng}}, \bibinfo {author} {\bibfnamefont {L.}~\bibnamefont {Zhou}}, \bibinfo {author} {\bibfnamefont {B.-H.}\ \bibnamefont {Liu}}, \bibinfo {author} {\bibfnamefont {Y.}~\bibnamefont {Guo}}, \bibinfo {author} {\bibfnamefont {C.}~\bibnamefont {Zhang}}, \bibinfo {author} {\bibfnamefont {W.-B.}\ \bibnamefont {Xing}}, \bibinfo {author} {\bibfnamefont {Y.-F.}\ \bibnamefont {Huang}}, \bibinfo {author} {\bibfnamefont {C.-F.}\ \bibnamefont {Li}},\ and\ \bibinfo {author} {\bibfnamefont {G.-C.}\ \bibnamefont {Guo}},\ }\bibfield  {title} {\bibinfo {title} {Long-distance entanglement purification for quantum communication},\ }\href {https://doi.org/10.1103/PhysRevLett.126.010503} {\bibfield  {journal} {\bibinfo  {journal} {Phys. Rev. Lett.}\ }\textbf {\bibinfo {volume} {126}},\ \bibinfo {pages} {010503} (\bibinfo
  {year} {2021})}\BibitemShut {NoStop}%
\bibitem [{\citenamefont {Ecker}\ \emph {et~al.}(2021)\citenamefont {Ecker}, \citenamefont {Sohr}, \citenamefont {Bulla}, \citenamefont {Huber}, \citenamefont {Bohmann},\ and\ \citenamefont {Ursin}}]{Ecker2021}%
  \BibitemOpen
  \bibfield  {author} {\bibinfo {author} {\bibfnamefont {S.}~\bibnamefont {Ecker}}, \bibinfo {author} {\bibfnamefont {P.}~\bibnamefont {Sohr}}, \bibinfo {author} {\bibfnamefont {L.}~\bibnamefont {Bulla}}, \bibinfo {author} {\bibfnamefont {M.}~\bibnamefont {Huber}}, \bibinfo {author} {\bibfnamefont {M.}~\bibnamefont {Bohmann}},\ and\ \bibinfo {author} {\bibfnamefont {R.}~\bibnamefont {Ursin}},\ }\bibfield  {title} {\bibinfo {title} {Experimental single-copy entanglement distillation},\ }\href {https://doi.org/10.1103/PhysRevLett.127.040506} {\bibfield  {journal} {\bibinfo  {journal} {Phys. Rev. Lett.}\ }\textbf {\bibinfo {volume} {127}},\ \bibinfo {pages} {040506} (\bibinfo {year} {2021})}\BibitemShut {NoStop}%
\bibitem [{\citenamefont {Huang}\ \emph {et~al.}(2022)\citenamefont {Huang}, \citenamefont {Hu}, \citenamefont {Liu}, \citenamefont {Zhou}, \citenamefont {Sheng}, \citenamefont {Li},\ and\ \citenamefont {Guo}}]{Huang2022}%
  \BibitemOpen
  \bibfield  {author} {\bibinfo {author} {\bibfnamefont {C.-X.}\ \bibnamefont {Huang}}, \bibinfo {author} {\bibfnamefont {X.-M.}\ \bibnamefont {Hu}}, \bibinfo {author} {\bibfnamefont {B.-H.}\ \bibnamefont {Liu}}, \bibinfo {author} {\bibfnamefont {L.}~\bibnamefont {Zhou}}, \bibinfo {author} {\bibfnamefont {Y.-B.}\ \bibnamefont {Sheng}}, \bibinfo {author} {\bibfnamefont {C.-F.}\ \bibnamefont {Li}},\ and\ \bibinfo {author} {\bibfnamefont {G.-C.}\ \bibnamefont {Guo}},\ }\bibfield  {title} {\bibinfo {title} {Experimental one-step deterministic polarization entanglement purification},\ }\href {https://doi.org/https://doi.org/10.1016/j.scib.2021.12.018} {\bibfield  {journal} {\bibinfo  {journal} {Science Bulletin}\ }\textbf {\bibinfo {volume} {67}},\ \bibinfo {pages} {593} (\bibinfo {year} {2022})}\BibitemShut {NoStop}%
\bibitem [{\citenamefont {Liu}\ \emph {et~al.}(2022)\citenamefont {Liu}, \citenamefont {Tu}, \citenamefont {Li}, \citenamefont {Liu}, \citenamefont {Zhu}, \citenamefont {Zhou}, \citenamefont {Li},\ and\ \citenamefont {Guo}}]{Liu2022}%
  \BibitemOpen
  \bibfield  {author} {\bibinfo {author} {\bibfnamefont {C.}~\bibnamefont {Liu}}, \bibinfo {author} {\bibfnamefont {T.}~\bibnamefont {Tu}}, \bibinfo {author} {\bibfnamefont {P.-Y.}\ \bibnamefont {Li}}, \bibinfo {author} {\bibfnamefont {X.}~\bibnamefont {Liu}}, \bibinfo {author} {\bibfnamefont {X.-Y.}\ \bibnamefont {Zhu}}, \bibinfo {author} {\bibfnamefont {Z.-Q.}\ \bibnamefont {Zhou}}, \bibinfo {author} {\bibfnamefont {C.-F.}\ \bibnamefont {Li}},\ and\ \bibinfo {author} {\bibfnamefont {G.-C.}\ \bibnamefont {Guo}},\ }\bibfield  {title} {\bibinfo {title} {Towards entanglement distillation between atomic ensembles using high-fidelity spin operations},\ }\href {https://doi.org/10.1038/s42005-022-00835-0} {\bibfield  {journal} {\bibinfo  {journal} {Communications Physics}\ }\textbf {\bibinfo {volume} {5}},\ \bibinfo {pages} {67} (\bibinfo {year} {2022})}\BibitemShut {NoStop}%
\bibitem [{\citenamefont {Yan}\ \emph {et~al.}(2022)\citenamefont {Yan}, \citenamefont {Zhong}, \citenamefont {Chang}, \citenamefont {Bienfait}, \citenamefont {Chou}, \citenamefont {Conner}, \citenamefont {Dumur}, \citenamefont {Grebel}, \citenamefont {Povey},\ and\ \citenamefont {Cleland}}]{Yan2022}%
  \BibitemOpen
  \bibfield  {author} {\bibinfo {author} {\bibfnamefont {H.}~\bibnamefont {Yan}}, \bibinfo {author} {\bibfnamefont {Y.}~\bibnamefont {Zhong}}, \bibinfo {author} {\bibfnamefont {H.-S.}\ \bibnamefont {Chang}}, \bibinfo {author} {\bibfnamefont {A.}~\bibnamefont {Bienfait}}, \bibinfo {author} {\bibfnamefont {M.-H.}\ \bibnamefont {Chou}}, \bibinfo {author} {\bibfnamefont {C.~R.}\ \bibnamefont {Conner}}, \bibinfo {author} {\bibfnamefont {E.}~\bibnamefont {Dumur}}, \bibinfo {author} {\bibfnamefont {J.}~\bibnamefont {Grebel}}, \bibinfo {author} {\bibfnamefont {R.~G.}\ \bibnamefont {Povey}},\ and\ \bibinfo {author} {\bibfnamefont {A.~N.}\ \bibnamefont {Cleland}},\ }\bibfield  {title} {\bibinfo {title} {Entanglement purification and protection in a superconducting quantum network},\ }\href {https://doi.org/10.1103/PhysRevLett.128.080504} {\bibfield  {journal} {\bibinfo  {journal} {Phys. Rev. Lett.}\ }\textbf {\bibinfo {volume} {128}},\ \bibinfo {pages} {080504} (\bibinfo {year} {2022})}\BibitemShut {NoStop}%
\bibitem [{\citenamefont {Cai}\ \emph {et~al.}(2024)\citenamefont {Cai}, \citenamefont {Mu}, \citenamefont {Wang}, \citenamefont {Zhou}, \citenamefont {Ma}, \citenamefont {Pan}, \citenamefont {Hua}, \citenamefont {Liu}, \citenamefont {Xue}, \citenamefont {Yu}, \citenamefont {Wang}, \citenamefont {Song}, \citenamefont {Zou},\ and\ \citenamefont {Sun}}]{Cai2024}%
  \BibitemOpen
  \bibfield  {author} {\bibinfo {author} {\bibfnamefont {W.}~\bibnamefont {Cai}}, \bibinfo {author} {\bibfnamefont {X.}~\bibnamefont {Mu}}, \bibinfo {author} {\bibfnamefont {W.}~\bibnamefont {Wang}}, \bibinfo {author} {\bibfnamefont {J.}~\bibnamefont {Zhou}}, \bibinfo {author} {\bibfnamefont {Y.}~\bibnamefont {Ma}}, \bibinfo {author} {\bibfnamefont {X.}~\bibnamefont {Pan}}, \bibinfo {author} {\bibfnamefont {Z.}~\bibnamefont {Hua}}, \bibinfo {author} {\bibfnamefont {X.}~\bibnamefont {Liu}}, \bibinfo {author} {\bibfnamefont {G.}~\bibnamefont {Xue}}, \bibinfo {author} {\bibfnamefont {H.}~\bibnamefont {Yu}}, \bibinfo {author} {\bibfnamefont {H.}~\bibnamefont {Wang}}, \bibinfo {author} {\bibfnamefont {Y.}~\bibnamefont {Song}}, \bibinfo {author} {\bibfnamefont {C.-L.}\ \bibnamefont {Zou}},\ and\ \bibinfo {author} {\bibfnamefont {L.}~\bibnamefont {Sun}},\ }\bibfield  {title} {\bibinfo {title} {Protecting entanglement between logical qubits via quantum error correction},\ }\bibfield  {journal} {\bibinfo  {journal}
  {Nature Physics}\ }\href {https://doi.org/10.1038/s41567-024-02446-8} {10.1038/s41567-024-02446-8} (\bibinfo {year} {2024})\BibitemShut {NoStop}%
\bibitem [{\citenamefont {Murao}\ \emph {et~al.}(1998)\citenamefont {Murao}, \citenamefont {Plenio}, \citenamefont {Popescu}, \citenamefont {Vedral},\ and\ \citenamefont {Knight}}]{Murao1998}%
  \BibitemOpen
  \bibfield  {author} {\bibinfo {author} {\bibfnamefont {M.}~\bibnamefont {Murao}}, \bibinfo {author} {\bibfnamefont {M.~B.}\ \bibnamefont {Plenio}}, \bibinfo {author} {\bibfnamefont {S.}~\bibnamefont {Popescu}}, \bibinfo {author} {\bibfnamefont {V.}~\bibnamefont {Vedral}},\ and\ \bibinfo {author} {\bibfnamefont {P.~L.}\ \bibnamefont {Knight}},\ }\bibfield  {title} {\bibinfo {title} {Multiparticle entanglement purification protocols},\ }\href {https://doi.org/10.1103/PhysRevA.57.R4075} {\bibfield  {journal} {\bibinfo  {journal} {Phys. Rev. A}\ }\textbf {\bibinfo {volume} {57}},\ \bibinfo {pages} {R4075} (\bibinfo {year} {1998})}\BibitemShut {NoStop}%
\bibitem [{\citenamefont {D\"ur}\ \emph {et~al.}(2003)\citenamefont {D\"ur}, \citenamefont {Aschauer},\ and\ \citenamefont {Briegel}}]{Dur2003a}%
  \BibitemOpen
  \bibfield  {author} {\bibinfo {author} {\bibfnamefont {W.}~\bibnamefont {D\"ur}}, \bibinfo {author} {\bibfnamefont {H.}~\bibnamefont {Aschauer}},\ and\ \bibinfo {author} {\bibfnamefont {H.-J.}\ \bibnamefont {Briegel}},\ }\bibfield  {title} {\bibinfo {title} {Multiparticle entanglement purification for graph states},\ }\href {https://doi.org/10.1103/PhysRevLett.91.107903} {\bibfield  {journal} {\bibinfo  {journal} {Phys. Rev. Lett.}\ }\textbf {\bibinfo {volume} {91}},\ \bibinfo {pages} {107903} (\bibinfo {year} {2003})}\BibitemShut {NoStop}%
\bibitem [{\citenamefont {Aschauer}\ \emph {et~al.}(2005)\citenamefont {Aschauer}, \citenamefont {D\"ur},\ and\ \citenamefont {Briegel}}]{Aschauer2005a}%
  \BibitemOpen
  \bibfield  {author} {\bibinfo {author} {\bibfnamefont {H.}~\bibnamefont {Aschauer}}, \bibinfo {author} {\bibfnamefont {W.}~\bibnamefont {D\"ur}},\ and\ \bibinfo {author} {\bibfnamefont {H.-J.}\ \bibnamefont {Briegel}},\ }\bibfield  {title} {\bibinfo {title} {Multiparticle entanglement purification for two-colorable graph states},\ }\href {https://doi.org/10.1103/PhysRevA.71.012319} {\bibfield  {journal} {\bibinfo  {journal} {Phys. Rev. A}\ }\textbf {\bibinfo {volume} {71}},\ \bibinfo {pages} {012319} (\bibinfo {year} {2005})}\BibitemShut {NoStop}%
\bibitem [{\citenamefont {Hein}\ \emph {et~al.}(2006)\citenamefont {Hein}, \citenamefont {Dür}, \citenamefont {Eisert}, \citenamefont {Raussendorf}, \citenamefont {den Nest},\ and\ \citenamefont {Briegel}}]{Hein2006}%
  \BibitemOpen
  \bibfield  {author} {\bibinfo {author} {\bibfnamefont {M.}~\bibnamefont {Hein}}, \bibinfo {author} {\bibfnamefont {W.}~\bibnamefont {Dür}}, \bibinfo {author} {\bibfnamefont {J.}~\bibnamefont {Eisert}}, \bibinfo {author} {\bibfnamefont {R.}~\bibnamefont {Raussendorf}}, \bibinfo {author} {\bibfnamefont {M.~V.}\ \bibnamefont {den Nest}},\ and\ \bibinfo {author} {\bibfnamefont {H.~J.}\ \bibnamefont {Briegel}},\ }\href@noop {} {\bibinfo {title} {Entanglement in graph states and its applications}} (\bibinfo {year} {2006}),\ \Eprint {https://arxiv.org/abs/quant-ph/0602096} {arXiv:quant-ph/0602096 [quant-ph]} \BibitemShut {NoStop}%
\bibitem [{\citenamefont {Kruszynska}\ \emph {et~al.}(2006)\citenamefont {Kruszynska}, \citenamefont {Miyake}, \citenamefont {Briegel},\ and\ \citenamefont {D\"ur}}]{Kruszynska2006}%
  \BibitemOpen
  \bibfield  {author} {\bibinfo {author} {\bibfnamefont {C.}~\bibnamefont {Kruszynska}}, \bibinfo {author} {\bibfnamefont {A.}~\bibnamefont {Miyake}}, \bibinfo {author} {\bibfnamefont {H.~J.}\ \bibnamefont {Briegel}},\ and\ \bibinfo {author} {\bibfnamefont {W.}~\bibnamefont {D\"ur}},\ }\bibfield  {title} {\bibinfo {title} {Entanglement purification protocols for all graph states},\ }\href {https://doi.org/10.1103/PhysRevA.74.052316} {\bibfield  {journal} {\bibinfo  {journal} {Phys. Rev. A}\ }\textbf {\bibinfo {volume} {74}},\ \bibinfo {pages} {052316} (\bibinfo {year} {2006})}\BibitemShut {NoStop}%
\bibitem [{\citenamefont {Sheng}\ \emph {et~al.}(2015)\citenamefont {Sheng}, \citenamefont {Pan}, \citenamefont {Guo}, \citenamefont {Zhou},\ and\ \citenamefont {Wang}}]{Sheng2015}%
  \BibitemOpen
  \bibfield  {author} {\bibinfo {author} {\bibfnamefont {Y.}~\bibnamefont {Sheng}}, \bibinfo {author} {\bibfnamefont {J.}~\bibnamefont {Pan}}, \bibinfo {author} {\bibfnamefont {R.}~\bibnamefont {Guo}}, \bibinfo {author} {\bibfnamefont {L.}~\bibnamefont {Zhou}},\ and\ \bibinfo {author} {\bibfnamefont {L.}~\bibnamefont {Wang}},\ }\bibfield  {title} {\bibinfo {title} {Efficient n-particle w state concentration with different parity check gates},\ }\href {https://doi.org/10.1007/s11433-015-5672-9} {\bibfield  {journal} {\bibinfo  {journal} {Science China Physics, Mechanics {\&} Astronomy}\ }\textbf {\bibinfo {volume} {58}},\ \bibinfo {pages} {1} (\bibinfo {year} {2015})}\BibitemShut {NoStop}%
\bibitem [{\citenamefont {de~Bone}\ \emph {et~al.}(2020)\citenamefont {de~Bone}, \citenamefont {Ouyang}, \citenamefont {Goodenough},\ and\ \citenamefont {Elkouss}}]{deBone2020}%
  \BibitemOpen
  \bibfield  {author} {\bibinfo {author} {\bibfnamefont {S.}~\bibnamefont {de~Bone}}, \bibinfo {author} {\bibfnamefont {R.}~\bibnamefont {Ouyang}}, \bibinfo {author} {\bibfnamefont {K.}~\bibnamefont {Goodenough}},\ and\ \bibinfo {author} {\bibfnamefont {D.}~\bibnamefont {Elkouss}},\ }\bibfield  {title} {\bibinfo {title} {Protocols for creating and distilling multipartite ghz states with bell pairs},\ }\href {https://doi.org/10.1109/TQE.2020.3044179} {\bibfield  {journal} {\bibinfo  {journal} {IEEE Transactions on Quantum Engineering}\ }\textbf {\bibinfo {volume} {1}},\ \bibinfo {pages} {1} (\bibinfo {year} {2020})}\BibitemShut {NoStop}%
\bibitem [{\citenamefont {Zhou}\ \emph {et~al.}(2021)\citenamefont {Zhou}, \citenamefont {Liu}, \citenamefont {Xu}, \citenamefont {Cui}, \citenamefont {Ran},\ and\ \citenamefont {Sheng}}]{Zhou2021}%
  \BibitemOpen
  \bibfield  {author} {\bibinfo {author} {\bibfnamefont {L.}~\bibnamefont {Zhou}}, \bibinfo {author} {\bibfnamefont {Z.-K.}\ \bibnamefont {Liu}}, \bibinfo {author} {\bibfnamefont {Z.-X.}\ \bibnamefont {Xu}}, \bibinfo {author} {\bibfnamefont {Y.-L.}\ \bibnamefont {Cui}}, \bibinfo {author} {\bibfnamefont {H.-J.}\ \bibnamefont {Ran}},\ and\ \bibinfo {author} {\bibfnamefont {Y.-B.}\ \bibnamefont {Sheng}},\ }\bibfield  {title} {\bibinfo {title} {Economical multi-photon polarization entanglement purification with bell state},\ }\href {https://doi.org/10.1007/s11128-021-03192-z} {\bibfield  {journal} {\bibinfo  {journal} {Quantum Information Processing}\ }\textbf {\bibinfo {volume} {20}},\ \bibinfo {pages} {257} (\bibinfo {year} {2021})}\BibitemShut {NoStop}%
\bibitem [{\citenamefont {Miguel-Ramiro}\ \emph {et~al.}(2023)\citenamefont {Miguel-Ramiro}, \citenamefont {Riera-S\`abat},\ and\ \citenamefont {D\"ur}}]{Miguel2023}%
  \BibitemOpen
  \bibfield  {author} {\bibinfo {author} {\bibfnamefont {J.}~\bibnamefont {Miguel-Ramiro}}, \bibinfo {author} {\bibfnamefont {F.}~\bibnamefont {Riera-S\`abat}},\ and\ \bibinfo {author} {\bibfnamefont {W.}~\bibnamefont {D\"ur}},\ }\bibfield  {title} {\bibinfo {title} {Quantum repeater for $w$ states},\ }\href {https://doi.org/10.1103/PRXQuantum.4.040323} {\bibfield  {journal} {\bibinfo  {journal} {PRX Quantum}\ }\textbf {\bibinfo {volume} {4}},\ \bibinfo {pages} {040323} (\bibinfo {year} {2023})}\BibitemShut {NoStop}%
\bibitem [{\citenamefont {Vandr\'e}\ and\ \citenamefont {G\"uhne}(2023)}]{Vandre2023}%
  \BibitemOpen
  \bibfield  {author} {\bibinfo {author} {\bibfnamefont {L.}~\bibnamefont {Vandr\'e}}\ and\ \bibinfo {author} {\bibfnamefont {O.}~\bibnamefont {G\"uhne}},\ }\bibfield  {title} {\bibinfo {title} {Entanglement purification of hypergraph states},\ }\href {https://doi.org/10.1103/PhysRevA.108.062417} {\bibfield  {journal} {\bibinfo  {journal} {Phys. Rev. A}\ }\textbf {\bibinfo {volume} {108}},\ \bibinfo {pages} {062417} (\bibinfo {year} {2023})}\BibitemShut {NoStop}%
\bibitem [{\citenamefont {Zwerger}\ \emph {et~al.}(2014)\citenamefont {Zwerger}, \citenamefont {Briegel},\ and\ \citenamefont {D\"ur}}]{Zwerger2014}%
  \BibitemOpen
  \bibfield  {author} {\bibinfo {author} {\bibfnamefont {M.}~\bibnamefont {Zwerger}}, \bibinfo {author} {\bibfnamefont {H.~J.}\ \bibnamefont {Briegel}},\ and\ \bibinfo {author} {\bibfnamefont {W.}~\bibnamefont {D\"ur}},\ }\bibfield  {title} {\bibinfo {title} {Robustness of hashing protocols for entanglement purification},\ }\href {https://doi.org/10.1103/PhysRevA.90.012314} {\bibfield  {journal} {\bibinfo  {journal} {Phys. Rev. A}\ }\textbf {\bibinfo {volume} {90}},\ \bibinfo {pages} {012314} (\bibinfo {year} {2014})}\BibitemShut {NoStop}%
\bibitem [{\citenamefont {Horodecki}\ and\ \citenamefont {Horodecki}(1999)}]{Horodecki1999}%
  \BibitemOpen
  \bibfield  {author} {\bibinfo {author} {\bibfnamefont {M.}~\bibnamefont {Horodecki}}\ and\ \bibinfo {author} {\bibfnamefont {P.}~\bibnamefont {Horodecki}},\ }\bibfield  {title} {\bibinfo {title} {Reduction criterion of separability and limits for a class of distillation protocols},\ }\href {https://doi.org/10.1103/PhysRevA.59.4206} {\bibfield  {journal} {\bibinfo  {journal} {Phys. Rev. A}\ }\textbf {\bibinfo {volume} {59}},\ \bibinfo {pages} {4206} (\bibinfo {year} {1999})}\BibitemShut {NoStop}%
\bibitem [{\citenamefont {Alber}\ \emph {et~al.}(2001)\citenamefont {Alber}, \citenamefont {Delgado}, \citenamefont {Gisin},\ and\ \citenamefont {Jex}}]{Alber2001}%
  \BibitemOpen
  \bibfield  {author} {\bibinfo {author} {\bibfnamefont {G.}~\bibnamefont {Alber}}, \bibinfo {author} {\bibfnamefont {A.}~\bibnamefont {Delgado}}, \bibinfo {author} {\bibfnamefont {N.}~\bibnamefont {Gisin}},\ and\ \bibinfo {author} {\bibfnamefont {I.}~\bibnamefont {Jex}},\ }\bibfield  {title} {\bibinfo {title} {Efficient bipartite quantum state purification in arbitrary dimensional hilbert spaces},\ }\href {https://doi.org/10.1088/0305-4470/34/42/307} {\bibfield  {journal} {\bibinfo  {journal} {Journal of Physics A: Mathematical and General}\ }\textbf {\bibinfo {volume} {34}},\ \bibinfo {pages} {8821} (\bibinfo {year} {2001})}\BibitemShut {NoStop}%
\bibitem [{\citenamefont {Vollbrecht}\ and\ \citenamefont {Wolf}(2003)}]{Vollbrecht2003}%
  \BibitemOpen
  \bibfield  {author} {\bibinfo {author} {\bibfnamefont {K.~G.~H.}\ \bibnamefont {Vollbrecht}}\ and\ \bibinfo {author} {\bibfnamefont {M.~M.}\ \bibnamefont {Wolf}},\ }\bibfield  {title} {\bibinfo {title} {Efficient distillation beyond qubits},\ }\href {https://doi.org/10.1103/PhysRevA.67.012303} {\bibfield  {journal} {\bibinfo  {journal} {Phys. Rev. A}\ }\textbf {\bibinfo {volume} {67}},\ \bibinfo {pages} {012303} (\bibinfo {year} {2003})}\BibitemShut {NoStop}%
\bibitem [{\citenamefont {Cheong}\ \emph {et~al.}(2007)\citenamefont {Cheong}, \citenamefont {Lee}, \citenamefont {Lee},\ and\ \citenamefont {Lee}}]{Cheong2007}%
  \BibitemOpen
  \bibfield  {author} {\bibinfo {author} {\bibfnamefont {Y.~W.}\ \bibnamefont {Cheong}}, \bibinfo {author} {\bibfnamefont {S.-W.}\ \bibnamefont {Lee}}, \bibinfo {author} {\bibfnamefont {J.}~\bibnamefont {Lee}},\ and\ \bibinfo {author} {\bibfnamefont {H.-W.}\ \bibnamefont {Lee}},\ }\bibfield  {title} {\bibinfo {title} {Entanglement purification for high-dimensional multipartite systems},\ }\href {https://doi.org/10.1103/PhysRevA.76.042314} {\bibfield  {journal} {\bibinfo  {journal} {Phys. Rev. A}\ }\textbf {\bibinfo {volume} {76}},\ \bibinfo {pages} {042314} (\bibinfo {year} {2007})}\BibitemShut {NoStop}%
\bibitem [{\citenamefont {Sheng}\ and\ \citenamefont {Deng}(2010)}]{Sheng2010}%
  \BibitemOpen
  \bibfield  {author} {\bibinfo {author} {\bibfnamefont {Y.-B.}\ \bibnamefont {Sheng}}\ and\ \bibinfo {author} {\bibfnamefont {F.-G.}\ \bibnamefont {Deng}},\ }\bibfield  {title} {\bibinfo {title} {Deterministic entanglement purification and complete nonlocal bell-state analysis with hyperentanglement},\ }\href {https://doi.org/10.1103/PhysRevA.81.032307} {\bibfield  {journal} {\bibinfo  {journal} {Phys. Rev. A}\ }\textbf {\bibinfo {volume} {81}},\ \bibinfo {pages} {032307} (\bibinfo {year} {2010})}\BibitemShut {NoStop}%
\bibitem [{\citenamefont {Miguel-Ramiro}\ and\ \citenamefont {D\"ur}(2018)}]{Miguel2018}%
  \BibitemOpen
  \bibfield  {author} {\bibinfo {author} {\bibfnamefont {J.}~\bibnamefont {Miguel-Ramiro}}\ and\ \bibinfo {author} {\bibfnamefont {W.}~\bibnamefont {D\"ur}},\ }\bibfield  {title} {\bibinfo {title} {Efficient entanglement purification protocols for $d$-level systems},\ }\href {https://doi.org/10.1103/PhysRevA.98.042309} {\bibfield  {journal} {\bibinfo  {journal} {Phys. Rev. A}\ }\textbf {\bibinfo {volume} {98}},\ \bibinfo {pages} {042309} (\bibinfo {year} {2018})}\BibitemShut {NoStop}%
\bibitem [{\citenamefont {Vollbrecht}\ and\ \citenamefont {Verstraete}(2005)}]{Vollbrecht2005}%
  \BibitemOpen
  \bibfield  {author} {\bibinfo {author} {\bibfnamefont {K.~G.~H.}\ \bibnamefont {Vollbrecht}}\ and\ \bibinfo {author} {\bibfnamefont {F.}~\bibnamefont {Verstraete}},\ }\bibfield  {title} {\bibinfo {title} {Interpolation of recurrence and hashing entanglement distillation protocols},\ }\href {https://doi.org/10.1103/PhysRevA.71.062325} {\bibfield  {journal} {\bibinfo  {journal} {Phys. Rev. A}\ }\textbf {\bibinfo {volume} {71}},\ \bibinfo {pages} {062325} (\bibinfo {year} {2005})}\BibitemShut {NoStop}%
\bibitem [{\citenamefont {Luo}\ and\ \citenamefont {Devetak}(2007)}]{Luo2007}%
  \BibitemOpen
  \bibfield  {author} {\bibinfo {author} {\bibfnamefont {Z.}~\bibnamefont {Luo}}\ and\ \bibinfo {author} {\bibfnamefont {I.}~\bibnamefont {Devetak}},\ }\bibfield  {title} {\bibinfo {title} {Efficiently implementable codes for quantum key expansion},\ }\href {https://doi.org/10.1103/PhysRevA.75.010303} {\bibfield  {journal} {\bibinfo  {journal} {Phys. Rev. A}\ }\textbf {\bibinfo {volume} {75}},\ \bibinfo {pages} {010303} (\bibinfo {year} {2007})}\BibitemShut {NoStop}%
\bibitem [{\citenamefont {Wilde}\ \emph {et~al.}(2010)\citenamefont {Wilde}, \citenamefont {Krovi},\ and\ \citenamefont {Brun}}]{Wilde2010}%
  \BibitemOpen
  \bibfield  {author} {\bibinfo {author} {\bibfnamefont {M.~M.}\ \bibnamefont {Wilde}}, \bibinfo {author} {\bibfnamefont {H.}~\bibnamefont {Krovi}},\ and\ \bibinfo {author} {\bibfnamefont {T.~A.}\ \bibnamefont {Brun}},\ }\bibfield  {title} {\bibinfo {title} {Convolutional entanglement distillation},\ }in\ \href {https://doi.org/10.1109/ISIT.2010.5513666} {\emph {\bibinfo {booktitle} {2010 IEEE International Symposium on Information Theory}}}\ (\bibinfo {year} {2010})\ pp.\ \bibinfo {pages} {2657--2661}\BibitemShut {NoStop}%
\bibitem [{\citenamefont {Riera-S\`abat}\ \emph {et~al.}(2021{\natexlab{a}})\citenamefont {Riera-S\`abat}, \citenamefont {Sekatski}, \citenamefont {Pirker},\ and\ \citenamefont {D\"ur}}]{Riera2021}%
  \BibitemOpen
  \bibfield  {author} {\bibinfo {author} {\bibfnamefont {F.}~\bibnamefont {Riera-S\`abat}}, \bibinfo {author} {\bibfnamefont {P.}~\bibnamefont {Sekatski}}, \bibinfo {author} {\bibfnamefont {A.}~\bibnamefont {Pirker}},\ and\ \bibinfo {author} {\bibfnamefont {W.}~\bibnamefont {D\"ur}},\ }\bibfield  {title} {\bibinfo {title} {Entanglement-assisted entanglement purification},\ }\href {https://doi.org/10.1103/PhysRevLett.127.040502} {\bibfield  {journal} {\bibinfo  {journal} {Phys. Rev. Lett.}\ }\textbf {\bibinfo {volume} {127}},\ \bibinfo {pages} {040502} (\bibinfo {year} {2021}{\natexlab{a}})}\BibitemShut {NoStop}%
\bibitem [{\citenamefont {Riera-S\`abat}\ \emph {et~al.}(2021{\natexlab{b}})\citenamefont {Riera-S\`abat}, \citenamefont {Sekatski}, \citenamefont {Pirker},\ and\ \citenamefont {D\"ur}}]{Riera2021a}%
  \BibitemOpen
  \bibfield  {author} {\bibinfo {author} {\bibfnamefont {F.}~\bibnamefont {Riera-S\`abat}}, \bibinfo {author} {\bibfnamefont {P.}~\bibnamefont {Sekatski}}, \bibinfo {author} {\bibfnamefont {A.}~\bibnamefont {Pirker}},\ and\ \bibinfo {author} {\bibfnamefont {W.}~\bibnamefont {D\"ur}},\ }\bibfield  {title} {\bibinfo {title} {Entanglement purification by counting and locating errors with entangling measurements},\ }\href {https://doi.org/10.1103/PhysRevA.104.012419} {\bibfield  {journal} {\bibinfo  {journal} {Phys. Rev. A}\ }\textbf {\bibinfo {volume} {104}},\ \bibinfo {pages} {012419} (\bibinfo {year} {2021}{\natexlab{b}})}\BibitemShut {NoStop}%
\bibitem [{\citenamefont {Dehaene}\ \emph {et~al.}(2003)\citenamefont {Dehaene}, \citenamefont {Van~den Nest}, \citenamefont {De~Moor},\ and\ \citenamefont {Verstraete}}]{Dehaene2003}%
  \BibitemOpen
  \bibfield  {author} {\bibinfo {author} {\bibfnamefont {J.}~\bibnamefont {Dehaene}}, \bibinfo {author} {\bibfnamefont {M.}~\bibnamefont {Van~den Nest}}, \bibinfo {author} {\bibfnamefont {B.}~\bibnamefont {De~Moor}},\ and\ \bibinfo {author} {\bibfnamefont {F.}~\bibnamefont {Verstraete}},\ }\bibfield  {title} {\bibinfo {title} {Local permutations of products of bell states and entanglement distillation},\ }\href {https://doi.org/10.1103/PhysRevA.67.022310} {\bibfield  {journal} {\bibinfo  {journal} {Phys. Rev. A}\ }\textbf {\bibinfo {volume} {67}},\ \bibinfo {pages} {022310} (\bibinfo {year} {2003})}\BibitemShut {NoStop}%
\bibitem [{\citenamefont {Jansen}\ \emph {et~al.}(2022)\citenamefont {Jansen}, \citenamefont {Goodenough}, \citenamefont {de~Bone}, \citenamefont {Gijswijt},\ and\ \citenamefont {Elkouss}}]{Jansen2022}%
  \BibitemOpen
  \bibfield  {author} {\bibinfo {author} {\bibfnamefont {S.}~\bibnamefont {Jansen}}, \bibinfo {author} {\bibfnamefont {K.}~\bibnamefont {Goodenough}}, \bibinfo {author} {\bibfnamefont {S.}~\bibnamefont {de~Bone}}, \bibinfo {author} {\bibfnamefont {D.}~\bibnamefont {Gijswijt}},\ and\ \bibinfo {author} {\bibfnamefont {D.}~\bibnamefont {Elkouss}},\ }\bibfield  {title} {\bibinfo {title} {Enumerating all bilocal {C}lifford distillation protocols through symmetry reduction},\ }\href {https://doi.org/10.22331/q-2022-05-19-715} {\bibfield  {journal} {\bibinfo  {journal} {{Quantum}}\ }\textbf {\bibinfo {volume} {6}},\ \bibinfo {pages} {715} (\bibinfo {year} {2022})}\BibitemShut {NoStop}%
\bibitem [{\citenamefont {Rozpedek}\ \emph {et~al.}(2018)\citenamefont {Rozpedek}, \citenamefont {Schiet}, \citenamefont {Thinh}, \citenamefont {Elkouss}, \citenamefont {Doherty},\ and\ \citenamefont {Wehner}}]{Rozpedek2018}%
  \BibitemOpen
  \bibfield  {author} {\bibinfo {author} {\bibfnamefont {F.}~\bibnamefont {Rozpedek}}, \bibinfo {author} {\bibfnamefont {T.}~\bibnamefont {Schiet}}, \bibinfo {author} {\bibfnamefont {L.~P.}\ \bibnamefont {Thinh}}, \bibinfo {author} {\bibfnamefont {D.}~\bibnamefont {Elkouss}}, \bibinfo {author} {\bibfnamefont {A.~C.}\ \bibnamefont {Doherty}},\ and\ \bibinfo {author} {\bibfnamefont {S.}~\bibnamefont {Wehner}},\ }\bibfield  {title} {\bibinfo {title} {Optimizing practical entanglement distillation},\ }\href {https://doi.org/10.1103/PhysRevA.97.062333} {\bibfield  {journal} {\bibinfo  {journal} {Phys. Rev. A}\ }\textbf {\bibinfo {volume} {97}},\ \bibinfo {pages} {062333} (\bibinfo {year} {2018})}\BibitemShut {NoStop}%
\bibitem [{\citenamefont {Krastanov}\ \emph {et~al.}(2019)\citenamefont {Krastanov}, \citenamefont {Albert},\ and\ \citenamefont {Jiang}}]{Krastanov2019}%
  \BibitemOpen
  \bibfield  {author} {\bibinfo {author} {\bibfnamefont {S.}~\bibnamefont {Krastanov}}, \bibinfo {author} {\bibfnamefont {V.~V.}\ \bibnamefont {Albert}},\ and\ \bibinfo {author} {\bibfnamefont {L.}~\bibnamefont {Jiang}},\ }\bibfield  {title} {\bibinfo {title} {Optimized {E}ntanglement {P}urification},\ }\href {https://doi.org/10.22331/q-2019-02-18-123} {\bibfield  {journal} {\bibinfo  {journal} {{Quantum}}\ }\textbf {\bibinfo {volume} {3}},\ \bibinfo {pages} {123} (\bibinfo {year} {2019})}\BibitemShut {NoStop}%
\bibitem [{\citenamefont {Walln\"ofer}\ \emph {et~al.}(2020)\citenamefont {Walln\"ofer}, \citenamefont {Melnikov}, \citenamefont {D\"ur},\ and\ \citenamefont {Briegel}}]{Wallnofer2020}%
  \BibitemOpen
  \bibfield  {author} {\bibinfo {author} {\bibfnamefont {J.}~\bibnamefont {Walln\"ofer}}, \bibinfo {author} {\bibfnamefont {A.~A.}\ \bibnamefont {Melnikov}}, \bibinfo {author} {\bibfnamefont {W.}~\bibnamefont {D\"ur}},\ and\ \bibinfo {author} {\bibfnamefont {H.~J.}\ \bibnamefont {Briegel}},\ }\bibfield  {title} {\bibinfo {title} {Machine learning for long-distance quantum communication},\ }\href {https://doi.org/10.1103/PRXQuantum.1.010301} {\bibfield  {journal} {\bibinfo  {journal} {PRX Quantum}\ }\textbf {\bibinfo {volume} {1}},\ \bibinfo {pages} {010301} (\bibinfo {year} {2020})}\BibitemShut {NoStop}%
\bibitem [{\citenamefont {Zhao}\ \emph {et~al.}(2021)\citenamefont {Zhao}, \citenamefont {Zhao}, \citenamefont {Wang}, \citenamefont {Song},\ and\ \citenamefont {Wang}}]{Zhao2021}%
  \BibitemOpen
  \bibfield  {author} {\bibinfo {author} {\bibfnamefont {X.}~\bibnamefont {Zhao}}, \bibinfo {author} {\bibfnamefont {B.}~\bibnamefont {Zhao}}, \bibinfo {author} {\bibfnamefont {Z.}~\bibnamefont {Wang}}, \bibinfo {author} {\bibfnamefont {Z.}~\bibnamefont {Song}},\ and\ \bibinfo {author} {\bibfnamefont {X.}~\bibnamefont {Wang}},\ }\bibfield  {title} {\bibinfo {title} {Practical distributed quantum information processing with loccnet},\ }\href {https://doi.org/10.1038/s41534-021-00496-x} {\bibfield  {journal} {\bibinfo  {journal} {npj Quantum Information}\ }\textbf {\bibinfo {volume} {7}},\ \bibinfo {pages} {159} (\bibinfo {year} {2021})}\BibitemShut {NoStop}%
\bibitem [{\citenamefont {Krastanov}\ \emph {et~al.}(2021)\citenamefont {Krastanov}, \citenamefont {de~la Cerda},\ and\ \citenamefont {Narang}}]{Krastanov2021}%
  \BibitemOpen
  \bibfield  {author} {\bibinfo {author} {\bibfnamefont {S.}~\bibnamefont {Krastanov}}, \bibinfo {author} {\bibfnamefont {A.~S.}\ \bibnamefont {de~la Cerda}},\ and\ \bibinfo {author} {\bibfnamefont {P.}~\bibnamefont {Narang}},\ }\bibfield  {title} {\bibinfo {title} {Heterogeneous multipartite entanglement purification for size-constrained quantum devices},\ }\href {https://doi.org/10.1103/PhysRevResearch.3.033164} {\bibfield  {journal} {\bibinfo  {journal} {Phys. Rev. Res.}\ }\textbf {\bibinfo {volume} {3}},\ \bibinfo {pages} {033164} (\bibinfo {year} {2021})}\BibitemShut {NoStop}%
\bibitem [{\citenamefont {Goodenough}\ \emph {et~al.}(2024)\citenamefont {Goodenough}, \citenamefont {de~Bone}, \citenamefont {Addala}, \citenamefont {Krastanov}, \citenamefont {Jansen}, \citenamefont {Gijswijt},\ and\ \citenamefont {Elkouss}}]{Goodenough2024}%
  \BibitemOpen
  \bibfield  {author} {\bibinfo {author} {\bibfnamefont {K.}~\bibnamefont {Goodenough}}, \bibinfo {author} {\bibfnamefont {S.}~\bibnamefont {de~Bone}}, \bibinfo {author} {\bibfnamefont {V.}~\bibnamefont {Addala}}, \bibinfo {author} {\bibfnamefont {S.}~\bibnamefont {Krastanov}}, \bibinfo {author} {\bibfnamefont {S.}~\bibnamefont {Jansen}}, \bibinfo {author} {\bibfnamefont {D.}~\bibnamefont {Gijswijt}},\ and\ \bibinfo {author} {\bibfnamefont {D.}~\bibnamefont {Elkouss}},\ }\bibfield  {title} {\bibinfo {title} {Near-term n to k distillation protocols using graph codes},\ }\href {https://doi.org/10.1109/JSAC.2024.3380094} {\bibfield  {journal} {\bibinfo  {journal} {IEEE Journal on Selected Areas in Communications}\ }\textbf {\bibinfo {volume} {42}},\ \bibinfo {pages} {1830} (\bibinfo {year} {2024})}\BibitemShut {NoStop}%
\bibitem [{\citenamefont {Addala}\ \emph {et~al.}(2025)\citenamefont {Addala}, \citenamefont {Ge},\ and\ \citenamefont {Krastanov}}]{Addala2025}%
  \BibitemOpen
  \bibfield  {author} {\bibinfo {author} {\bibfnamefont {V.~L.}\ \bibnamefont {Addala}}, \bibinfo {author} {\bibfnamefont {S.}~\bibnamefont {Ge}},\ and\ \bibinfo {author} {\bibfnamefont {S.}~\bibnamefont {Krastanov}},\ }\bibfield  {title} {\bibinfo {title} {Faster-than-clifford simulations of entanglement purification circuits and their full-stack optimization},\ }\href {https://doi.org/10.1038/s41534-024-00948-0} {\bibfield  {journal} {\bibinfo  {journal} {npj Quantum Information}\ }\textbf {\bibinfo {volume} {11}},\ \bibinfo {pages} {12} (\bibinfo {year} {2025})}\BibitemShut {NoStop}%
\bibitem [{\citenamefont {Aschauer}(2005)}]{Aschauer2005}%
  \BibitemOpen
  \bibfield  {author} {\bibinfo {author} {\bibfnamefont {H.}~\bibnamefont {Aschauer}},\ }\href {http://nbn-resolving.de/urn:nbn:de:bvb:19-35882} {\bibinfo {title} {Quantum communication in noisy environments}} (\bibinfo {year} {2005})\BibitemShut {NoStop}%
\bibitem [{\citenamefont {Dür}\ and\ \citenamefont {Briegel}(2007)}]{Dur2007}%
  \BibitemOpen
  \bibfield  {author} {\bibinfo {author} {\bibfnamefont {W.}~\bibnamefont {Dür}}\ and\ \bibinfo {author} {\bibfnamefont {H.~J.}\ \bibnamefont {Briegel}},\ }\bibfield  {title} {\bibinfo {title} {Entanglement purification and quantum error correction},\ }\href {https://doi.org/10.1088/0034-4885/70/8/R03} {\bibfield  {journal} {\bibinfo  {journal} {Reports on Progress in Physics}\ }\textbf {\bibinfo {volume} {70}},\ \bibinfo {pages} {1381} (\bibinfo {year} {2007})}\BibitemShut {NoStop}%
\bibitem [{\citenamefont {Fujii}\ and\ \citenamefont {Yamamoto}(2009)}]{Fujii2009}%
  \BibitemOpen
  \bibfield  {author} {\bibinfo {author} {\bibfnamefont {K.}~\bibnamefont {Fujii}}\ and\ \bibinfo {author} {\bibfnamefont {K.}~\bibnamefont {Yamamoto}},\ }\bibfield  {title} {\bibinfo {title} {Entanglement purification with double selection},\ }\href {https://doi.org/10.1103/PhysRevA.80.042308} {\bibfield  {journal} {\bibinfo  {journal} {Phys. Rev. A}\ }\textbf {\bibinfo {volume} {80}},\ \bibinfo {pages} {042308} (\bibinfo {year} {2009})}\BibitemShut {NoStop}%
\bibitem [{\citenamefont {Ruan}\ \emph {et~al.}(2018)\citenamefont {Ruan}, \citenamefont {Dai},\ and\ \citenamefont {Win}}]{Ruan2018}%
  \BibitemOpen
  \bibfield  {author} {\bibinfo {author} {\bibfnamefont {L.}~\bibnamefont {Ruan}}, \bibinfo {author} {\bibfnamefont {W.}~\bibnamefont {Dai}},\ and\ \bibinfo {author} {\bibfnamefont {M.~Z.}\ \bibnamefont {Win}},\ }\bibfield  {title} {\bibinfo {title} {Adaptive recurrence quantum entanglement distillation for two-kraus-operator channels},\ }\href {https://doi.org/10.1103/PhysRevA.97.052332} {\bibfield  {journal} {\bibinfo  {journal} {Phys. Rev. A}\ }\textbf {\bibinfo {volume} {97}},\ \bibinfo {pages} {052332} (\bibinfo {year} {2018})}\BibitemShut {NoStop}%
\bibitem [{\citenamefont {Jing}\ \emph {et~al.}(2020)\citenamefont {Jing}, \citenamefont {Alsina},\ and\ \citenamefont {Razavi}}]{Jing2020}%
  \BibitemOpen
  \bibfield  {author} {\bibinfo {author} {\bibfnamefont {Y.}~\bibnamefont {Jing}}, \bibinfo {author} {\bibfnamefont {D.}~\bibnamefont {Alsina}},\ and\ \bibinfo {author} {\bibfnamefont {M.}~\bibnamefont {Razavi}},\ }\bibfield  {title} {\bibinfo {title} {Quantum key distribution over quantum repeaters with encoding: Using error detection as an effective postselection tool},\ }\href {https://doi.org/10.1103/PhysRevApplied.14.064037} {\bibfield  {journal} {\bibinfo  {journal} {Phys. Rev. Appl.}\ }\textbf {\bibinfo {volume} {14}},\ \bibinfo {pages} {064037} (\bibinfo {year} {2020})}\BibitemShut {NoStop}%
\bibitem [{\citenamefont {Rengaswamy}\ \emph {et~al.}(2023)\citenamefont {Rengaswamy}, \citenamefont {Raina}, \citenamefont {Raveendran},\ and\ \citenamefont {Vasić}}]{Rengaswamy2023}%
  \BibitemOpen
  \bibfield  {author} {\bibinfo {author} {\bibfnamefont {N.}~\bibnamefont {Rengaswamy}}, \bibinfo {author} {\bibfnamefont {A.}~\bibnamefont {Raina}}, \bibinfo {author} {\bibfnamefont {N.}~\bibnamefont {Raveendran}},\ and\ \bibinfo {author} {\bibfnamefont {B.}~\bibnamefont {Vasić}},\ }\bibfield  {title} {\bibinfo {title} {Ghz distillation using quantum ldpc codes},\ }in\ \href {https://doi.org/10.1109/ISTC57237.2023.10273456} {\emph {\bibinfo {booktitle} {2023 12th International Symposium on Topics in Coding (ISTC)}}}\ (\bibinfo {year} {2023})\ pp.\ \bibinfo {pages} {1--5}\BibitemShut {NoStop}%
\bibitem [{\citenamefont {Rengaswamy}\ \emph {et~al.}(2024)\citenamefont {Rengaswamy}, \citenamefont {Raveendran}, \citenamefont {Raina},\ and\ \citenamefont {Vasi{\'{c}}}}]{Rengaswamy2024}%
  \BibitemOpen
  \bibfield  {author} {\bibinfo {author} {\bibfnamefont {N.}~\bibnamefont {Rengaswamy}}, \bibinfo {author} {\bibfnamefont {N.}~\bibnamefont {Raveendran}}, \bibinfo {author} {\bibfnamefont {A.}~\bibnamefont {Raina}},\ and\ \bibinfo {author} {\bibfnamefont {B.}~\bibnamefont {Vasi{\'{c}}}},\ }\bibfield  {title} {\bibinfo {title} {Entanglement {P}urification with {Q}uantum {LDPC} {C}odes and {I}terative {D}ecoding},\ }\href {https://doi.org/10.22331/q-2024-01-24-1233} {\bibfield  {journal} {\bibinfo  {journal} {{Quantum}}\ }\textbf {\bibinfo {volume} {8}},\ \bibinfo {pages} {1233} (\bibinfo {year} {2024})}\BibitemShut {NoStop}%
\bibitem [{\citenamefont {Wilde}\ \emph {et~al.}(2007)\citenamefont {Wilde}, \citenamefont {Krovi},\ and\ \citenamefont {Brun}}]{Wilde2007_Arxiv}%
  \BibitemOpen
  \bibfield  {author} {\bibinfo {author} {\bibfnamefont {M.~M.}\ \bibnamefont {Wilde}}, \bibinfo {author} {\bibfnamefont {H.}~\bibnamefont {Krovi}},\ and\ \bibinfo {author} {\bibfnamefont {T.~A.}\ \bibnamefont {Brun}},\ }\href@noop {} {\bibinfo {title} {Convolutional entanglement distillation}} (\bibinfo {year} {2007}),\ \Eprint {https://arxiv.org/abs/0708.3699} {arXiv:0708.3699 [quant-ph]} \BibitemShut {NoStop}%
\bibitem [{\citenamefont {Breuckmann}\ and\ \citenamefont {Eberhardt}(2021)}]{Breuckmann2021}%
  \BibitemOpen
  \bibfield  {author} {\bibinfo {author} {\bibfnamefont {N.~P.}\ \bibnamefont {Breuckmann}}\ and\ \bibinfo {author} {\bibfnamefont {J.~N.}\ \bibnamefont {Eberhardt}},\ }\bibfield  {title} {\bibinfo {title} {Quantum low-density parity-check codes},\ }\href {https://doi.org/10.1103/PRXQuantum.2.040101} {\bibfield  {journal} {\bibinfo  {journal} {PRX Quantum}\ }\textbf {\bibinfo {volume} {2}},\ \bibinfo {pages} {040101} (\bibinfo {year} {2021})}\BibitemShut {NoStop}%
\bibitem [{\citenamefont {Dennis}\ \emph {et~al.}(2002)\citenamefont {Dennis}, \citenamefont {Kitaev}, \citenamefont {Landahl},\ and\ \citenamefont {Preskill}}]{Dennis2002}%
  \BibitemOpen
  \bibfield  {author} {\bibinfo {author} {\bibfnamefont {E.}~\bibnamefont {Dennis}}, \bibinfo {author} {\bibfnamefont {A.}~\bibnamefont {Kitaev}}, \bibinfo {author} {\bibfnamefont {A.}~\bibnamefont {Landahl}},\ and\ \bibinfo {author} {\bibfnamefont {J.}~\bibnamefont {Preskill}},\ }\bibfield  {title} {\bibinfo {title} {{Topological quantum memory}},\ }\href {https://doi.org/10.1063/1.1499754} {\bibfield  {journal} {\bibinfo  {journal} {Journal of Mathematical Physics}\ }\textbf {\bibinfo {volume} {43}},\ \bibinfo {pages} {4452} (\bibinfo {year} {2002})}\BibitemShut {NoStop}%
\bibitem [{\citenamefont {Kitaev}(2003)}]{Kitaev2003}%
  \BibitemOpen
  \bibfield  {author} {\bibinfo {author} {\bibfnamefont {A.}~\bibnamefont {Kitaev}},\ }\bibfield  {title} {\bibinfo {title} {Fault-tolerant quantum computation by anyons},\ }\href {https://doi.org/https://doi.org/10.1016/S0003-4916(02)00018-0} {\bibfield  {journal} {\bibinfo  {journal} {Annals of Physics}\ }\textbf {\bibinfo {volume} {303}},\ \bibinfo {pages} {2} (\bibinfo {year} {2003})}\BibitemShut {NoStop}%
\bibitem [{\citenamefont {Bravyi}\ and\ \citenamefont {Kitaev}(1998)}]{Bravyi1998}%
  \BibitemOpen
  \bibfield  {author} {\bibinfo {author} {\bibfnamefont {S.~B.}\ \bibnamefont {Bravyi}}\ and\ \bibinfo {author} {\bibfnamefont {A.~Y.}\ \bibnamefont {Kitaev}},\ }\href {https://arxiv.org/abs/quant-ph/9811052} {\bibinfo {title} {Quantum codes on a lattice with boundary}} (\bibinfo {year} {1998}),\ \Eprint {https://arxiv.org/abs/quant-ph/9811052} {arXiv:quant-ph/9811052 [quant-ph]} \BibitemShut {NoStop}%
\bibitem [{\citenamefont {Gottesman}(1997)}]{Gottesman1997}%
  \BibitemOpen
  \bibfield  {author} {\bibinfo {author} {\bibfnamefont {D.}~\bibnamefont {Gottesman}},\ }\href@noop {} {\bibinfo {title} {Stabilizer codes and quantum error correction}} (\bibinfo {year} {1997}),\ \Eprint {https://arxiv.org/abs/quant-ph/9705052} {arXiv:quant-ph/9705052 [quant-ph]} \BibitemShut {NoStop}%
\bibitem [{\citenamefont {Bravyi}\ \emph {et~al.}(2006)\citenamefont {Bravyi}, \citenamefont {Hastings},\ and\ \citenamefont {Verstraete}}]{Bravyi2006}%
  \BibitemOpen
  \bibfield  {author} {\bibinfo {author} {\bibfnamefont {S.}~\bibnamefont {Bravyi}}, \bibinfo {author} {\bibfnamefont {M.~B.}\ \bibnamefont {Hastings}},\ and\ \bibinfo {author} {\bibfnamefont {F.}~\bibnamefont {Verstraete}},\ }\bibfield  {title} {\bibinfo {title} {Lieb-robinson bounds and the generation of correlations and topological quantum order},\ }\href {https://doi.org/10.1103/PhysRevLett.97.050401} {\bibfield  {journal} {\bibinfo  {journal} {Phys. Rev. Lett.}\ }\textbf {\bibinfo {volume} {97}},\ \bibinfo {pages} {050401} (\bibinfo {year} {2006})}\BibitemShut {NoStop}%
\bibitem [{\citenamefont {Aharonov}\ and\ \citenamefont {Touati}(2018)}]{Aharonov2018}%
  \BibitemOpen
  \bibfield  {author} {\bibinfo {author} {\bibfnamefont {D.}~\bibnamefont {Aharonov}}\ and\ \bibinfo {author} {\bibfnamefont {Y.}~\bibnamefont {Touati}},\ }\href@noop {} {\bibinfo {title} {Quantum circuit depth lower bounds for homological codes}} (\bibinfo {year} {2018}),\ \Eprint {https://arxiv.org/abs/1810.03912} {arXiv:1810.03912 [quant-ph]} \BibitemShut {NoStop}%
\bibitem [{\citenamefont {Forney}\ and\ \citenamefont {Guha}(2005)}]{Forney2005}%
  \BibitemOpen
  \bibfield  {author} {\bibinfo {author} {\bibfnamefont {G.}~\bibnamefont {Forney}}\ and\ \bibinfo {author} {\bibfnamefont {S.}~\bibnamefont {Guha}},\ }\bibfield  {title} {\bibinfo {title} {Simple rate-1/3 convolutional and tail-biting quantum error-correcting codes},\ }in\ \href {https://doi.org/10.1109/ISIT.2005.1523495} {\emph {\bibinfo {booktitle} {Proceedings. International Symposium on Information Theory, 2005. ISIT 2005.}}}\ (\bibinfo {year} {2005})\ pp.\ \bibinfo {pages} {1028--1032}\BibitemShut {NoStop}%
\bibitem [{\citenamefont {Forney}\ \emph {et~al.}(2007)\citenamefont {Forney}, \citenamefont {Grassl},\ and\ \citenamefont {Guha}}]{Forney2007}%
  \BibitemOpen
  \bibfield  {author} {\bibinfo {author} {\bibfnamefont {G.~D.}\ \bibnamefont {Forney}}, \bibinfo {author} {\bibfnamefont {M.}~\bibnamefont {Grassl}},\ and\ \bibinfo {author} {\bibfnamefont {S.}~\bibnamefont {Guha}},\ }\bibfield  {title} {\bibinfo {title} {Convolutional and tail-biting quantum error-correcting codes},\ }\href {https://doi.org/10.1109/TIT.2006.890698} {\bibfield  {journal} {\bibinfo  {journal} {IEEE Transactions on Information Theory}\ }\textbf {\bibinfo {volume} {53}},\ \bibinfo {pages} {865} (\bibinfo {year} {2007})}\BibitemShut {NoStop}%
\bibitem [{\citenamefont {Grassl}\ and\ \citenamefont {Rotteler}(2006)}]{Grassl2006}%
  \BibitemOpen
  \bibfield  {author} {\bibinfo {author} {\bibfnamefont {M.}~\bibnamefont {Grassl}}\ and\ \bibinfo {author} {\bibfnamefont {M.}~\bibnamefont {Rotteler}},\ }\bibfield  {title} {\bibinfo {title} {Non-catastrophic encoders and encoder inverses for quantum convolutional codes},\ }in\ \href {https://doi.org/10.1109/ISIT.2006.261956} {\emph {\bibinfo {booktitle} {2006 IEEE International Symposium on Information Theory}}}\ (\bibinfo {year} {2006})\ pp.\ \bibinfo {pages} {1109--1113}\BibitemShut {NoStop}%
\bibitem [{\citenamefont {Houshmand}\ and\ \citenamefont {Wilde}(2013)}]{Houshmand2013}%
  \BibitemOpen
  \bibfield  {author} {\bibinfo {author} {\bibfnamefont {M.}~\bibnamefont {Houshmand}}\ and\ \bibinfo {author} {\bibfnamefont {M.~M.}\ \bibnamefont {Wilde}},\ }\bibfield  {title} {\bibinfo {title} {Recursive quantum convolutional encoders are catastrophic: A simple proof},\ }\href {https://doi.org/10.1109/TIT.2013.2272932} {\bibfield  {journal} {\bibinfo  {journal} {IEEE Transactions on Information Theory}\ }\textbf {\bibinfo {volume} {59}},\ \bibinfo {pages} {6724} (\bibinfo {year} {2013})}\BibitemShut {NoStop}%
\bibitem [{\citenamefont {Bluvstein}\ \emph {et~al.}(2024)\citenamefont {Bluvstein}, \citenamefont {Evered}, \citenamefont {Geim}, \citenamefont {Li}, \citenamefont {Zhou}, \citenamefont {Manovitz}, \citenamefont {Ebadi}, \citenamefont {Cain}, \citenamefont {Kalinowski}, \citenamefont {Hangleiter}, \citenamefont {Bonilla~Ataides}, \citenamefont {Maskara}, \citenamefont {Cong}, \citenamefont {Gao}, \citenamefont {Sales~Rodriguez}, \citenamefont {Karolyshyn}, \citenamefont {Semeghini}, \citenamefont {Gullans}, \citenamefont {Greiner}, \citenamefont {Vuleti{\'{c}}},\ and\ \citenamefont {Lukin}}]{Bluvstein2024}%
  \BibitemOpen
  \bibfield  {author} {\bibinfo {author} {\bibfnamefont {D.}~\bibnamefont {Bluvstein}}, \bibinfo {author} {\bibfnamefont {S.~J.}\ \bibnamefont {Evered}}, \bibinfo {author} {\bibfnamefont {A.~A.}\ \bibnamefont {Geim}}, \bibinfo {author} {\bibfnamefont {S.~H.}\ \bibnamefont {Li}}, \bibinfo {author} {\bibfnamefont {H.}~\bibnamefont {Zhou}}, \bibinfo {author} {\bibfnamefont {T.}~\bibnamefont {Manovitz}}, \bibinfo {author} {\bibfnamefont {S.}~\bibnamefont {Ebadi}}, \bibinfo {author} {\bibfnamefont {M.}~\bibnamefont {Cain}}, \bibinfo {author} {\bibfnamefont {M.}~\bibnamefont {Kalinowski}}, \bibinfo {author} {\bibfnamefont {D.}~\bibnamefont {Hangleiter}}, \bibinfo {author} {\bibfnamefont {J.~P.}\ \bibnamefont {Bonilla~Ataides}}, \bibinfo {author} {\bibfnamefont {N.}~\bibnamefont {Maskara}}, \bibinfo {author} {\bibfnamefont {I.}~\bibnamefont {Cong}}, \bibinfo {author} {\bibfnamefont {X.}~\bibnamefont {Gao}}, \bibinfo {author} {\bibfnamefont {P.}~\bibnamefont {Sales~Rodriguez}}, \bibinfo {author} {\bibfnamefont
  {T.}~\bibnamefont {Karolyshyn}}, \bibinfo {author} {\bibfnamefont {G.}~\bibnamefont {Semeghini}}, \bibinfo {author} {\bibfnamefont {M.~J.}\ \bibnamefont {Gullans}}, \bibinfo {author} {\bibfnamefont {M.}~\bibnamefont {Greiner}}, \bibinfo {author} {\bibfnamefont {V.}~\bibnamefont {Vuleti{\'{c}}}},\ and\ \bibinfo {author} {\bibfnamefont {M.~D.}\ \bibnamefont {Lukin}},\ }\bibfield  {title} {\bibinfo {title} {Logical quantum processor based on reconfigurable atom arrays},\ }\href {https://doi.org/10.1038/s41586-023-06927-3} {\bibfield  {journal} {\bibinfo  {journal} {Nature}\ }\textbf {\bibinfo {volume} {626}},\ \bibinfo {pages} {58} (\bibinfo {year} {2024})}\BibitemShut {NoStop}%
\bibitem [{\citenamefont {Xu}\ \emph {et~al.}(2024)\citenamefont {Xu}, \citenamefont {Bonilla~Ataides}, \citenamefont {Pattison}, \citenamefont {Raveendran}, \citenamefont {Bluvstein}, \citenamefont {Wurtz}, \citenamefont {Vasi{\'{c}}}, \citenamefont {Lukin}, \citenamefont {Jiang},\ and\ \citenamefont {Zhou}}]{Xu2024}%
  \BibitemOpen
  \bibfield  {author} {\bibinfo {author} {\bibfnamefont {Q.}~\bibnamefont {Xu}}, \bibinfo {author} {\bibfnamefont {J.~P.}\ \bibnamefont {Bonilla~Ataides}}, \bibinfo {author} {\bibfnamefont {C.~A.}\ \bibnamefont {Pattison}}, \bibinfo {author} {\bibfnamefont {N.}~\bibnamefont {Raveendran}}, \bibinfo {author} {\bibfnamefont {D.}~\bibnamefont {Bluvstein}}, \bibinfo {author} {\bibfnamefont {J.}~\bibnamefont {Wurtz}}, \bibinfo {author} {\bibfnamefont {B.}~\bibnamefont {Vasi{\'{c}}}}, \bibinfo {author} {\bibfnamefont {M.~D.}\ \bibnamefont {Lukin}}, \bibinfo {author} {\bibfnamefont {L.}~\bibnamefont {Jiang}},\ and\ \bibinfo {author} {\bibfnamefont {H.}~\bibnamefont {Zhou}},\ }\bibfield  {title} {\bibinfo {title} {Constant-overhead fault-tolerant quantum computation with reconfigurable atom arrays},\ }\bibfield  {journal} {\bibinfo  {journal} {Nature Physics}\ }\href {https://doi.org/10.1038/s41567-024-02479-z} {10.1038/s41567-024-02479-z} (\bibinfo {year} {2024})\BibitemShut {NoStop}%
\bibitem [{\citenamefont {Nielsen}\ and\ \citenamefont {Chuang}(2010)}]{Nielsen2010}%
  \BibitemOpen
  \bibfield  {author} {\bibinfo {author} {\bibfnamefont {M.}~\bibnamefont {Nielsen}}\ and\ \bibinfo {author} {\bibfnamefont {I.}~\bibnamefont {Chuang}},\ }\href {https://books.google.com/books?id=-s4DEy7o-a0C} {\emph {\bibinfo {title} {Quantum Computation and Quantum Information: 10th Anniversary Edition}}}\ (\bibinfo  {publisher} {Cambridge University Press},\ \bibinfo {year} {2010})\BibitemShut {NoStop}%
\bibitem [{\citenamefont {Bacon}\ and\ \citenamefont {Casaccino}(2006)}]{Bacon2006}%
  \BibitemOpen
  \bibfield  {author} {\bibinfo {author} {\bibfnamefont {D.}~\bibnamefont {Bacon}}\ and\ \bibinfo {author} {\bibfnamefont {A.}~\bibnamefont {Casaccino}},\ }\href {https://arxiv.org/abs/quant-ph/0610088} {\bibinfo {title} {Quantum error correcting subsystem codes from two classical linear codes}} (\bibinfo {year} {2006}),\ \Eprint {https://arxiv.org/abs/quant-ph/0610088} {arXiv:quant-ph/0610088 [quant-ph]} \BibitemShut {NoStop}%
\bibitem [{\citenamefont {Aaronson}\ and\ \citenamefont {Gottesman}(2004)}]{Aaronson2004}%
  \BibitemOpen
  \bibfield  {author} {\bibinfo {author} {\bibfnamefont {S.}~\bibnamefont {Aaronson}}\ and\ \bibinfo {author} {\bibfnamefont {D.}~\bibnamefont {Gottesman}},\ }\bibfield  {title} {\bibinfo {title} {Improved simulation of stabilizer circuits},\ }\href {https://doi.org/10.1103/PhysRevA.70.052328} {\bibfield  {journal} {\bibinfo  {journal} {Phys. Rev. A}\ }\textbf {\bibinfo {volume} {70}},\ \bibinfo {pages} {052328} (\bibinfo {year} {2004})}\BibitemShut {NoStop}%
\bibitem [{\citenamefont {Tillich}\ and\ \citenamefont {Zémor}(2014)}]{Tillich2014}%
  \BibitemOpen
  \bibfield  {author} {\bibinfo {author} {\bibfnamefont {J.-P.}\ \bibnamefont {Tillich}}\ and\ \bibinfo {author} {\bibfnamefont {G.}~\bibnamefont {Zémor}},\ }\bibfield  {title} {\bibinfo {title} {Quantum ldpc codes with positive rate and minimum distance proportional to the square root of the blocklength},\ }\href {https://doi.org/10.1109/TIT.2013.2292061} {\bibfield  {journal} {\bibinfo  {journal} {IEEE Transactions on Information Theory}\ }\textbf {\bibinfo {volume} {60}},\ \bibinfo {pages} {1193} (\bibinfo {year} {2014})}\BibitemShut {NoStop}%
\bibitem [{\citenamefont {Moehring}\ \emph {et~al.}(2007)\citenamefont {Moehring}, \citenamefont {Madsen}, \citenamefont {Younge}, \citenamefont {R.~N.~Kohn}, \citenamefont {Maunz}, \citenamefont {Duan}, \citenamefont {Monroe},\ and\ \citenamefont {Blinov}}]{Moehring2007}%
  \BibitemOpen
  \bibfield  {author} {\bibinfo {author} {\bibfnamefont {D.~L.}\ \bibnamefont {Moehring}}, \bibinfo {author} {\bibfnamefont {M.~J.}\ \bibnamefont {Madsen}}, \bibinfo {author} {\bibfnamefont {K.~C.}\ \bibnamefont {Younge}}, \bibinfo {author} {\bibfnamefont {J.}~\bibnamefont {R.~N.~Kohn}}, \bibinfo {author} {\bibfnamefont {P.}~\bibnamefont {Maunz}}, \bibinfo {author} {\bibfnamefont {L.-M.}\ \bibnamefont {Duan}}, \bibinfo {author} {\bibfnamefont {C.}~\bibnamefont {Monroe}},\ and\ \bibinfo {author} {\bibfnamefont {B.~B.}\ \bibnamefont {Blinov}},\ }\bibfield  {title} {\bibinfo {title} {Quantum networking with photons and trapped atoms (invited)},\ }\href {https://doi.org/10.1364/JOSAB.24.000300} {\bibfield  {journal} {\bibinfo  {journal} {J. Opt. Soc. Am. B}\ }\textbf {\bibinfo {volume} {24}},\ \bibinfo {pages} {300} (\bibinfo {year} {2007})}\BibitemShut {NoStop}%
\bibitem [{\citenamefont {Dhara}\ \emph {et~al.}(2023)\citenamefont {Dhara}, \citenamefont {Englund},\ and\ \citenamefont {Guha}}]{Dhara2023}%
  \BibitemOpen
  \bibfield  {author} {\bibinfo {author} {\bibfnamefont {P.}~\bibnamefont {Dhara}}, \bibinfo {author} {\bibfnamefont {D.}~\bibnamefont {Englund}},\ and\ \bibinfo {author} {\bibfnamefont {S.}~\bibnamefont {Guha}},\ }\bibfield  {title} {\bibinfo {title} {Entangling quantum memories via heralded photonic bell measurement},\ }\href {https://doi.org/10.1103/PhysRevResearch.5.033149} {\bibfield  {journal} {\bibinfo  {journal} {Phys. Rev. Res.}\ }\textbf {\bibinfo {volume} {5}},\ \bibinfo {pages} {033149} (\bibinfo {year} {2023})}\BibitemShut {NoStop}%
\bibitem [{\citenamefont {Shi}\ and\ \citenamefont {Waks}(2023)}]{Shi2023}%
  \BibitemOpen
  \bibfield  {author} {\bibinfo {author} {\bibfnamefont {Y.}~\bibnamefont {Shi}}\ and\ \bibinfo {author} {\bibfnamefont {E.}~\bibnamefont {Waks}},\ }\bibfield  {title} {\bibinfo {title} {Error metric for non-trace-preserving quantum operations},\ }\href {https://doi.org/10.1103/PhysRevA.108.032609} {\bibfield  {journal} {\bibinfo  {journal} {Phys. Rev. A}\ }\textbf {\bibinfo {volume} {108}},\ \bibinfo {pages} {032609} (\bibinfo {year} {2023})}\BibitemShut {NoStop}%
\bibitem [{\citenamefont {Ollivier}\ and\ \citenamefont {Tillich}(2003)}]{Ollivier2003}%
  \BibitemOpen
  \bibfield  {author} {\bibinfo {author} {\bibfnamefont {H.}~\bibnamefont {Ollivier}}\ and\ \bibinfo {author} {\bibfnamefont {J.-P.}\ \bibnamefont {Tillich}},\ }\bibfield  {title} {\bibinfo {title} {Description of a quantum convolutional code},\ }\href {https://doi.org/10.1103/PhysRevLett.91.177902} {\bibfield  {journal} {\bibinfo  {journal} {Phys. Rev. Lett.}\ }\textbf {\bibinfo {volume} {91}},\ \bibinfo {pages} {177902} (\bibinfo {year} {2003})}\BibitemShut {NoStop}%
\bibitem [{\citenamefont {Ollivier}\ and\ \citenamefont {Tillich}(2004)}]{Ollivier2004}%
  \BibitemOpen
  \bibfield  {author} {\bibinfo {author} {\bibfnamefont {H.}~\bibnamefont {Ollivier}}\ and\ \bibinfo {author} {\bibfnamefont {J.~P.}\ \bibnamefont {Tillich}},\ }\href@noop {} {\bibinfo {title} {Quantum convolutional codes: fundamentals}} (\bibinfo {year} {2004}),\ \Eprint {https://arxiv.org/abs/quant-ph/0401134} {arXiv:quant-ph/0401134 [quant-ph]} \BibitemShut {NoStop}%
\bibitem [{\citenamefont {Grassl}\ and\ \citenamefont {Rotteler}(2007)}]{Grassl2007}%
  \BibitemOpen
  \bibfield  {author} {\bibinfo {author} {\bibfnamefont {M.}~\bibnamefont {Grassl}}\ and\ \bibinfo {author} {\bibfnamefont {M.}~\bibnamefont {Rotteler}},\ }\bibfield  {title} {\bibinfo {title} {Constructions of quantum convolutional codes},\ }in\ \href {https://doi.org/10.1109/ISIT.2007.4557325} {\emph {\bibinfo {booktitle} {2007 IEEE International Symposium on Information Theory}}}\ (\bibinfo {year} {2007})\ pp.\ \bibinfo {pages} {816--820}\BibitemShut {NoStop}%
\bibitem [{\citenamefont {Johannesson}\ and\ \citenamefont {Zigangirov}(2015)}]{Johannesson2015}%
  \BibitemOpen
  \bibfield  {author} {\bibinfo {author} {\bibfnamefont {R.}~\bibnamefont {Johannesson}}\ and\ \bibinfo {author} {\bibfnamefont {K.~S.}\ \bibnamefont {Zigangirov}},\ }\href {https://doi.org/10.1002/9781119098799.fmatter} {\emph {\bibinfo {title} {Fundamentals of Convolutional Coding}}}\ (\bibinfo  {publisher} {John Wiley \& Sons},\ \bibinfo {year} {2015})\BibitemShut {NoStop}%
\bibitem [{\citenamefont {Poulin}\ \emph {et~al.}(2009)\citenamefont {Poulin}, \citenamefont {Tillich},\ and\ \citenamefont {Ollivier}}]{Poulin2009}%
  \BibitemOpen
  \bibfield  {author} {\bibinfo {author} {\bibfnamefont {D.}~\bibnamefont {Poulin}}, \bibinfo {author} {\bibfnamefont {J.-P.}\ \bibnamefont {Tillich}},\ and\ \bibinfo {author} {\bibfnamefont {H.}~\bibnamefont {Ollivier}},\ }\bibfield  {title} {\bibinfo {title} {Quantum serial turbo codes},\ }\href {https://doi.org/10.1109/TIT.2009.2018339} {\bibfield  {journal} {\bibinfo  {journal} {IEEE Transactions on Information Theory}\ }\textbf {\bibinfo {volume} {55}},\ \bibinfo {pages} {2776} (\bibinfo {year} {2009})}\BibitemShut {NoStop}%
\bibitem [{\citenamefont {Wilde}\ \emph {et~al.}(2014)\citenamefont {Wilde}, \citenamefont {Hsieh},\ and\ \citenamefont {Babar}}]{Wilde2014}%
  \BibitemOpen
  \bibfield  {author} {\bibinfo {author} {\bibfnamefont {M.~M.}\ \bibnamefont {Wilde}}, \bibinfo {author} {\bibfnamefont {M.-H.}\ \bibnamefont {Hsieh}},\ and\ \bibinfo {author} {\bibfnamefont {Z.}~\bibnamefont {Babar}},\ }\bibfield  {title} {\bibinfo {title} {Entanglement-assisted quantum turbo codes},\ }\href {https://doi.org/10.1109/TIT.2013.2292052} {\bibfield  {journal} {\bibinfo  {journal} {IEEE Transactions on Information Theory}\ }\textbf {\bibinfo {volume} {60}},\ \bibinfo {pages} {1203} (\bibinfo {year} {2014})}\BibitemShut {NoStop}%
\bibitem [{\citenamefont {Bolt}\ \emph {et~al.}(2016)\citenamefont {Bolt}, \citenamefont {Duclos-Cianci}, \citenamefont {Poulin},\ and\ \citenamefont {Stace}}]{Bolt2016}%
  \BibitemOpen
  \bibfield  {author} {\bibinfo {author} {\bibfnamefont {A.}~\bibnamefont {Bolt}}, \bibinfo {author} {\bibfnamefont {G.}~\bibnamefont {Duclos-Cianci}}, \bibinfo {author} {\bibfnamefont {D.}~\bibnamefont {Poulin}},\ and\ \bibinfo {author} {\bibfnamefont {T.~M.}\ \bibnamefont {Stace}},\ }\bibfield  {title} {\bibinfo {title} {Foliated quantum error-correcting codes},\ }\href {https://doi.org/10.1103/PhysRevLett.117.070501} {\bibfield  {journal} {\bibinfo  {journal} {Phys. Rev. Lett.}\ }\textbf {\bibinfo {volume} {117}},\ \bibinfo {pages} {070501} (\bibinfo {year} {2016})}\BibitemShut {NoStop}%
\bibitem [{\citenamefont {Schalkwijk}\ and\ \citenamefont {Vinck}(1975)}]{Schalkwijk1975}%
  \BibitemOpen
  \bibfield  {author} {\bibinfo {author} {\bibfnamefont {J.}~\bibnamefont {Schalkwijk}}\ and\ \bibinfo {author} {\bibfnamefont {A.}~\bibnamefont {Vinck}},\ }\bibfield  {title} {\bibinfo {title} {Syndrome decoding of convolutional codes},\ }\href {https://doi.org/10.1109/TCOM.1975.1092884} {\bibfield  {journal} {\bibinfo  {journal} {IEEE Transactions on Communications}\ }\textbf {\bibinfo {volume} {23}},\ \bibinfo {pages} {789} (\bibinfo {year} {1975})}\BibitemShut {NoStop}%
\bibitem [{\citenamefont {Bravyi}\ \emph {et~al.}(2020)\citenamefont {Bravyi}, \citenamefont {Gosset}, \citenamefont {K{\"o}nig},\ and\ \citenamefont {Tomamichel}}]{Bravyi2020}%
  \BibitemOpen
  \bibfield  {author} {\bibinfo {author} {\bibfnamefont {S.}~\bibnamefont {Bravyi}}, \bibinfo {author} {\bibfnamefont {D.}~\bibnamefont {Gosset}}, \bibinfo {author} {\bibfnamefont {R.}~\bibnamefont {K{\"o}nig}},\ and\ \bibinfo {author} {\bibfnamefont {M.}~\bibnamefont {Tomamichel}},\ }\bibfield  {title} {\bibinfo {title} {Quantum advantage with noisy shallow circuits},\ }\href {https://doi.org/10.1038/s41567-020-0948-z} {\bibfield  {journal} {\bibinfo  {journal} {Nature Physics}\ }\textbf {\bibinfo {volume} {16}},\ \bibinfo {pages} {1040} (\bibinfo {year} {2020})}\BibitemShut {NoStop}%
\bibitem [{\citenamefont {Fowler}\ \emph {et~al.}(2009)\citenamefont {Fowler}, \citenamefont {Stephens},\ and\ \citenamefont {Groszkowski}}]{Fowler2009}%
  \BibitemOpen
  \bibfield  {author} {\bibinfo {author} {\bibfnamefont {A.~G.}\ \bibnamefont {Fowler}}, \bibinfo {author} {\bibfnamefont {A.~M.}\ \bibnamefont {Stephens}},\ and\ \bibinfo {author} {\bibfnamefont {P.}~\bibnamefont {Groszkowski}},\ }\bibfield  {title} {\bibinfo {title} {High-threshold universal quantum computation on the surface code},\ }\href {https://doi.org/10.1103/PhysRevA.80.052312} {\bibfield  {journal} {\bibinfo  {journal} {Phys. Rev. A}\ }\textbf {\bibinfo {volume} {80}},\ \bibinfo {pages} {052312} (\bibinfo {year} {2009})}\BibitemShut {NoStop}%
\bibitem [{\citenamefont {Moussa}(2016)}]{Moussa2016}%
  \BibitemOpen
  \bibfield  {author} {\bibinfo {author} {\bibfnamefont {J.~E.}\ \bibnamefont {Moussa}},\ }\bibfield  {title} {\bibinfo {title} {Transversal clifford gates on folded surface codes},\ }\href {https://doi.org/10.1103/PhysRevA.94.042316} {\bibfield  {journal} {\bibinfo  {journal} {Phys. Rev. A}\ }\textbf {\bibinfo {volume} {94}},\ \bibinfo {pages} {042316} (\bibinfo {year} {2016})}\BibitemShut {NoStop}%
\bibitem [{\citenamefont {Mazurek}\ \emph {et~al.}(2014)\citenamefont {Mazurek}, \citenamefont {Grudka}, \citenamefont {Horodecki}, \citenamefont {Horodecki}, \citenamefont {\L{}odyga}, \citenamefont {Pankowski},\ and\ \citenamefont {Przysi\ifmmode \mbox{\k{e}}\else \k{e}\fi{}\ifmmode~\dot{z}\else \.{z}\fi{}na}}]{Mazurek2014}%
  \BibitemOpen
  \bibfield  {author} {\bibinfo {author} {\bibfnamefont {P.}~\bibnamefont {Mazurek}}, \bibinfo {author} {\bibfnamefont {A.}~\bibnamefont {Grudka}}, \bibinfo {author} {\bibfnamefont {M.}~\bibnamefont {Horodecki}}, \bibinfo {author} {\bibfnamefont {P.}~\bibnamefont {Horodecki}}, \bibinfo {author} {\bibfnamefont {J.}~\bibnamefont {\L{}odyga}}, \bibinfo {author} {\bibfnamefont {L.}~\bibnamefont {Pankowski}},\ and\ \bibinfo {author} {\bibfnamefont {A.}~\bibnamefont {Przysi\ifmmode \mbox{\k{e}}\else \k{e}\fi{}\ifmmode~\dot{z}\else \.{z}\fi{}na}},\ }\bibfield  {title} {\bibinfo {title} {Long-distance quantum communication over noisy networks without long-time quantum memory},\ }\href {https://doi.org/10.1103/PhysRevA.90.062311} {\bibfield  {journal} {\bibinfo  {journal} {Phys. Rev. A}\ }\textbf {\bibinfo {volume} {90}},\ \bibinfo {pages} {062311} (\bibinfo {year} {2014})}\BibitemShut {NoStop}%
\bibitem [{\citenamefont {Bomb\'{\i}n}(2015)}]{Bombin2015}%
  \BibitemOpen
  \bibfield  {author} {\bibinfo {author} {\bibfnamefont {H.}~\bibnamefont {Bomb\'{\i}n}},\ }\bibfield  {title} {\bibinfo {title} {Single-shot fault-tolerant quantum error correction},\ }\href {https://doi.org/10.1103/PhysRevX.5.031043} {\bibfield  {journal} {\bibinfo  {journal} {Phys. Rev. X}\ }\textbf {\bibinfo {volume} {5}},\ \bibinfo {pages} {031043} (\bibinfo {year} {2015})}\BibitemShut {NoStop}%
\bibitem [{\citenamefont {Shor}(1996)}]{Shor1996}%
  \BibitemOpen
  \bibfield  {author} {\bibinfo {author} {\bibfnamefont {P.~W.}\ \bibnamefont {Shor}},\ }\bibfield  {title} {\bibinfo {title} {Fault-tolerant quantum computation},\ }in\ \href@noop {} {\emph {\bibinfo {booktitle} {Proceedings of the 37th Annual Symposium on Foundations of Computer Science}}},\ \bibinfo {series and number} {FOCS '96}\ (\bibinfo  {publisher} {IEEE Computer Society},\ \bibinfo {address} {USA},\ \bibinfo {year} {1996})\ p.~\bibinfo {pages} {56}\BibitemShut {NoStop}%
\bibitem [{\citenamefont {Steane}(1997)}]{Steane1997}%
  \BibitemOpen
  \bibfield  {author} {\bibinfo {author} {\bibfnamefont {A.~M.}\ \bibnamefont {Steane}},\ }\bibfield  {title} {\bibinfo {title} {Active stabilization, quantum computation, and quantum state synthesis},\ }\href {https://doi.org/10.1103/PhysRevLett.78.2252} {\bibfield  {journal} {\bibinfo  {journal} {Phys. Rev. Lett.}\ }\textbf {\bibinfo {volume} {78}},\ \bibinfo {pages} {2252} (\bibinfo {year} {1997})}\BibitemShut {NoStop}%
\bibitem [{\citenamefont {Knill}(2005)}]{Knill2005}%
  \BibitemOpen
  \bibfield  {author} {\bibinfo {author} {\bibfnamefont {E.}~\bibnamefont {Knill}},\ }\bibfield  {title} {\bibinfo {title} {Scalable quantum computing in the presence of large detected-error rates},\ }\href {https://doi.org/10.1103/PhysRevA.71.042322} {\bibfield  {journal} {\bibinfo  {journal} {Phys. Rev. A}\ }\textbf {\bibinfo {volume} {71}},\ \bibinfo {pages} {042322} (\bibinfo {year} {2005})}\BibitemShut {NoStop}%
\bibitem [{\citenamefont {Gottesman}(2014)}]{Gottesman2014}%
  \BibitemOpen
  \bibfield  {author} {\bibinfo {author} {\bibfnamefont {D.}~\bibnamefont {Gottesman}},\ }\bibfield  {title} {\bibinfo {title} {Fault-tolerant quantum computation with constant overhead},\ }\href@noop {} {\bibfield  {journal} {\bibinfo  {journal} {Quantum Info. Comput.}\ }\textbf {\bibinfo {volume} {14}},\ \bibinfo {pages} {1338–1372} (\bibinfo {year} {2014})}\BibitemShut {NoStop}%
\bibitem [{\citenamefont {Bravyi}\ \emph {et~al.}(2010)\citenamefont {Bravyi}, \citenamefont {Poulin},\ and\ \citenamefont {Terhal}}]{Bravyi2010}%
  \BibitemOpen
  \bibfield  {author} {\bibinfo {author} {\bibfnamefont {S.}~\bibnamefont {Bravyi}}, \bibinfo {author} {\bibfnamefont {D.}~\bibnamefont {Poulin}},\ and\ \bibinfo {author} {\bibfnamefont {B.}~\bibnamefont {Terhal}},\ }\bibfield  {title} {\bibinfo {title} {Tradeoffs for reliable quantum information storage in 2d systems},\ }\href {https://doi.org/10.1103/PhysRevLett.104.050503} {\bibfield  {journal} {\bibinfo  {journal} {Phys. Rev. Lett.}\ }\textbf {\bibinfo {volume} {104}},\ \bibinfo {pages} {050503} (\bibinfo {year} {2010})}\BibitemShut {NoStop}%
\bibitem [{\citenamefont {Bravyi}\ and\ \citenamefont {Terhal}(2009)}]{Bravyi2009}%
  \BibitemOpen
  \bibfield  {author} {\bibinfo {author} {\bibfnamefont {S.}~\bibnamefont {Bravyi}}\ and\ \bibinfo {author} {\bibfnamefont {B.}~\bibnamefont {Terhal}},\ }\bibfield  {title} {\bibinfo {title} {A no-go theorem for a two-dimensional self-correcting quantum memory based on stabilizer codes},\ }\href {https://doi.org/10.1088/1367-2630/11/4/043029} {\bibfield  {journal} {\bibinfo  {journal} {New Journal of Physics}\ }\textbf {\bibinfo {volume} {11}},\ \bibinfo {pages} {043029} (\bibinfo {year} {2009})}\BibitemShut {NoStop}%
\bibitem [{\citenamefont {Haah}(2021)}]{Haah2021}%
  \BibitemOpen
  \bibfield  {author} {\bibinfo {author} {\bibfnamefont {J.}~\bibnamefont {Haah}},\ }\bibfield  {title} {\bibinfo {title} {{A degeneracy bound for homogeneous topological order}},\ }\href {https://doi.org/10.21468/SciPostPhys.10.1.011} {\bibfield  {journal} {\bibinfo  {journal} {SciPost Phys.}\ }\textbf {\bibinfo {volume} {10}},\ \bibinfo {pages} {011} (\bibinfo {year} {2021})}\BibitemShut {NoStop}%
\bibitem [{\citenamefont {Portnoy}(2023)}]{Portnoy2023}%
  \BibitemOpen
  \bibfield  {author} {\bibinfo {author} {\bibfnamefont {E.}~\bibnamefont {Portnoy}},\ }\href {https://arxiv.org/abs/2303.06755} {\bibinfo {title} {Local quantum codes from subdivided manifolds}} (\bibinfo {year} {2023}),\ \Eprint {https://arxiv.org/abs/2303.06755} {arXiv:2303.06755 [quant-ph]} \BibitemShut {NoStop}%
\bibitem [{\citenamefont {Lin}\ \emph {et~al.}(2024)\citenamefont {Lin}, \citenamefont {Wills},\ and\ \citenamefont {Hsieh}}]{Lin2024}%
  \BibitemOpen
  \bibfield  {author} {\bibinfo {author} {\bibfnamefont {T.-C.}\ \bibnamefont {Lin}}, \bibinfo {author} {\bibfnamefont {A.}~\bibnamefont {Wills}},\ and\ \bibinfo {author} {\bibfnamefont {M.-H.}\ \bibnamefont {Hsieh}},\ }\href {https://arxiv.org/abs/2309.16104} {\bibinfo {title} {Geometrically local quantum and classical codes from subdivision}} (\bibinfo {year} {2024}),\ \Eprint {https://arxiv.org/abs/2309.16104} {arXiv:2309.16104 [quant-ph]} \BibitemShut {NoStop}%
\bibitem [{\citenamefont {Freedman}\ and\ \citenamefont {Hastings}(2021)}]{Freedman2021}%
  \BibitemOpen
  \bibfield  {author} {\bibinfo {author} {\bibfnamefont {M.}~\bibnamefont {Freedman}}\ and\ \bibinfo {author} {\bibfnamefont {M.~B.}\ \bibnamefont {Hastings}},\ }\href {https://arxiv.org/abs/2012.02249} {\bibinfo {title} {Building manifolds from quantum codes}} (\bibinfo {year} {2021}),\ \Eprint {https://arxiv.org/abs/2012.02249} {arXiv:2012.02249 [math.DG]} \BibitemShut {NoStop}%
\bibitem [{\citenamefont {Bravyi}\ \emph {et~al.}(2024)\citenamefont {Bravyi}, \citenamefont {Cross}, \citenamefont {Gambetta}, \citenamefont {Maslov}, \citenamefont {Rall},\ and\ \citenamefont {Yoder}}]{Bravyi2024}%
  \BibitemOpen
  \bibfield  {author} {\bibinfo {author} {\bibfnamefont {S.}~\bibnamefont {Bravyi}}, \bibinfo {author} {\bibfnamefont {A.~W.}\ \bibnamefont {Cross}}, \bibinfo {author} {\bibfnamefont {J.~M.}\ \bibnamefont {Gambetta}}, \bibinfo {author} {\bibfnamefont {D.}~\bibnamefont {Maslov}}, \bibinfo {author} {\bibfnamefont {P.}~\bibnamefont {Rall}},\ and\ \bibinfo {author} {\bibfnamefont {T.~J.}\ \bibnamefont {Yoder}},\ }\bibfield  {title} {\bibinfo {title} {High-threshold and low-overhead fault-tolerant quantum memory},\ }\href {https://doi.org/10.1038/s41586-024-07107-7} {\bibfield  {journal} {\bibinfo  {journal} {Nature}\ }\textbf {\bibinfo {volume} {627}},\ \bibinfo {pages} {778} (\bibinfo {year} {2024})}\BibitemShut {NoStop}%
\bibitem [{\citenamefont {Tremblay}\ \emph {et~al.}(2022)\citenamefont {Tremblay}, \citenamefont {Delfosse},\ and\ \citenamefont {Beverland}}]{Tremblay2022}%
  \BibitemOpen
  \bibfield  {author} {\bibinfo {author} {\bibfnamefont {M.~A.}\ \bibnamefont {Tremblay}}, \bibinfo {author} {\bibfnamefont {N.}~\bibnamefont {Delfosse}},\ and\ \bibinfo {author} {\bibfnamefont {M.~E.}\ \bibnamefont {Beverland}},\ }\bibfield  {title} {\bibinfo {title} {Constant-overhead quantum error correction with thin planar connectivity},\ }\href {https://doi.org/10.1103/PhysRevLett.129.050504} {\bibfield  {journal} {\bibinfo  {journal} {Phys. Rev. Lett.}\ }\textbf {\bibinfo {volume} {129}},\ \bibinfo {pages} {050504} (\bibinfo {year} {2022})}\BibitemShut {NoStop}%
\bibitem [{\citenamefont {Quintavalle}\ \emph {et~al.}(2021)\citenamefont {Quintavalle}, \citenamefont {Vasmer}, \citenamefont {Roffe},\ and\ \citenamefont {Campbell}}]{Quintavalle2021}%
  \BibitemOpen
  \bibfield  {author} {\bibinfo {author} {\bibfnamefont {A.~O.}\ \bibnamefont {Quintavalle}}, \bibinfo {author} {\bibfnamefont {M.}~\bibnamefont {Vasmer}}, \bibinfo {author} {\bibfnamefont {J.}~\bibnamefont {Roffe}},\ and\ \bibinfo {author} {\bibfnamefont {E.~T.}\ \bibnamefont {Campbell}},\ }\bibfield  {title} {\bibinfo {title} {Single-shot error correction of three-dimensional homological product codes},\ }\href {https://doi.org/10.1103/PRXQuantum.2.020340} {\bibfield  {journal} {\bibinfo  {journal} {PRX Quantum}\ }\textbf {\bibinfo {volume} {2}},\ \bibinfo {pages} {020340} (\bibinfo {year} {2021})}\BibitemShut {NoStop}%
\bibitem [{\citenamefont {Fawzi}\ \emph {et~al.}(2018)\citenamefont {Fawzi}, \citenamefont {Grospellier},\ and\ \citenamefont {Leverrier}}]{Fawzi2018}%
  \BibitemOpen
  \bibfield  {author} {\bibinfo {author} {\bibfnamefont {O.}~\bibnamefont {Fawzi}}, \bibinfo {author} {\bibfnamefont {A.}~\bibnamefont {Grospellier}},\ and\ \bibinfo {author} {\bibfnamefont {A.}~\bibnamefont {Leverrier}},\ }\bibfield  {title} {\bibinfo {title} {Constant overhead quantum fault-tolerance with quantum expander codes},\ }in\ \href {https://doi.org/10.1109/FOCS.2018.00076} {\emph {\bibinfo {booktitle} {2018 IEEE 59th Annual Symposium on Foundations of Computer Science (FOCS)}}}\ (\bibinfo {year} {2018})\ pp.\ \bibinfo {pages} {743--754}\BibitemShut {NoStop}%
\bibitem [{\citenamefont {Gu}\ \emph {et~al.}(2024)\citenamefont {Gu}, \citenamefont {Tang}, \citenamefont {Caha}, \citenamefont {Choe}, \citenamefont {He},\ and\ \citenamefont {Kubica}}]{Gu2024}%
  \BibitemOpen
  \bibfield  {author} {\bibinfo {author} {\bibfnamefont {S.}~\bibnamefont {Gu}}, \bibinfo {author} {\bibfnamefont {E.}~\bibnamefont {Tang}}, \bibinfo {author} {\bibfnamefont {L.}~\bibnamefont {Caha}}, \bibinfo {author} {\bibfnamefont {S.~H.}\ \bibnamefont {Choe}}, \bibinfo {author} {\bibfnamefont {Z.}~\bibnamefont {He}},\ and\ \bibinfo {author} {\bibfnamefont {A.}~\bibnamefont {Kubica}},\ }\bibfield  {title} {\bibinfo {title} {Single-shot decoding of good quantum ldpc codes},\ }\href {https://doi.org/10.1007/s00220-024-04951-6} {\bibfield  {journal} {\bibinfo  {journal} {Communications in Mathematical Physics}\ }\textbf {\bibinfo {volume} {405}},\ \bibinfo {pages} {85} (\bibinfo {year} {2024})}\BibitemShut {NoStop}%
\bibitem [{\citenamefont {Krishna}\ and\ \citenamefont {Poulin}(2021)}]{Krishna2021}%
  \BibitemOpen
  \bibfield  {author} {\bibinfo {author} {\bibfnamefont {A.}~\bibnamefont {Krishna}}\ and\ \bibinfo {author} {\bibfnamefont {D.}~\bibnamefont {Poulin}},\ }\bibfield  {title} {\bibinfo {title} {Fault-tolerant gates on hypergraph product codes},\ }\href {https://doi.org/10.1103/PhysRevX.11.011023} {\bibfield  {journal} {\bibinfo  {journal} {Phys. Rev. X}\ }\textbf {\bibinfo {volume} {11}},\ \bibinfo {pages} {011023} (\bibinfo {year} {2021})}\BibitemShut {NoStop}%
\bibitem [{\citenamefont {Cohen}\ \emph {et~al.}(2022)\citenamefont {Cohen}, \citenamefont {Kim}, \citenamefont {Bartlett},\ and\ \citenamefont {Brown}}]{Cohen2022}%
  \BibitemOpen
  \bibfield  {author} {\bibinfo {author} {\bibfnamefont {L.~Z.}\ \bibnamefont {Cohen}}, \bibinfo {author} {\bibfnamefont {I.~H.}\ \bibnamefont {Kim}}, \bibinfo {author} {\bibfnamefont {S.~D.}\ \bibnamefont {Bartlett}},\ and\ \bibinfo {author} {\bibfnamefont {B.~J.}\ \bibnamefont {Brown}},\ }\bibfield  {title} {\bibinfo {title} {Low-overhead fault-tolerant quantum computing using long-range connectivity},\ }\href {https://doi.org/10.1126/sciadv.abn1717} {\bibfield  {journal} {\bibinfo  {journal} {Science Advances}\ }\textbf {\bibinfo {volume} {8}},\ \bibinfo {pages} {eabn1717} (\bibinfo {year} {2022})}\BibitemShut {NoStop}%
\bibitem [{\citenamefont {Poulsen~Nautrup}\ \emph {et~al.}(2017)\citenamefont {Poulsen~Nautrup}, \citenamefont {Friis},\ and\ \citenamefont {Briegel}}]{Nautrup2017}%
  \BibitemOpen
  \bibfield  {author} {\bibinfo {author} {\bibfnamefont {H.}~\bibnamefont {Poulsen~Nautrup}}, \bibinfo {author} {\bibfnamefont {N.}~\bibnamefont {Friis}},\ and\ \bibinfo {author} {\bibfnamefont {H.~J.}\ \bibnamefont {Briegel}},\ }\bibfield  {title} {\bibinfo {title} {Fault-tolerant interface between quantum memories and quantum processors},\ }\href {https://doi.org/10.1038/s41467-017-01418-2} {\bibfield  {journal} {\bibinfo  {journal} {Nature Communications}\ }\textbf {\bibinfo {volume} {8}},\ \bibinfo {pages} {1321} (\bibinfo {year} {2017})}\BibitemShut {NoStop}%
\bibitem [{\citenamefont {Horsman}\ \emph {et~al.}(2012)\citenamefont {Horsman}, \citenamefont {Fowler}, \citenamefont {Devitt},\ and\ \citenamefont {Meter}}]{Horsman2012}%
  \BibitemOpen
  \bibfield  {author} {\bibinfo {author} {\bibfnamefont {D.}~\bibnamefont {Horsman}}, \bibinfo {author} {\bibfnamefont {A.~G.}\ \bibnamefont {Fowler}}, \bibinfo {author} {\bibfnamefont {S.}~\bibnamefont {Devitt}},\ and\ \bibinfo {author} {\bibfnamefont {R.~V.}\ \bibnamefont {Meter}},\ }\bibfield  {title} {\bibinfo {title} {Surface code quantum computing by lattice surgery},\ }\href {https://doi.org/10.1088/1367-2630/14/12/123011} {\bibfield  {journal} {\bibinfo  {journal} {New Journal of Physics}\ }\textbf {\bibinfo {volume} {14}},\ \bibinfo {pages} {123011} (\bibinfo {year} {2012})}\BibitemShut {NoStop}%
\bibitem [{\citenamefont {Breuckmann}\ \emph {et~al.}(2017)\citenamefont {Breuckmann}, \citenamefont {Vuillot}, \citenamefont {Campbell}, \citenamefont {Krishna},\ and\ \citenamefont {Terhal}}]{Breuckmann2017}%
  \BibitemOpen
  \bibfield  {author} {\bibinfo {author} {\bibfnamefont {N.~P.}\ \bibnamefont {Breuckmann}}, \bibinfo {author} {\bibfnamefont {C.}~\bibnamefont {Vuillot}}, \bibinfo {author} {\bibfnamefont {E.}~\bibnamefont {Campbell}}, \bibinfo {author} {\bibfnamefont {A.}~\bibnamefont {Krishna}},\ and\ \bibinfo {author} {\bibfnamefont {B.~M.}\ \bibnamefont {Terhal}},\ }\bibfield  {title} {\bibinfo {title} {Hyperbolic and semi-hyperbolic surface codes for quantum storage},\ }\href {https://doi.org/10.1088/2058-9565/aa7d3b} {\bibfield  {journal} {\bibinfo  {journal} {Quantum Science and Technology}\ }\textbf {\bibinfo {volume} {2}},\ \bibinfo {pages} {035007} (\bibinfo {year} {2017})}\BibitemShut {NoStop}%
\bibitem [{\citenamefont {Gottesman}(2022)}]{Gottesman2022}%
  \BibitemOpen
  \bibfield  {author} {\bibinfo {author} {\bibfnamefont {D.}~\bibnamefont {Gottesman}},\ }\href {https://arxiv.org/abs/2210.15844} {\bibinfo {title} {Opportunities and challenges in fault-tolerant quantum computation}} (\bibinfo {year} {2022}),\ \Eprint {https://arxiv.org/abs/2210.15844} {arXiv:2210.15844 [quant-ph]} \BibitemShut {NoStop}%
\bibitem [{\citenamefont {{\L}odyga}\ \emph {et~al.}(2015)\citenamefont {{\L}odyga}, \citenamefont {Mazurek}, \citenamefont {Grudka},\ and\ \citenamefont {Horodecki}}]{Lodyga2015}%
  \BibitemOpen
  \bibfield  {author} {\bibinfo {author} {\bibfnamefont {J.}~\bibnamefont {{\L}odyga}}, \bibinfo {author} {\bibfnamefont {P.}~\bibnamefont {Mazurek}}, \bibinfo {author} {\bibfnamefont {A.}~\bibnamefont {Grudka}},\ and\ \bibinfo {author} {\bibfnamefont {M.}~\bibnamefont {Horodecki}},\ }\bibfield  {title} {\bibinfo {title} {Simple scheme for encoding and decoding a qubit in unknown state for various topological codes},\ }\href {https://doi.org/10.1038/srep08975} {\bibfield  {journal} {\bibinfo  {journal} {Scientific Reports}\ }\textbf {\bibinfo {volume} {5}},\ \bibinfo {pages} {8975} (\bibinfo {year} {2015})}\BibitemShut {NoStop}%
\bibitem [{\citenamefont {Raussendorf}\ \emph {et~al.}(2005)\citenamefont {Raussendorf}, \citenamefont {Bravyi},\ and\ \citenamefont {Harrington}}]{Raussendorf2005}%
  \BibitemOpen
  \bibfield  {author} {\bibinfo {author} {\bibfnamefont {R.}~\bibnamefont {Raussendorf}}, \bibinfo {author} {\bibfnamefont {S.}~\bibnamefont {Bravyi}},\ and\ \bibinfo {author} {\bibfnamefont {J.}~\bibnamefont {Harrington}},\ }\bibfield  {title} {\bibinfo {title} {Long-range quantum entanglement in noisy cluster states},\ }\href {https://doi.org/10.1103/PhysRevA.71.062313} {\bibfield  {journal} {\bibinfo  {journal} {Phys. Rev. A}\ }\textbf {\bibinfo {volume} {71}},\ \bibinfo {pages} {062313} (\bibinfo {year} {2005})}\BibitemShut {NoStop}%
\bibitem [{\citenamefont {Patil}\ \emph {et~al.}(2024)\citenamefont {Patil}, \citenamefont {Pacenti}, \citenamefont {Vasić}, \citenamefont {Guha},\ and\ \citenamefont {Rengaswamy}}]{Patil2024}%
  \BibitemOpen
  \bibfield  {author} {\bibinfo {author} {\bibfnamefont {A.}~\bibnamefont {Patil}}, \bibinfo {author} {\bibfnamefont {M.}~\bibnamefont {Pacenti}}, \bibinfo {author} {\bibfnamefont {B.}~\bibnamefont {Vasić}}, \bibinfo {author} {\bibfnamefont {S.}~\bibnamefont {Guha}},\ and\ \bibinfo {author} {\bibfnamefont {N.}~\bibnamefont {Rengaswamy}},\ }\href@noop {} {\bibinfo {title} {Entanglement routing using quantum error correction for distillation}} (\bibinfo {year} {2024}),\ \Eprint {https://arxiv.org/abs/2405.00849} {arXiv:2405.00849 [quant-ph]} \BibitemShut {NoStop}%
\bibitem [{\citenamefont {Milligen}\ \emph {et~al.}(2024)\citenamefont {Milligen}, \citenamefont {Jacobson}, \citenamefont {Patil}, \citenamefont {Vardoyan}, \citenamefont {Towsley},\ and\ \citenamefont {Guha}}]{Milligen2024}%
  \BibitemOpen
  \bibfield  {author} {\bibinfo {author} {\bibfnamefont {E.~A.~V.}\ \bibnamefont {Milligen}}, \bibinfo {author} {\bibfnamefont {E.}~\bibnamefont {Jacobson}}, \bibinfo {author} {\bibfnamefont {A.}~\bibnamefont {Patil}}, \bibinfo {author} {\bibfnamefont {G.}~\bibnamefont {Vardoyan}}, \bibinfo {author} {\bibfnamefont {D.}~\bibnamefont {Towsley}},\ and\ \bibinfo {author} {\bibfnamefont {S.}~\bibnamefont {Guha}},\ }\href {https://arxiv.org/abs/2308.15028} {\bibinfo {title} {Entanglement routing over networks with time multiplexed repeaters}} (\bibinfo {year} {2024}),\ \Eprint {https://arxiv.org/abs/2308.15028} {arXiv:2308.15028 [quant-ph]} \BibitemShut {NoStop}%
\end{thebibliography}%

\end{document}